\documentclass[journal=jctcce,manuscript=article]{achemso}
\setkeys{acs}{usetitle=true}

\usepackage[version=3]{mhchem} 

\usepackage{color, xcolor, colortbl}
\usepackage{graphicx,epstopdf}
\usepackage{geometry}
\usepackage{amsmath,amssymb,amsthm}
\usepackage{algorithm}
\usepackage{algorithmic}
\usepackage{bm}
\usepackage[caption=false]{subfig}
\usepackage{appendix}
\usepackage{multirow}
\usepackage{braket}
\usepackage[english]{babel}
\usepackage{hyperref}
\usepackage[capitalize]{cleveref}
\usepackage{adjustbox}
\usepackage{xspace}
\usepackage[roman]{complexity}
\usepackage{comment}
\usepackage[normalem]{ulem}

\newcommand{\bvec}[1]{\mathbf{#1}}

\newcommand{\vk}{\bvec{k}}

\newcommand{\vq}{\bvec{q}}
\newcommand{\vr}{\bvec{r}}

\newcommand{\vG}{\bvec{G}}

\newcommand{\conj}[1]{#1^*}

\newcommand{\xsum}{\mathop{\sum\nolimits'}}

\newcommand{\I}{\mathrm{i}}

\newcommand{\mc}[1]{\mathcal{#1}}

\newcommand{\abs}[1]{\left\lvert#1\right\rvert}

\newcommand{\ud}{\,\mathrm{d}}
\newcommand{\Or}{\mathcal{O}}

\newcommand{\REV}[1]{{#1}}

\title{Staggered mesh method for correlation energy
calculations of solids: Random phase approximation in direct ring coupled cluster doubles and adiabatic connection formalisms}

\author{Xin Xing}
\affiliation{Department of Mathematics, University of California,
Berkeley, CA 94720, USA}

\author{Lin Lin}
\email{linlin@math.berkeley.edu}
\affiliation{Department of Mathematics, University of California,
Berkeley, CA 94720, USA}
\alsoaffiliation{Computational Research Division, Lawrence Berkeley
National Laboratory, Berkeley, Berkeley, CA 94720, USA}
\alsoaffiliation{Challenge Institute for Quantum Computation, University of California, Berkeley, CA 94720, USA}

\begin{document}

\begin{abstract}
We propose a staggered mesh method for correlation energy calculations of periodic systems under the random phase approximation (RPA), which generalizes the recently developed staggered mesh method for periodic second order M{\o}ller--Plesset perturbation theory (MP2) calculations [Xing, Li, Lin, JCTC 2021]. 
Compared to standard RPA calculations, the staggered mesh method
introduces negligible additional computational cost. 
It avoids a significant portion of the finite-size error, and can be asymptotically advantageous for quasi-1D  systems and certain quasi-2D and 3D systems with high symmetries. 
We demonstrate the applicability of the method using two different formalisms: the direct ring coupled cluster doubles (drCCD) theory, and the adiabatic-connection (AC) fluctuation-dissipation theory.
In the drCCD formalism, the second order screened exchange (SOSEX) correction can also be readily obtained using the staggered mesh method. 
In the AC formalism, the staggered mesh method naturally avoids the need of performing ``head/wing'' corrections to the dielectric operator. 
The effectiveness of the staggered mesh method  for insulating systems  is theoretically justified by investigating the finite-size error of each individual perturbative term  in the RPA correlation energy, expanded as an infinite series of terms associated with ring diagrams.
As a side contribution, our analysis provides a proof that the finite-size error of each perturbative term of standard RPA and SOSEX calculations scales as $\mathcal{O}(N_\bvec{k}^{-1})$, where $N_\bvec{k}$ is the number of grid points in a Monkhorst--Pack mesh.
\end{abstract}

\section{Introduction}

Correlated wavefunction based methods are nowadays increasingly widely used for calculations of periodic systems\cite{MarsmanGruneisPaierKresse2009, GruneisMarsmanKresse2010, MuellerPaulus2012, McClainSunChanEtAl2017}, thanks to the improvement of numerical algorithms and the increase of computational powers. 
Since physical observables are defined in the thermodynamic limit (TDL), one key issue is to understand and to correct the finite-size error. 
In the past few decades, significant progresses have been achieved in the understanding of the finite-size error in density functional theory (DFT) and Hartree--Fock theory\cite{CarrierRohraGorling2007, GygiBaldereschi1986, MakovPayne1995}, wavefunction methods\cite{LiaoGrueneis2016, GruberLiaoTsatsoulisEtAl2018, MihmMcIsaacShepherd2019}, and quantum Monte Carlo methods~\cite{ChiesaCeperleyMartinEtAl2006, FraserFoulkesRajagopalEtAl1996,LinZongCeperley2001, DrummondNeedsSorouriEtAl2008, SpencerAlavi2008,SundararamanArias2013}.
Despite such progresses, there has been very limited rigorous understanding of the scaling of finite-size errors for general systems.  To our knowledge, our recent work~\cite{XingLiLin2021_2} accounts for the first rigorous treatment of finite-size errors in periodic Hartree--Fock exchange energy calculations, and periodic second order M{\o}ller--Plesset correlation energy (MP2) calculations (which are among the simpest wavefunction methods) for general insulating systems.
The key technical difficulty of analyzing the convergence towards the thermodynamic limit boils down to an analysis of quadrature error for a special class of nonsmooth integrands (Corollary 16 in Ref.~\citenum{XingLiLin2021_2}).  
The analysis also suggests a new algorithm for reducing the finite-size error, which employs \emph{two} staggered Monkhorst--Pack meshes for occupied and virtual orbitals, respectively. This naturally avoids the error contribution due to the zero momentum transfer between the occupied and virtual orbitals, and can significantly reduce the finite-size errors in MP2 calculations~\cite{XingLiLin2021}.

The main contribution of this paper is to generalize the staggered mesh method and the numerical analysis to correlation energy calculations based on the random phase approximation (RPA).  In the past decade, there has been a revival interest in RPA and its variants for post-Hartree--Fock/post-Kohn--Sham correlation energy calculations \cite{EshuisBatesFurche2012, ChenVooraAgeeEtAl2017, RenRinkeJoasScheffler2012, HarlSchimkaKresse2010,HarlKresse2009, HarlKresse2008, GruneisMarsmanHarlEtAl2009}. The calculation of the RPA correlation energy has a long history dating back to 1950s\cite{BohmPines1951}. 
Since then, the concept of the ``RPA correlation energy'' has arisen in several different contexts with very different expressions, including  adiabatic-connection (AC) fluctuation-dissipation theory\cite{LangrethPerdew1975, LangrethPerdew1977,GunnarssonLundqvist1976},  time-dependent Hartree--Fock theory and time-dependent density functional theory\cite{Furche2008}, many-body Green's function theory\cite{DahlenLeeuwenBarth2006}, and direct ring coupled cluster doubles theory (drCCD)~\cite{Freeman1977,ScuseriaHendersonSorensen2008,ScuseriaHendersonBulik2013}.
Despite significant progresses in reducing the computational cost of RPA calculations\cite{Moussa2014,LuThicke2017,Kallay2015,RenRinkeBlumEtAl2012, GruneisMarsmanHarlEta2009, Hesselmann2012}, it can still be very expensive to approach the RPA correlation energy in the TDL by brute force. 
To our knowledge, the only finite-size correction schemes available for RPA calculations are the power-law extrapolation and the structure factor interpolation methods\cite{GruberLiaoTsatsoulisEtAl2018, LiaoGrueneis2016}.


Throughout this paper, we only focus on insulating systems.
We consider two equivalent formalisms (drCCD and AC) for RPA correlation energy calculations.
In both formalisms, the staggered mesh method can be directly employed, i.e., the occupied orbitals belong to one Monkhorst--Pack grid, and the virtual orbitals belong to a shifted Monkhorst--Pack grid.
Compared to standard RPA calculations, the additional cost introduced by the use of the staggered mesh is negligible. 
In the drCCD formalism, once the amplitude equations are solved, we can also readily obtain the second order screened exchange (SOSEX) correction\cite{GruneisMarsmanHarlEtAl2009} to RPA using the staggered mesh.
We compare the numerical performance of the standard and the staggered mesh methods for RPA and RPA-SOSEX correlation energy calculations on periodic hydrogen dimer, lithium hydride, silicon, and diamond in the quasi-1D, quasi-2D, and 3D bulk settings using the PySCF software package \cite{SunBerkelbachEtAl2018}.
Numerical results indicate that the use of staggered meshes can significantly accelerate the convergence towards the TDL in two scenarios: quasi-1D systems, and certain quasi-2D or 3D bulk systems with high symmetries. Such performance is consistent with that in MP2 energy calculations~\cite{XingLiLin2021}.
In the AC formalism, one additional advantage of the staggered mesh method is that it naturally avoids the need of the so-called ``head/wing'' corrections to the dielectric operator~\cite{ZhuChan2021,WilhelmHutter2017,FreysoldtEggertEta2007}.

Our analysis of the finite-size error in RPA calculations using the standard and the staggered mesh methods is mainly based on the drCCD formalism. 
Using a fixed point iteration of the amplitude equations, the RPA correlation energy can be expressed as an infinite series of terms associated with rings diagrams.
For each individual term, the nonsmooth integrand satisfies the requirement of Corollary 16 in Ref.~\citenum{XingLiLin2021_2}, and hence the associated finite-size error can be analyzed to be $\Or(N_\vk^{-1})$ for general systems. 
Furthermore, for quasi-1D systems and certain quasi-2D or 3D bulk systems with high symmetries, the finite-size error due to the nonsmooth integrand can be $o(N_\vk^{-1})$. However, the standard RPA calculation always places certain quadrature nodes at the points of discontinuity of the integrand, resulting another $\Or(N_\vk^{-1})$ quadrature error. 
Thus, the overall quadrature error of each energy term in standard RPA always scales as $\Or(N_\vk^{-1})$.
Such a scaling is also numerically observed in \cref{sec:numerical}.
The staggered mesh method completely avoids the error due to the improper placement of quadrature nodes, and the finite-size error can be improved to $o(N_\vk^{-1})$. This explains its superior performance over the standard method observed in the numerical results. 
Similar error analysis is also applicable to the second order screened exchange (SOSEX) correction.

\section{Background}
Throughout the paper, we consider the spin-restricted setting, and all the discussions can be generalized straightforwardly to the spin-unrestricted case.
Let $\Omega$ be the unit cell, $|\Omega|$ be its volume, and  $\Omega^*$ be the associated Brillouin zone (BZ). 
Let $\mathbb{L}$ and $\mathbb{L}^*$ be the Bravais lattice and its associated reciprocal lattice, respectively. 
In numerical calculations, we sample a uniform mesh $\mathcal{K}$ in $\Omega^*$ (called the Monkhorst--Pack (MP) mesh), and denote $N_\vk$ as the number of $\vk$-points in $\mathcal{K}$.
For a mean-field calculation with $\mathcal{K}$, each spatial molecular orbital is characterized by a $\vk$-point and a band index $n$, and written as 
\[
\psi_{n\vk}(\vr) = \dfrac{1}{\sqrt{N_\vk}} e^{\I \vk\cdot\vr} u_{n\vk}(\vr)=
\frac{1}{\abs{\Omega}\sqrt{N_{\vk}}} \sum_{\mathbf{G}\in\mathbb{L}^*} \hat{u}_{n\vk}(\mathbf{G}) e^{\I (\vk+\mathbf{G}) \cdot \mathbf{r}},
\]
 associated with an orbital energy $\varepsilon_{n\vk}$.
The pair product is defined as 
\[
\varrho_{n'\vk',n\vk}(\vr)=\conj{u}_{n'\vk'}(\vr)  u_{n\vk}(\vr) =\frac{1}{\abs{\Omega}} \sum_{\mathbf{G}\in\mathbb{L}^*} \hat{\varrho}_{n'\vk',n\vk}(\mathbf{G}) e^{\I \mathbf{G} \cdot \mathbf{r}},
\]
and a two-electron repulsion integral (ERI) is then computed as 
\begin{equation}\label{eqn:eri}
\braket{n_1\vk_1,n_2\vk_2|n_3\vk_3,n_4\vk_4}
=  \frac{1}{\abs{\Omega}N_\vk} \xsum_{\vG\in\mathbb{L}^*}
\frac{4\pi}{\abs{\vk_3 - \vk_1+\vG}^2}  
\hat{\varrho}_{n_1\vk_1,n_3\vk_3}(\mathbf{G}) \hat{\varrho}_{n_2\vk_2,n_4\vk_4}(\vG_{\vk_1,\vk_2}^{\vk_3,\vk_4}-\mathbf{G}),
\end{equation}
where $\vG_{\vk_1,\vk_2}^{\vk_3,\vk_4} = \vk_1+\vk_2-\vk_3-\vk_4$ and $\xsum_{\vG\in\mathbb{L}^*}$ excludes the possible term with $\vk_3-\vk_1+\vG=\bm{0}$. 
Such an ERI can be non-zero only when $\vG_{\vk_1,\vk_2}^{\vk_3,\vk_4} \in \mathbb{L}^*$, corresponding to the crystal momentum conservation. 
We use band indices $i,j,k,l$ ($a,b,c,d$) to refer to the occupied (virtual) bands, respectively.
In this paper, we only focus on systems with a direct gap between occupied and virtual orbital energies, i.e., $\varepsilon_{i\vk_i}  - \varepsilon_{a\vk_a} \leqslant -\varepsilon_g < 0$ for any $i,a$ and $\vk_i,\vk_a$.


In the drCCD formalism, RPA correlation energy calculations keep all the particle-hole ring contractions in the standard coupled cluster double amplitude equation, and remove all the exchange terms. 
Specifically, the drCCD amplitude equation can be written as  the following algebraic Riccati equation
\begin{align}
        &(\epsilon_{i\vk_i} + \epsilon_{j\vk_j}-\epsilon_{a\vk_a}- \epsilon_{b\vk_b} ) t_{i\vk_i,j\vk_j}^{a\vk_a,b\vk_b} =\braket{a\vk_a, b\vk_b|i\vk_i, j\vk_j}  + 
        2\sum_{kc}\sum_{\vk_k\in\mathcal{K}} \braket{k\vk_k,b\vk_b|c\vk_c,j\vk_j} t_{i\vk_i,k\vk_k}^{a\vk_a,c\vk_c}
        \nonumber \\
        & 
         + 2\sum_{kc}\sum_{\vk_k \in \mathcal{K}} \braket{a\vk_a,k\vk_k|i\vk_i,c\vk_c} t_{k\vk_k,j\vk_j}^{c\vk_c, b\vk_b}
        + 4\sum_{klcd}\sum_{\vk_k,\vk_l \in\mathcal{K}} \braket{k\vk_k,l\vk_l|c\vk_c,d\vk_d} t_{i\vk_i,k\vk_k}^{a\vk_a,c\vk_c} t_{l\vk_l,k\vk_k}^{d\vk_d, b\vk_b},
        \label{eqn:rdccd_amplitude}
\end{align}
for each set of band indices $i,j,a,b$, and momentum vectors $\vk_i,\vk_j,\vk_a, \vk_b \in \mathcal{K}$ that satisfy the crystal momentum conservation relation, i.e., $\vk_i+\vk_j-\vk_a-\vk_b \in\mathbb{L}^*$. 
Momentum vectors $\vk_c, \vk_d$ in the equation are restricted to be in $\mathcal{K}$ and are uniquely determined by other related momentum vectors using crystal momentum conservation. 
After solving the double amplitudes $t_{i\vk_i,j\vk_j}^{a\vk_a,b\vk_b}$, the drCCD correlation energy (which is also the  RPA correlation energy, or more precisely the direct RPA correlation energy~\cite{ScuseriaHendersonSorensen2008}) per unit cell is computed as
\begin{equation}\label{eqn:rdccd_energy}
        E_\text{rpa}(N_\vk) = \dfrac{1}{N_\vk}\sum_{ijab}\sum_{\vk_i,\vk_j,\vk_a\in \mathcal{K}}2\braket{i\vk_i, j\vk_j|a\vk_a, b\vk_b} t_{i\vk_i,j\vk_j}^{a\vk_a,b\vk_b}. 
\end{equation}
One advantage of the drCCD formalism is that once the $t$-amplitudes are computed, the second order screened exchange (SOSEX) correction can be readily obtained, so that the RPA-SOSEX correlation energy takes the form~\cite{GruneisMarsmanHarlEta2009}
        \begin{equation}\label{eqn:rdccd_sosex_energy}
                E_\text{rpa-sosex}(N_\vk) = \dfrac{1}{N_\vk}\sum_{ijab}\sum_{\vk_i,\vk_j,\vk_a\in \mathcal{K}}
                \left(
                2\braket{i\vk_i, j\vk_j|a\vk_a, b\vk_b} - \braket{i\vk_i, j\vk_j|b\vk_b, a\vk_a}
                \right)
                t_{i\vk_i,j\vk_j}^{a\vk_a,b\vk_b}.
        \end{equation}
Compared to the RPA correlation energy, the SOSEX correction adds an exchange energy term, and the additional computational cost is small. 

In the AC formalism, the RPA correlation energy is computed in terms of an integral along the imaginary energy axis as
\begin{equation}\label{eqn:acfdt_energy}
        E_\text{rpa}(N_\vk)=\dfrac{1}{N_\vk}\frac{1}{4 \pi} \int_{-\infty}^{\infty} \operatorname{Tr}\left(\log \left(1-\chi_{0}(\mathrm{i} \omega) v_{C}\right)+\chi_{0}(\mathrm{i} \omega) v_{C}\right) \mathrm{d} \omega. 
\end{equation}
Here $v_\text{C}$ is the Coulomb operator and $\chi_{0}(\I\omega)$ is the (independent particle) polarizability operator along the imaginary energy axis. In the real space, $\chi_{0}(\I\omega)$ takes the form
\begin{equation}\label{eqn:chi0}
        \chi_{0}\left(\mathbf{r}, \mathbf{r}^{\prime} ; \mathrm{i} \omega\right)=4 
        \sum_{ia} \sum_{\vk_i,\vk_a \in \mathcal{K}} \psi^*_{i \mathbf{k}_{i}}(\mathbf{r}) \psi^*_{a \mathbf{k}_{a}}\left(\mathbf{r}^{\prime}\right) \psi_{i \mathbf{k}_{i}}\left(\mathbf{r}^{\prime}\right) \psi_{a \mathbf{k}_{a}}(\mathbf{r}) \frac{\epsilon_{i \mathbf{k}_{i}}-\epsilon_{a \mathbf{k}_{a}}}{\omega^{2}+\left(\epsilon_{a \mathbf{k}_{a}}-\epsilon_{i \mathbf{k}_{i}}\right)^{2}}.
\end{equation}
In numerical calculation, this formalism requires discretizing the integral over $\omega$ by a numerical quadrature scheme and also discretizing the two operators in a finite basis set, such as a planewave basis set, for the trace computation. 
The drCCD  and the AC formalisms are equivalent in terms of the computation of $E_\text{rpa}$~\cite{JansenLiuAngyan2010}.
\REV{
Meanwhile, it is also possible to use the computation in the AC formalism to retrieve the exact $t$-amplitude in the drCCD formalism\cite{EshuisYarkonyFurche2010} and thus to compute $E_\text{rpa-sosex}$ according to \cref{eqn:rdccd_sosex_energy}.
}

\section{Staggered mesh method for RPA and RPA-SOSEX}\label{sec:method}

\begin{figure}[htbp]
        \centering
        \includegraphics[width=0.5\textwidth]{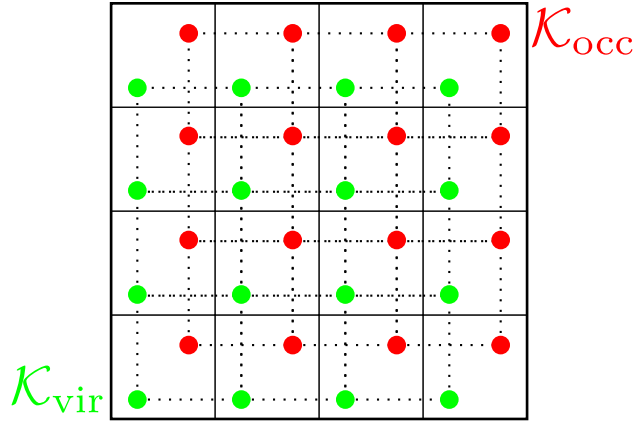}
        \caption{2D illustration of two staggered meshes $\mathcal{K}_\text{occ}$ and $\mathcal{K}_\text{vir}$.}
        \label{fig:stagger}
\end{figure}

So far all the momentum vectors are sampled on the same MP mesh $\mc{K}$.
A significant amount of the finite-size error is due to the zero momentum transfer between occupied and virtual orbitals. 
To reduce this finite-size error, 
in a recent work~\cite{XingLiLin2021} for MP2 correlation energy, we proposed to use two different but same-sized MP meshes $\mathcal{K}_\text{occ}$ and $\mathcal{K}_\text{vir}$ for the occupied and virtual orbitals, respectively. Here $\mathcal{K}_\text{occ}$ can be an arbitrarily chosen MP mesh, and $\mathcal{K}_\text{vir}$ is obtained by shifting $\mathcal{K}_\text{occ}$ in all extended directions by half mesh size (see \Cref{fig:stagger} for a 2D illustration). 
In the following, we apply the same staggered mesh method to RPA and RPA-SOSEX correlation energy calculations in both drCCD and AC formalisms. 
We first detail the procedure below. 
The analysis of this method will be given in \cref{sec:analysis}. 

The staggered mesh version of the drCCD amplitude equation  samples all occupied momentum vectors on $\mathcal{K}_\text{occ}$ and all virtual ones on $\mathcal{K}_\text{vir}$: 
\begin{align}
        &(\epsilon_{i\vk_i} + \epsilon_{j\vk_j}-\epsilon_{a\vk_a}- \epsilon_{b\vk_b} ) t_{i\vk_i,j\vk_j}^{a\vk_a,b\vk_b} =\braket{a\vk_a, b\vk_b|i\vk_i, j\vk_j}  + 
        2\sum_{kc}\sum_{\vk_k\in\mathcal{K}_\text{occ}} \braket{k\vk_k,b\vk_b|c\vk_c,j\vk_j} t_{i\vk_i,k\vk_k}^{a\vk_a,c\vk_c}
        \nonumber \\
        & 
        + 2\sum_{kc}\sum_{\vk_k \in \mathcal{K}_\text{occ}} \braket{a\vk_a,k\vk_k|i\vk_i,c\vk_c} t_{k\vk_k,j\vk_j}^{c\vk_c, b\vk_b}
        + 4\sum_{klcd}\sum_{\vk_k,\vk_l \in\mathcal{K}_\text{occ}} \braket{k\vk_k,l\vk_l|c\vk_c,d\vk_d} t_{i\vk_i,k\vk_k}^{a\vk_a,c\vk_c} t_{l\vk_l,k\vk_k}^{d\vk_d, b\vk_b},
        \label{eqn:rdccd_amplitude_stagger}
\end{align}
for each set of band indices $i,j,a,b$ and momentum vectors $\vk_i,\vk_j \in \mathcal{K}_\text{occ},\vk_a, \vk_b \in \mathcal{K}_\text{vir}$.
It is important to note that all the virtual momentum vectors  $\vk_c, \vk_d$ in the equation are restricted to be in $\mathcal{K}_\text{vir}$ and are
uniquely determined using other related momentum vectors by crystal momentum conservation. 
For example, both the ERI term $\braket{k\vk_k,b\vk_b|c\vk_c,j\vk_j}$ and the $t$-amplitude $t_{k\vk_k,j\vk_j}^{c\vk_c, b\vk_b}$ above have $\vk_k, \vk_j\in\mathcal{K}_\text{occ}$ and $\vk_b \in \mathcal{K}_\text{vir}$. 
Therefore the minimum image of $\vk_k+\vk_b - \vk_j$ defining $\vk_c$ always lies in $\mathcal{K}_\text{vir}$. 
The staggered mesh versions of the RPA and RPA-SOSEX correlation energies per unit cell are then calculated as
\begin{align}
        E_\text{rpa}^\text{stagger}(N_\vk) 
        & = \dfrac{1}{N_\vk}
        \sum_{ijab}\sum_{\vk_i,\vk_j\in\mathcal{K}_\text{occ}, \vk_a \in \mathcal{K}_\text{vir}} 
       2\braket{i\vk_i, j\vk_j|a\vk_a, b\vk_b} t_{i\vk_i,j\vk_j}^{a\vk_a,b\vk_b},
        \label{eqn:rpa_stagger}\\
        E_\text{rpa-sosex}^\text{stagger}(N_\vk) 
        & = 
        \dfrac{1}{N_\vk}
                \sum_{ijab}\sum_{\vk_i,\vk_j\in\mathcal{K}_\text{occ}, \vk_a \in \mathcal{K}_\text{vir}} 
        \left(2\braket{i\vk_i,j\vk_j|a\vk_a,b\vk_b} - \braket{i\vk_i,j\vk_j|b\vk_b, a\vk_a}\right)t_{i\vk_ij\vk_j}^{a\vk_ab\vk_b},
        \label{eqn:rpa_sosex_stagger}
\end{align}
where momentum vector $\vk_b$ is in $\mathcal{K}_\text{vir}$ and uniquely determined by $\vk_i,\vk_j,\vk_a$.

Similarly, the staggered mesh version of the AC formalism can be written as 
\begin{equation}\label{eqn:acfdt_energy_stagger}
        E_\text{rpa}^\text{stagger}(N_\vk)=\dfrac{1}{N_\vk}\frac{1}{4 \pi} \int_{-\infty}^{\infty} \operatorname{Tr}\left(\log \left(1-\chi_{0}^\text{stagger}(\mathrm{i} \omega) v_{C}\right)+\chi_{0}^\text{stagger}(\mathrm{i} \omega) v_{C}\right) \mathrm{d} \omega. 
\end{equation}
Here the only difference is that the polarizability operator is computed with two staggered meshes, i.e., the occupied momentum vectors are taken from $\mathcal{K}_\text{occ}$ and all virtual ones from $\mathcal{K}_\text{vir}$ as 
\begin{equation*}\label{eqn:chi0_stagger}
        \chi_{0}^\text{stagger}\left(\mathbf{r}, \mathbf{r}^{\prime} ; \mathrm{i} \omega\right)=4 
        \sum_{ia} \sum_{\vk_i\in\mathcal{K}_\text{occ},\vk_a \in \mathcal{K}_\text{vir}} \psi^*_{i \mathbf{k}_{i}}(\mathbf{r}) \psi^*_{a \mathbf{k}_{a}}\left(\mathbf{r}^{\prime}\right) \psi_{i \mathbf{k}_{i}}\left(\mathbf{r}^{\prime}\right) \psi_{a \mathbf{k}_{a}}(\mathbf{r}) \frac{\epsilon_{i \mathbf{k}_{i}}-\epsilon_{a \mathbf{k}_{a}}}{\omega^{2}+\left(\epsilon_{a \mathbf{k}_{a}}-\epsilon_{i \mathbf{k}_{i}}\right)^{2}}.
\end{equation*}

To implement  the staggered mesh method, we solve the self-consistent mean-field equation using one MP mesh, say $\mathcal{K}_\text{occ}$, and then evaluate the orbitals and orbital energies on the shifted mesh $\mathcal{K}_\text{vir}$  by solving the mean-field equations non-self-consistently on $\mathcal{K}_\text{vir}$. 
Another option is to solve the self-consistent mean-field equation using a larger MP mesh that contains $\mathcal{K}_\text{occ}$ and $\mathcal{K}_\text{vir}$ as two sub-meshes. 
The additional cost from the non-self-consistent calculation or the self-consistent calculation with a larger MP mesh can be negligible, compared to the computational cost of the RPA/RPA-SOSEX correlation energy calculations. 
The remaining computational cost of the staggered  mesh method is the same as that of the standard method.

For both the standard and the staggered mesh methods (using either drCCD or AC formalisms), direct calculation of the RPA and RPA-SOSEX correlation energies has $\Or(N^6)$  computational cost with respect to system size $N$. 
It is worth pointing out that the staggered mesh method can also be combined with existing acceleration techniques developed for the standard method, e.g., resolution of identity\cite{LuThicke2017, Moussa2014, GruneisMarsmanHarlEtAl2009,SunBerkelbachMcClainEtAl2017, ZhuChan2021, Hesselmann2012,RenRinkeBlumEtAl2012}, to reduce the computational cost to $\Or(N^4)$ or even less.

\section{Numerical examples}\label{sec:numerical}
In practical calculations, the finite-size error in molecular orbitals and orbital energies at the Hartree--Fock level also contributes to the overall finite-size errors in the standard and the staggered mesh methods.
This error can be reduced using other correction methods \cite{McClainSunChanEtAl2017, XingLiLin2021_2, SpencerAlavi2008,SundararamanArias2013} or a large MP mesh in mean-field calculations. 
We will not consider these improvements, and focus on numerical comparisons between the standard and the staggered mesh methods for RPA and RPA-SOSEX energy calculations.
For a given system, we first perform a self-consistent HF calculation with a fixed $\vk$-point mesh. 
All orbitals and orbital energies used in RPA and RPA-SOSEX energy calculations are then evaluated via non-self-consistent HF\ calculations at any required $\vk$ points and mesh sizes.
In this way, orbitals and orbital energies are generated from a fixed effective (non-local) potential field, and do not require further correction to the finite-size errors.

We consider four sets of periodic systems: hydrogen dimer, lithium hydride, silicon, and diamond.
The hydrogen dimer is placed at the center of a cubic unit cell of edge (the lattice constant is $6$ Bohr) pointing in the $x$-direction, and the bond distance is $1.8$ Bohr. 
For lithium hydride, silicon, and diamond systems, we use primitive unit cells each containing 2 atoms.
The reference HF calculations for all the tests use a $3\times 3\times 3$ $\Gamma$-centered MP mesh. 
All the tested systems are insulating systems with a direct gap between occupied and virtual orbitals.

We implement the staggered mesh method in the PySCF \cite{SunBerkelbachEtAl2018} software package.
Our implementation of the drCCD method using the standard and staggered meshes follow that of the ``kccsd'' module in PySCF. Our implementation of the staggered mesh AC method uses the PyRPA package~\cite{Zhu_pyrpa}, which is modified from the recent implementation of all-electron G$_0$W$_0$ method in PySCF \cite{ZhuChan2021}.
Our AC implementation by PyRPA is accelerated by Gaussian density fitting~\cite{SunBerkelbachMcClainEtAl2017}, which scales as $\Or(N^4)$. This is asymptotically faster than our current drCCD implementation, which scales as $\Or(N^6)$.
In all the AC calculations, a modified Gauss-Legendre grid\cite{RenRinkeBlumEtAl2012} with $40$ grid points are used for the integration over the imaginary energy axis. 
In all the tests below, we employ the gth-szv basis set and the gth-pade pseudopotential. 
Additional results with the larger gth-dzvp basis set are given in Appendix.
 The kinetic energy cutoff for plane-wave calculations is set to $100$ a.u.
We employ the spherical cutoff method \cite{SpencerAlavi2008} (given by the option exxdiv=`vcut\_sph' in PySCF) in the reference HF calculations to reduce the finite-size error due to the Fock exchange operator.
We always use a $\Gamma$-centered MP mesh for virtual orbitals. The standard method uses the same MP mesh for occupied orbitals.
The staggered mesh method shifts the MP mesh by half mesh size for occupied orbitals. For quasi-1D, quasi-2D, and 3D systems, the MP meshes are  of size $1\times 1\times N_\vk$, $1\times N_\vk^{1/2} \times N_\vk^{1/2}$, and 
$N_\vk^{1/3}\times N_\vk^{1/3} \times N_\vk^{1/3}$, respectively.
Atomic units are used in all the tests.

\begin{figure}[htbp]
        \centering
        \captionsetup[subfigure]{labelformat=empty}
        \subfloat[Diamond (quasi-1D)]{
                \includegraphics[width=0.4\textwidth]{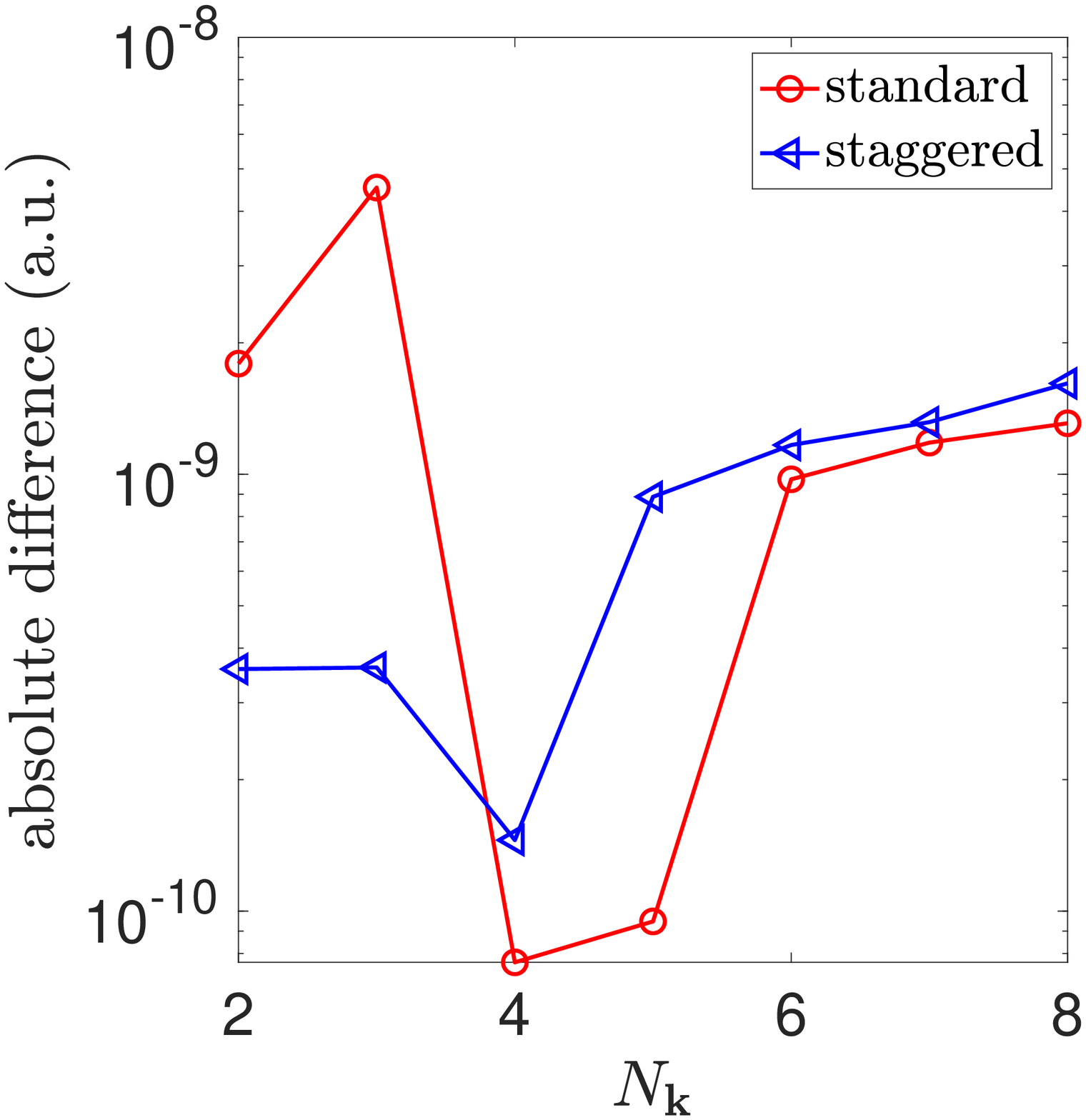}
        }
        \subfloat[H2 (quasi-2D)]{
                \includegraphics[width=0.4\textwidth]{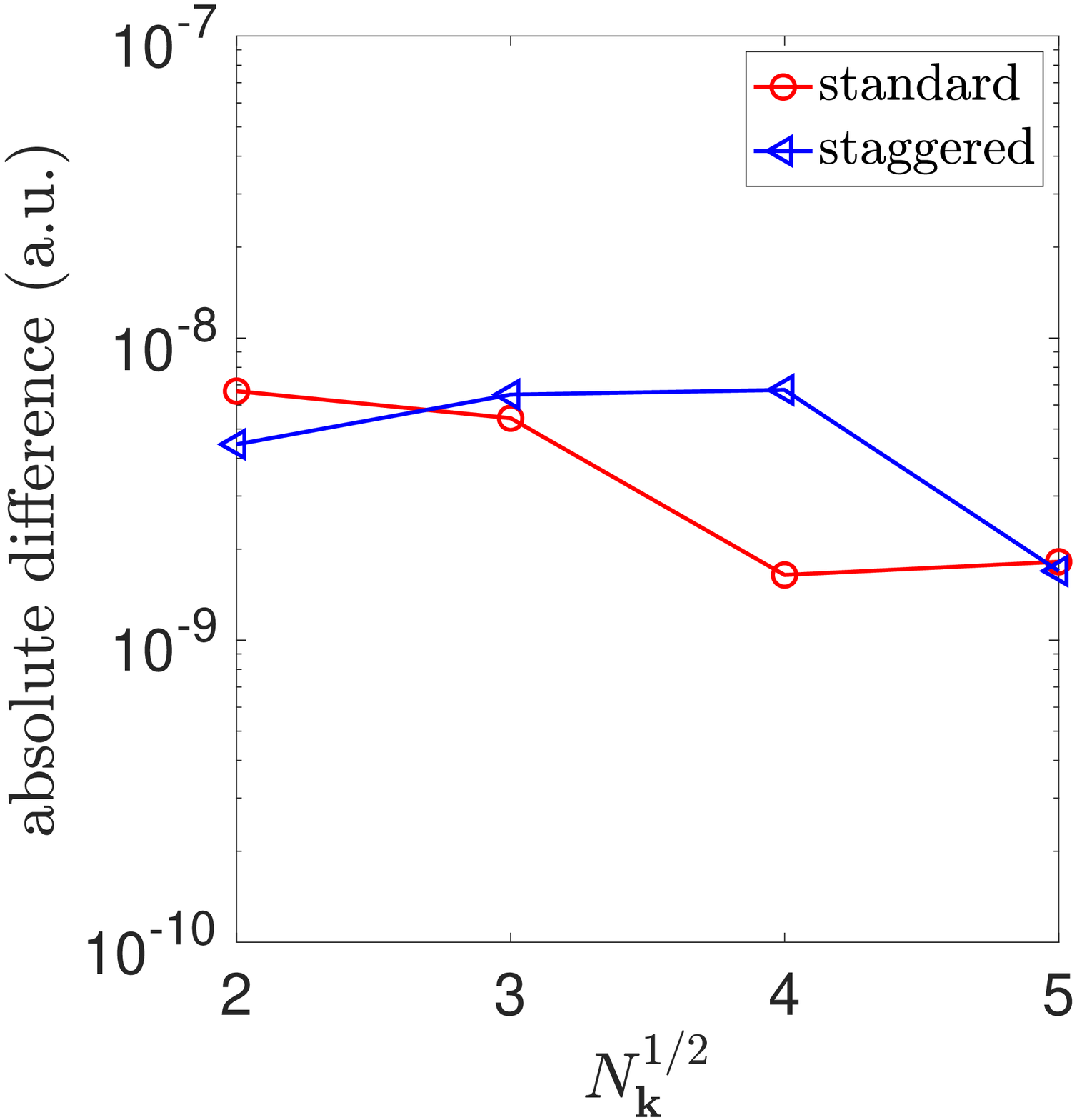}
        }
        
        \caption{Difference between RPA correlation energies computed using the drCCD and AC formalisms in both the standard and the staggered mesh methods.}
        \label{fig:diff}
\end{figure}

\Cref{fig:diff} plots the difference between RPA correlation energies computed using the drCCD and AC formalisms. 
The Gaussian density fitting is also used to evaluate the ERIs in the drCCD-based calculation for the comparison.
Despite the very different implementations, the difference between the results from drCCD and AC is negligible. This is the case in both the standard and the staggered mesh methods.
Having established the numerical equivalence of the two methods, in the numerical tests below, the RPA correlation energy is always computed using the AC formalism due to lower computational complexity in our implementation.
But the RPA-SOSEX correlation energy will still be computed using the drCCD formalism, which restricts our calculations to systems with relatively small sizes.

\begin{figure}[htbp]
        \centering
        \captionsetup[subfigure]{labelformat=empty}
        \subfloat[H2 quasi-1D]{
                \includegraphics[width=0.24\textwidth]{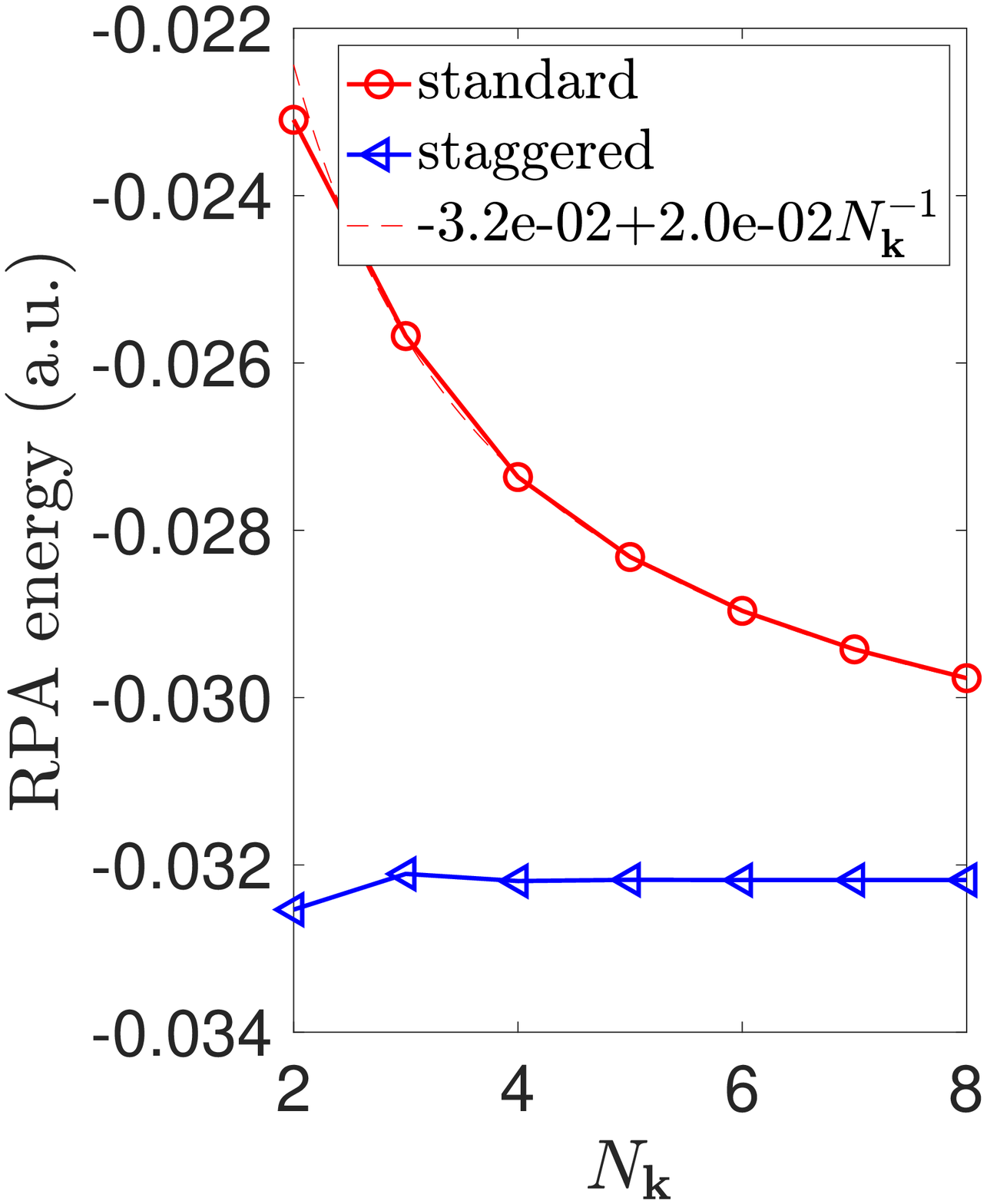}
        }
        \subfloat[LiH quasi-1D]{
                \includegraphics[width=0.24\textwidth]{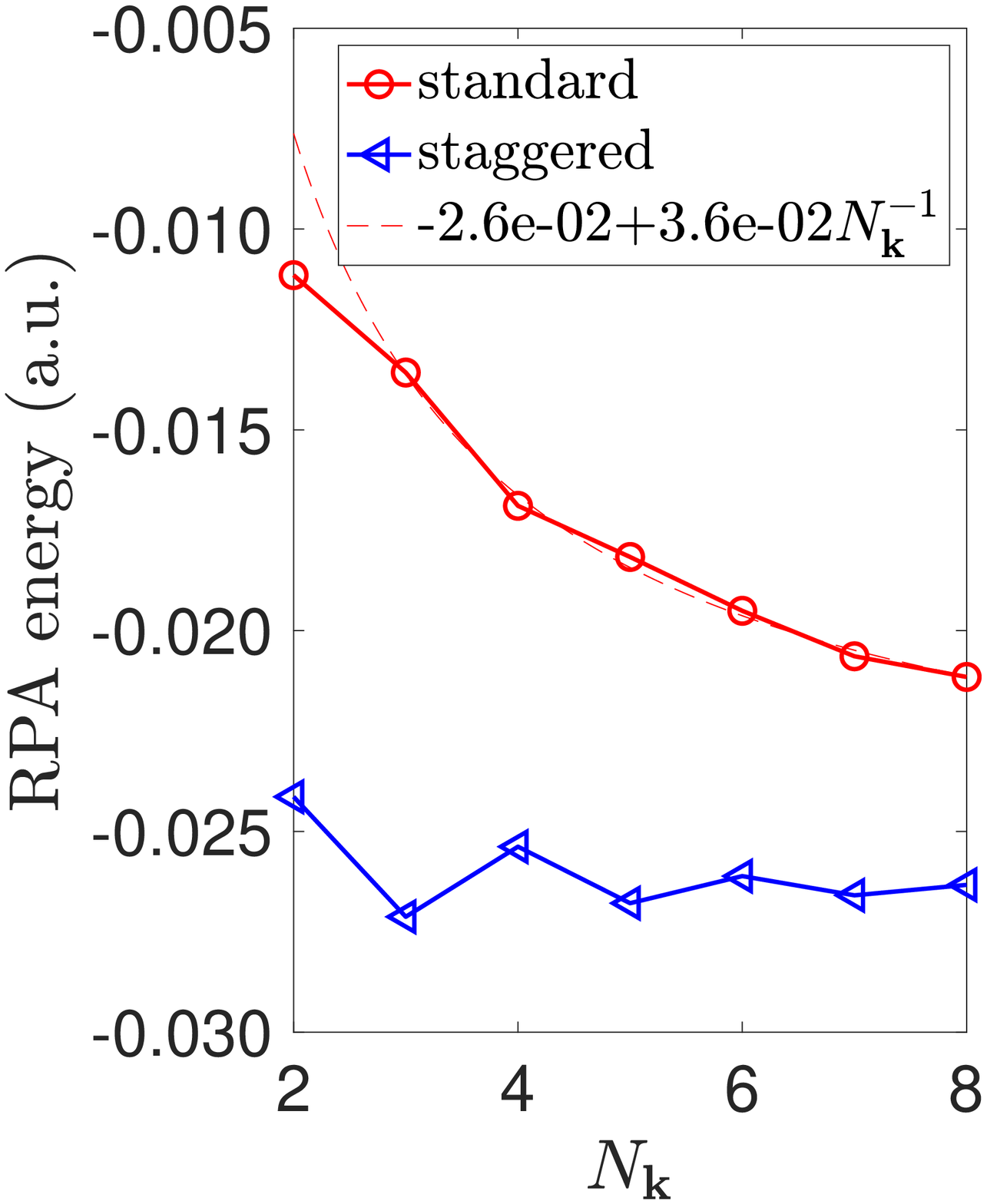}
        }
        \subfloat[Si quasi-1D]{
                \includegraphics[width=0.24\textwidth]{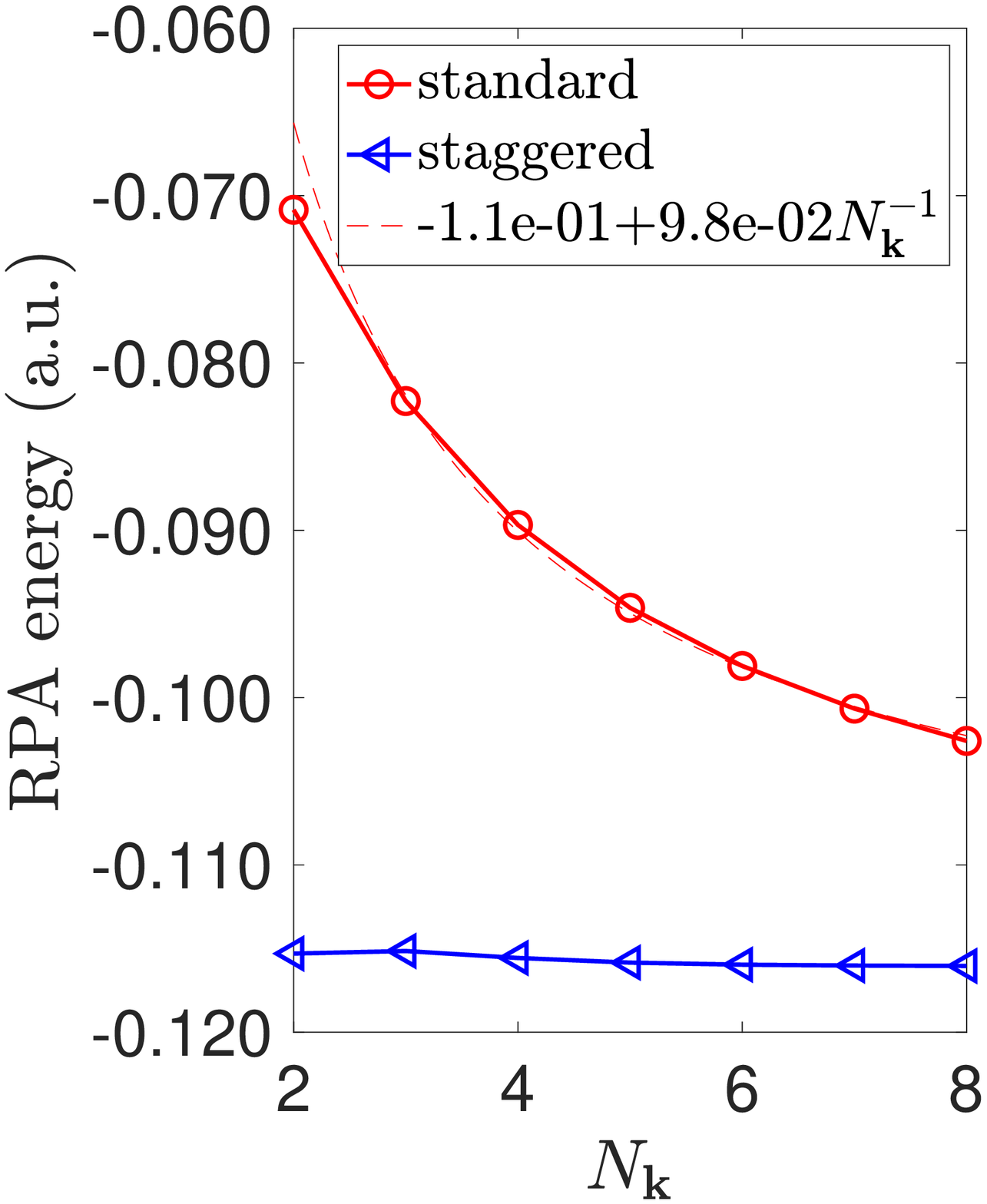}
        }
        \subfloat[Diamond quasi-1D]{
                \includegraphics[width=0.24\textwidth]{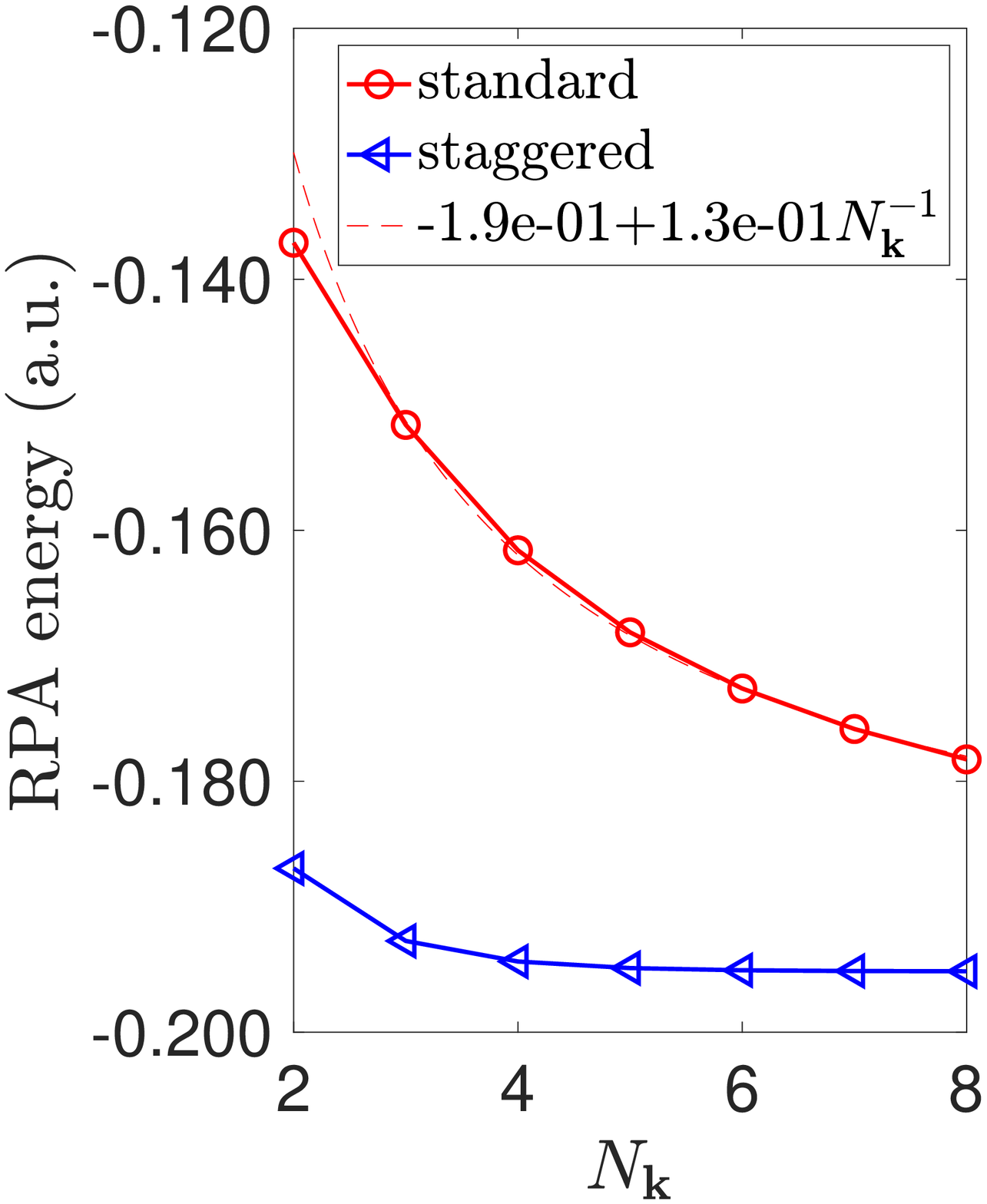}
        }

        \subfloat[H2 quasi-2D]{
        \includegraphics[width=0.24\textwidth]{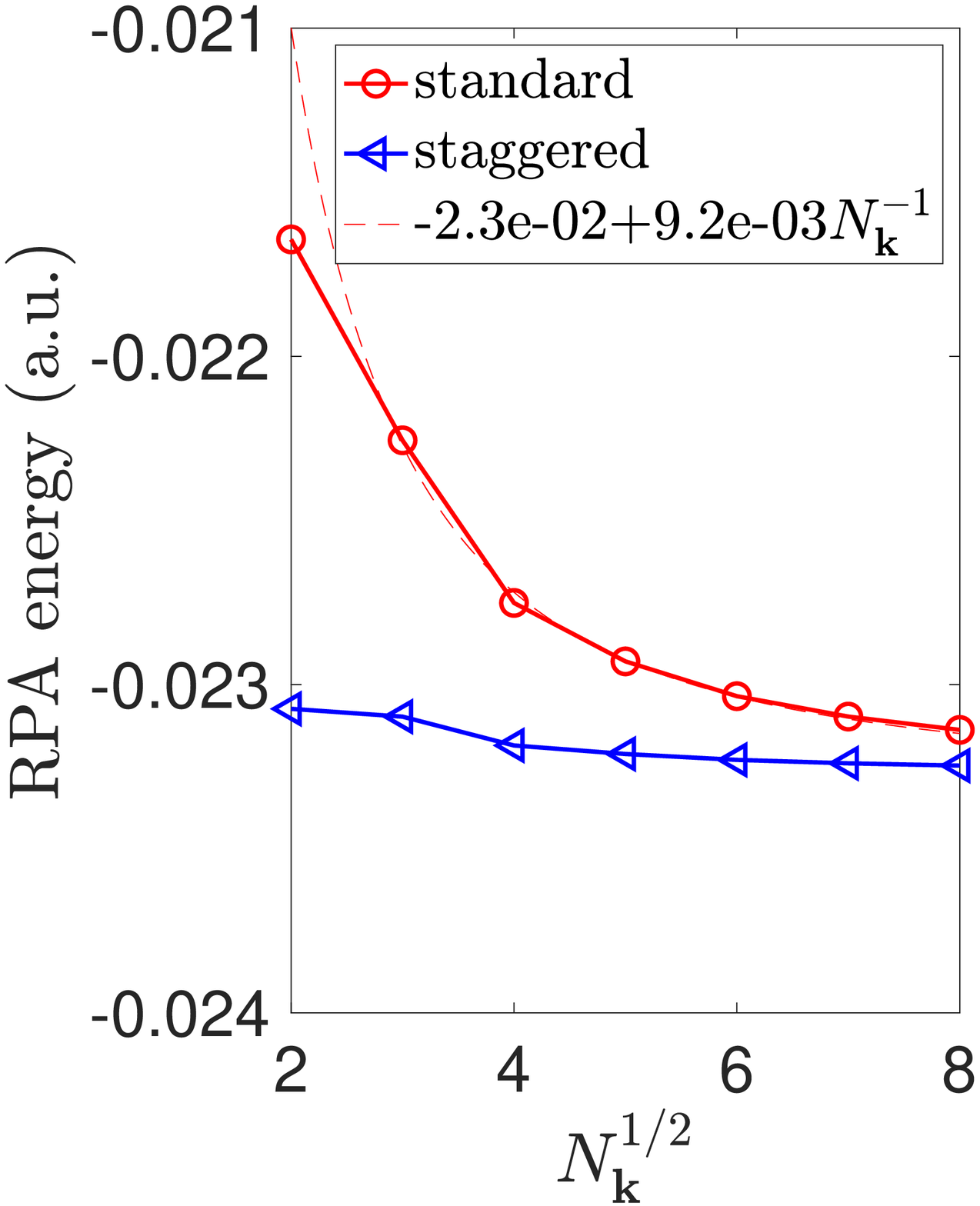}
        }
        \subfloat[LiH quasi-2D]{
                \includegraphics[width=0.24\textwidth]{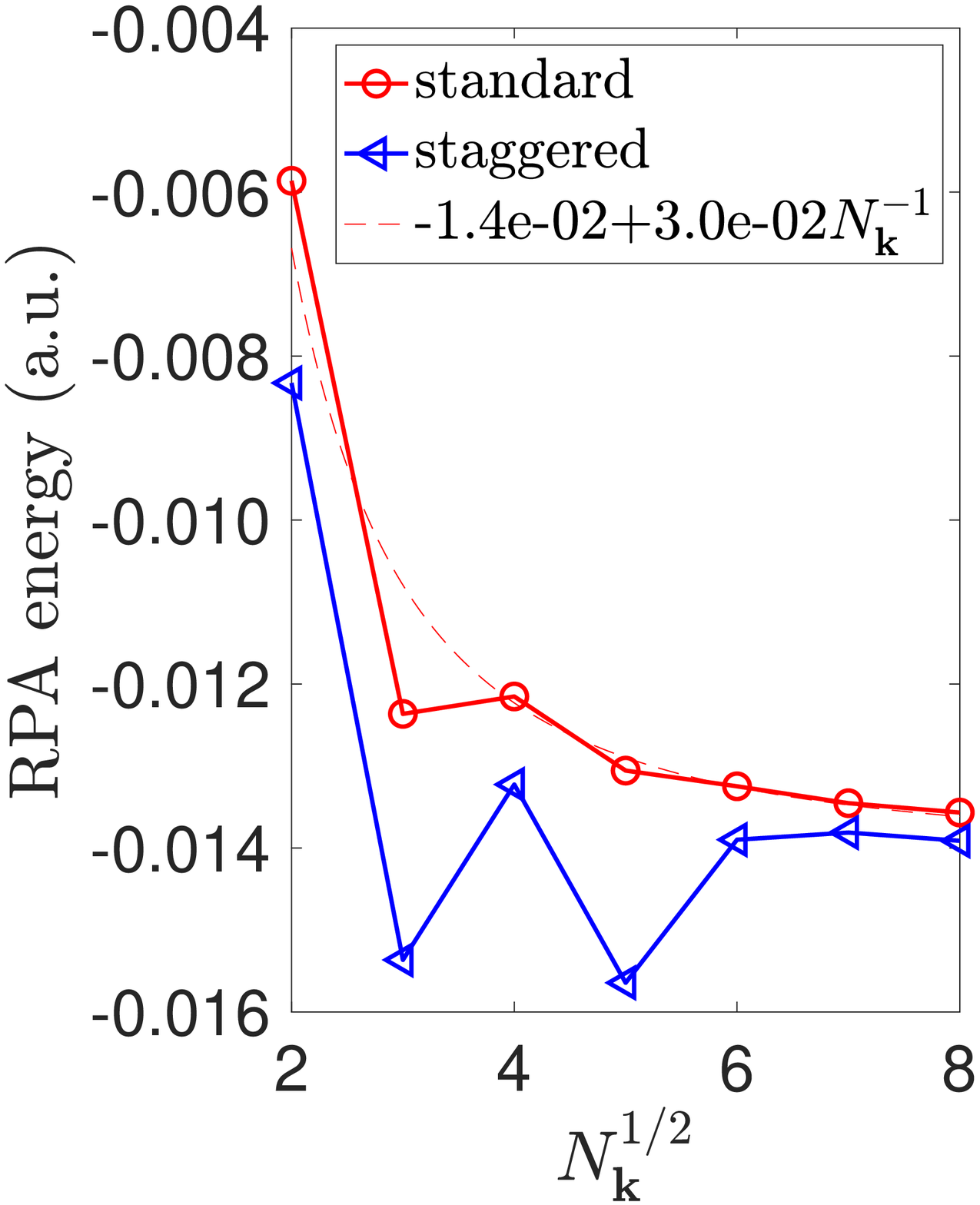}
        }
        \subfloat[Si quasi-2D]{
                \includegraphics[width=0.24\textwidth]{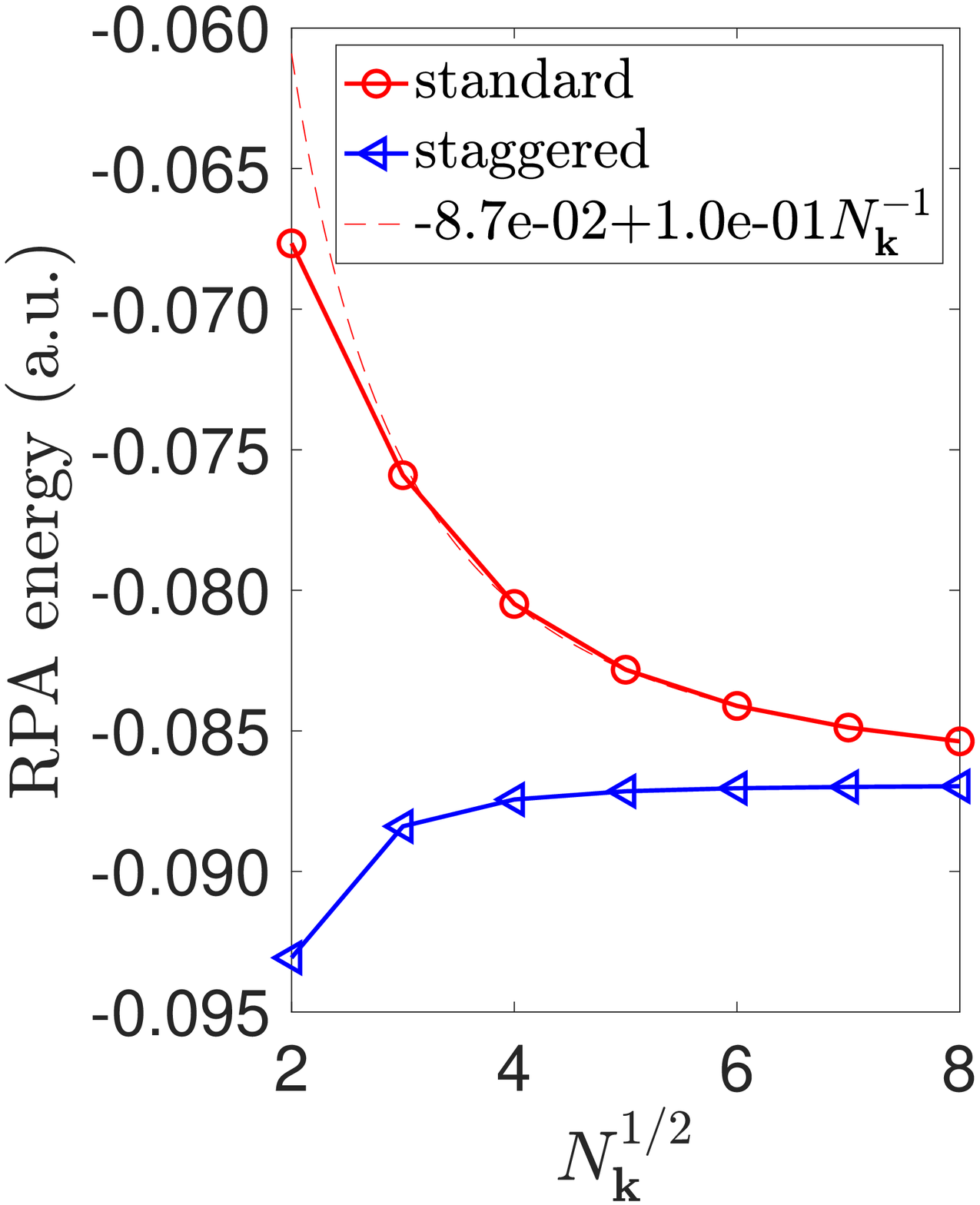}
        }
        \subfloat[Diamond quasi-2D]{
                \includegraphics[width=0.24\textwidth]{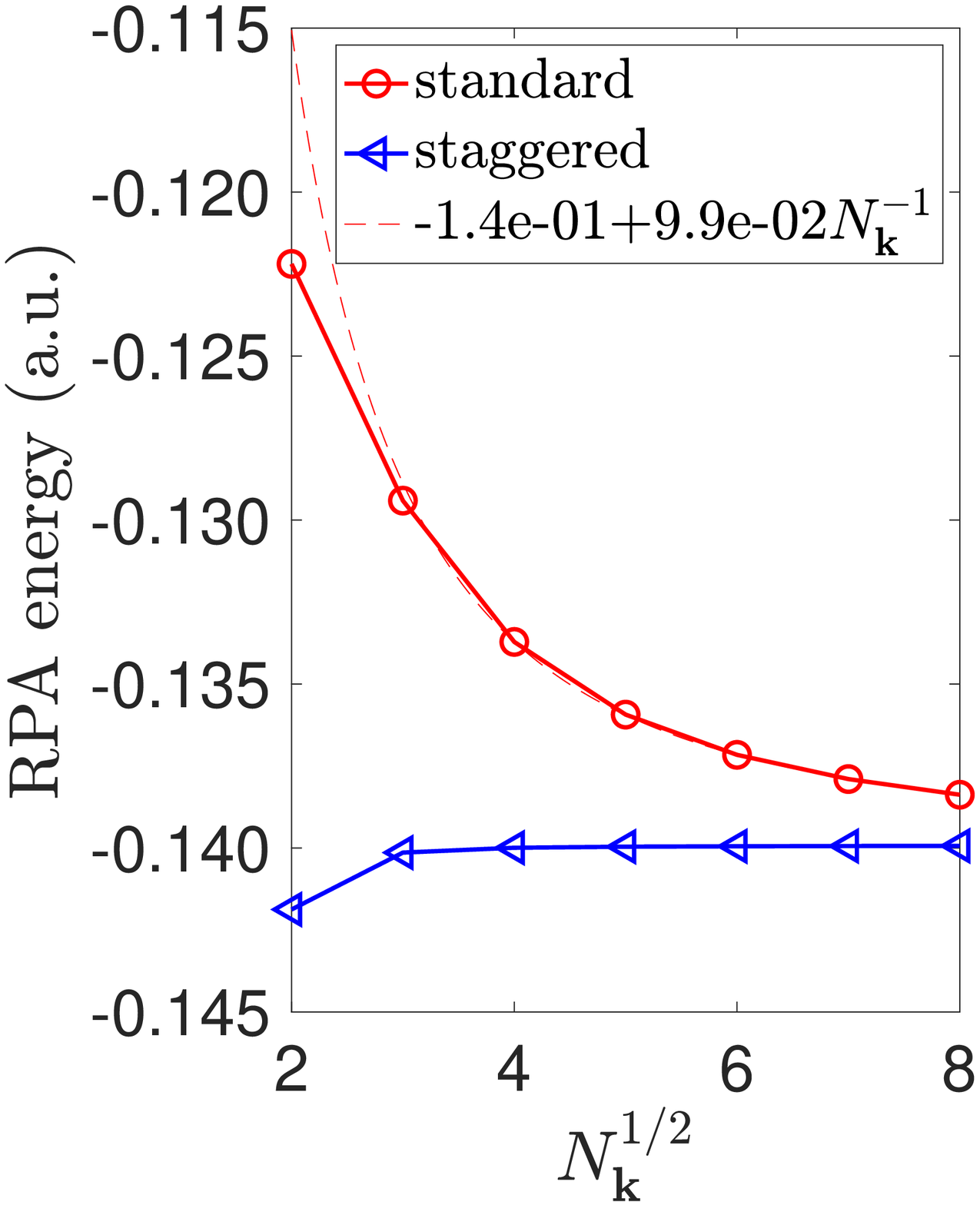}
        }

        \subfloat[H2 3D]{
        \includegraphics[width=0.24\textwidth]{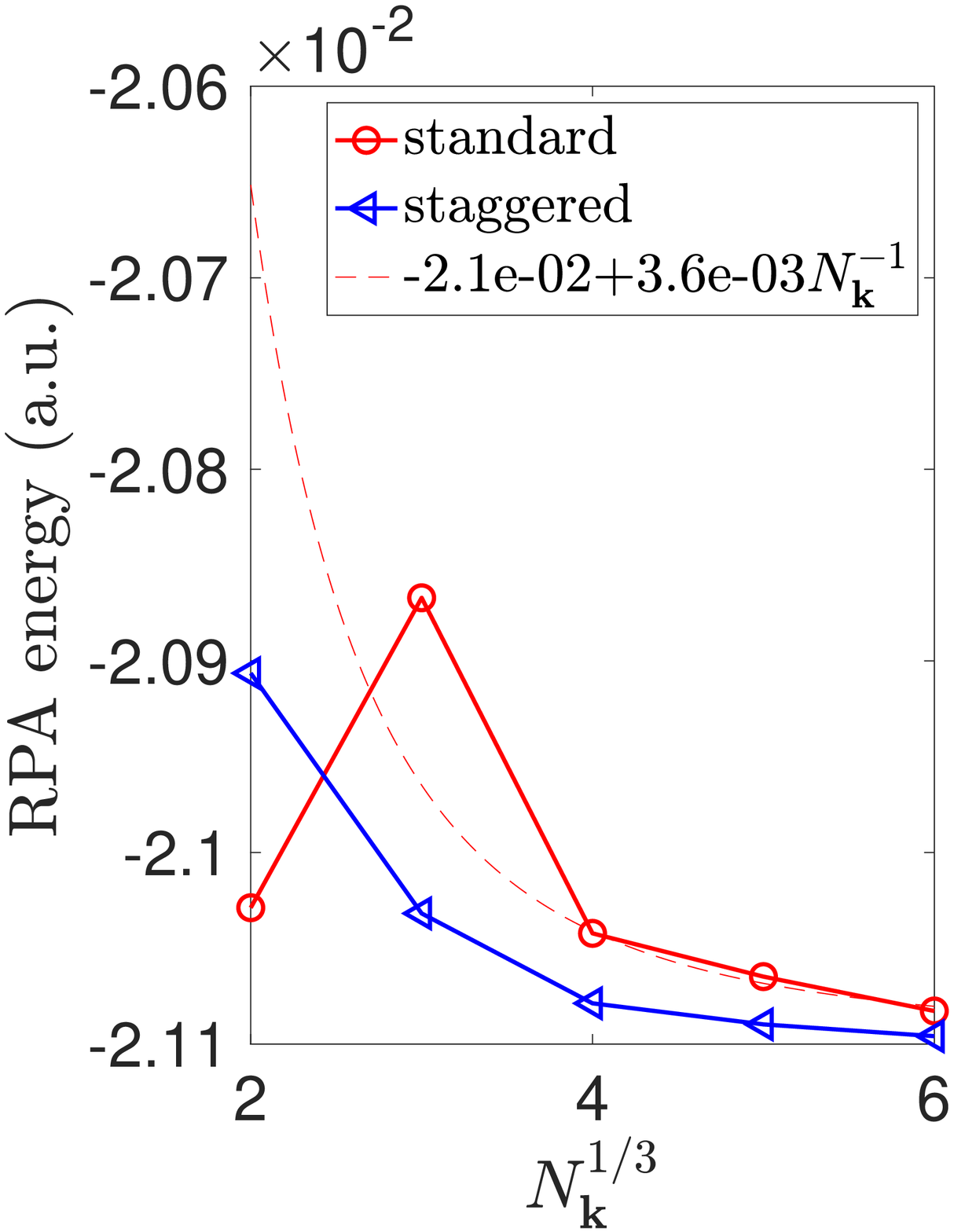}
                }
                \subfloat[LiH 3D]{
                        \includegraphics[width=0.24\textwidth]{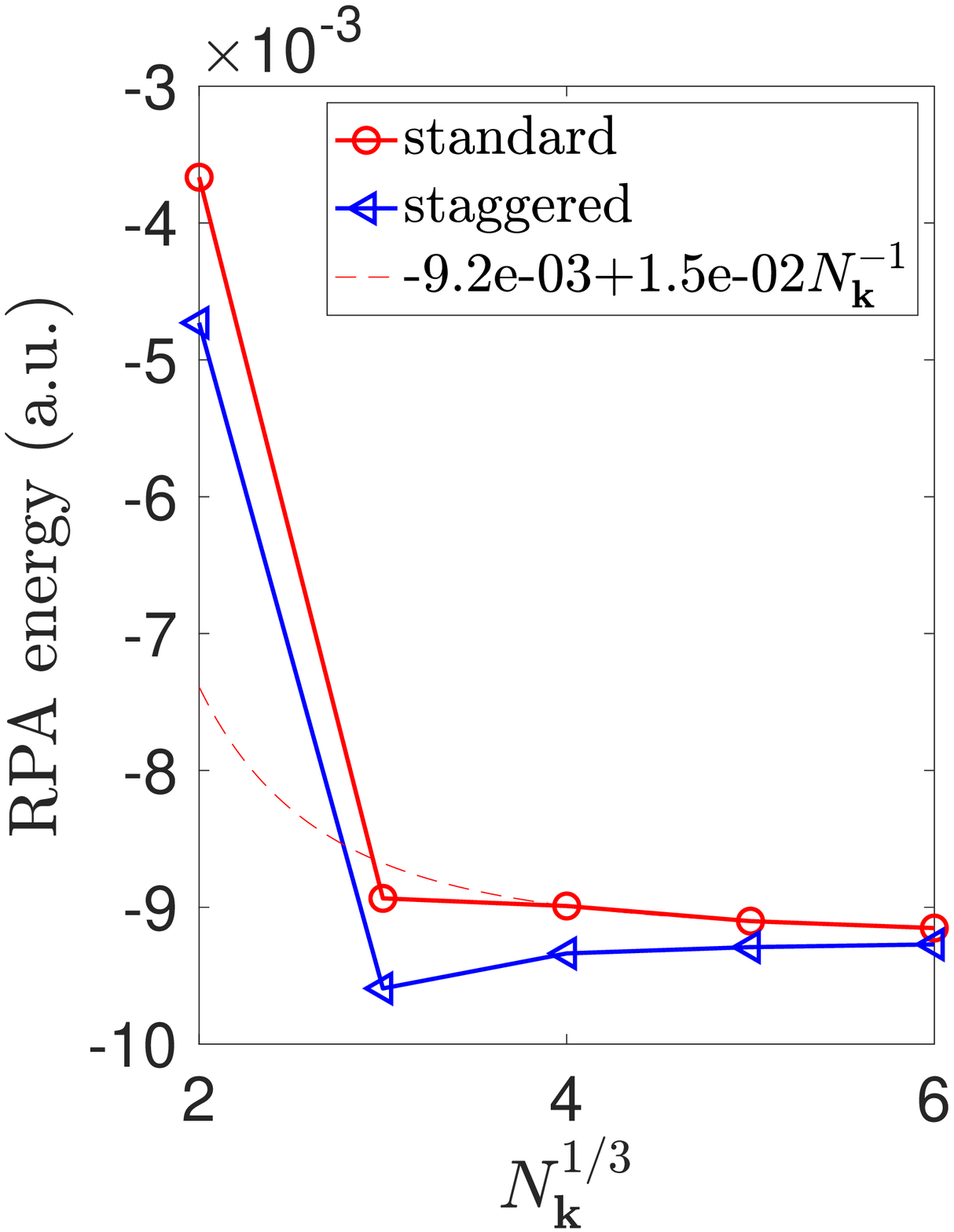}
                }
        \subfloat[Si 3D]{
                        \includegraphics[width=0.24\textwidth]{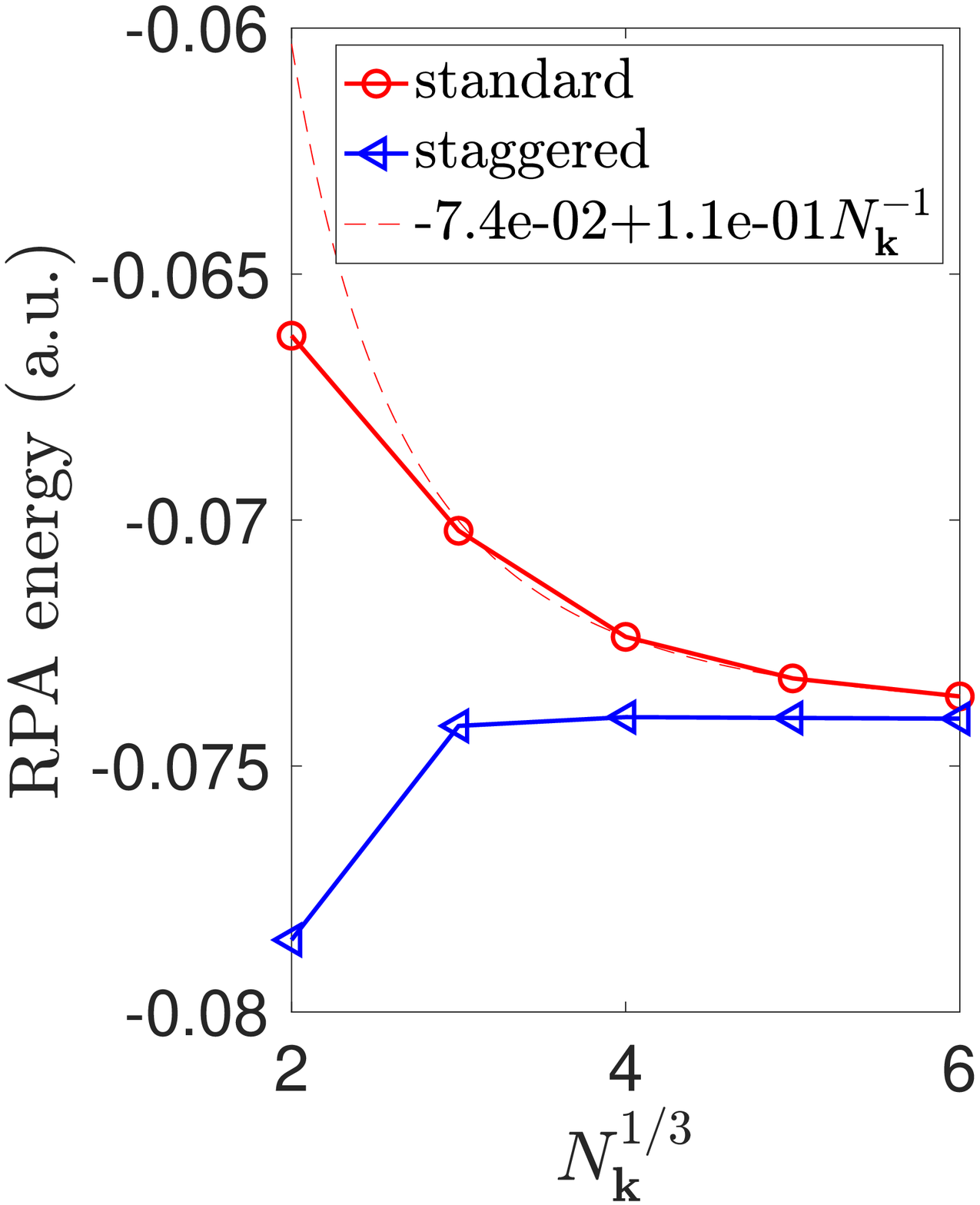}
                }
                \subfloat[Diamond 3D]{
                        \includegraphics[width=0.24\textwidth]{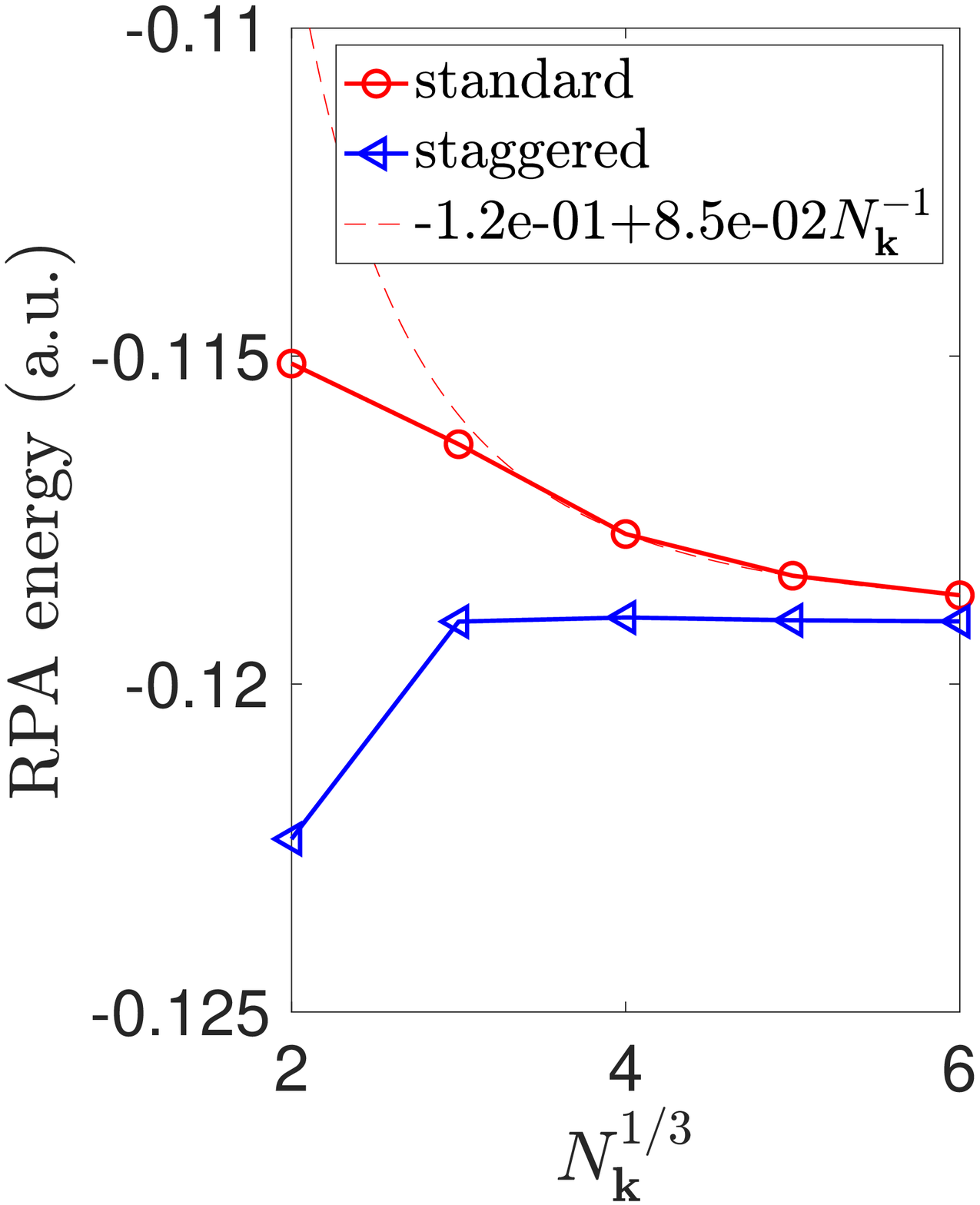}
                }

        \caption{RPA correlation energy per unit cell computed by the standard and the staggered mesh methods using the AC formalism.
        The fluctuation in LiH results is likely due to the small size of the basis set. This is confirmed by the results in \cref{fig:rpa_dzvp} using a larger gth-dzvp basis set.}
        \label{fig:rpa}
\end{figure}

\begin{figure}[htbp]
        \centering
        \captionsetup[subfigure]{labelformat=empty}
        \subfloat[H2 quasi-1D]{
                \includegraphics[width=0.24\textwidth]{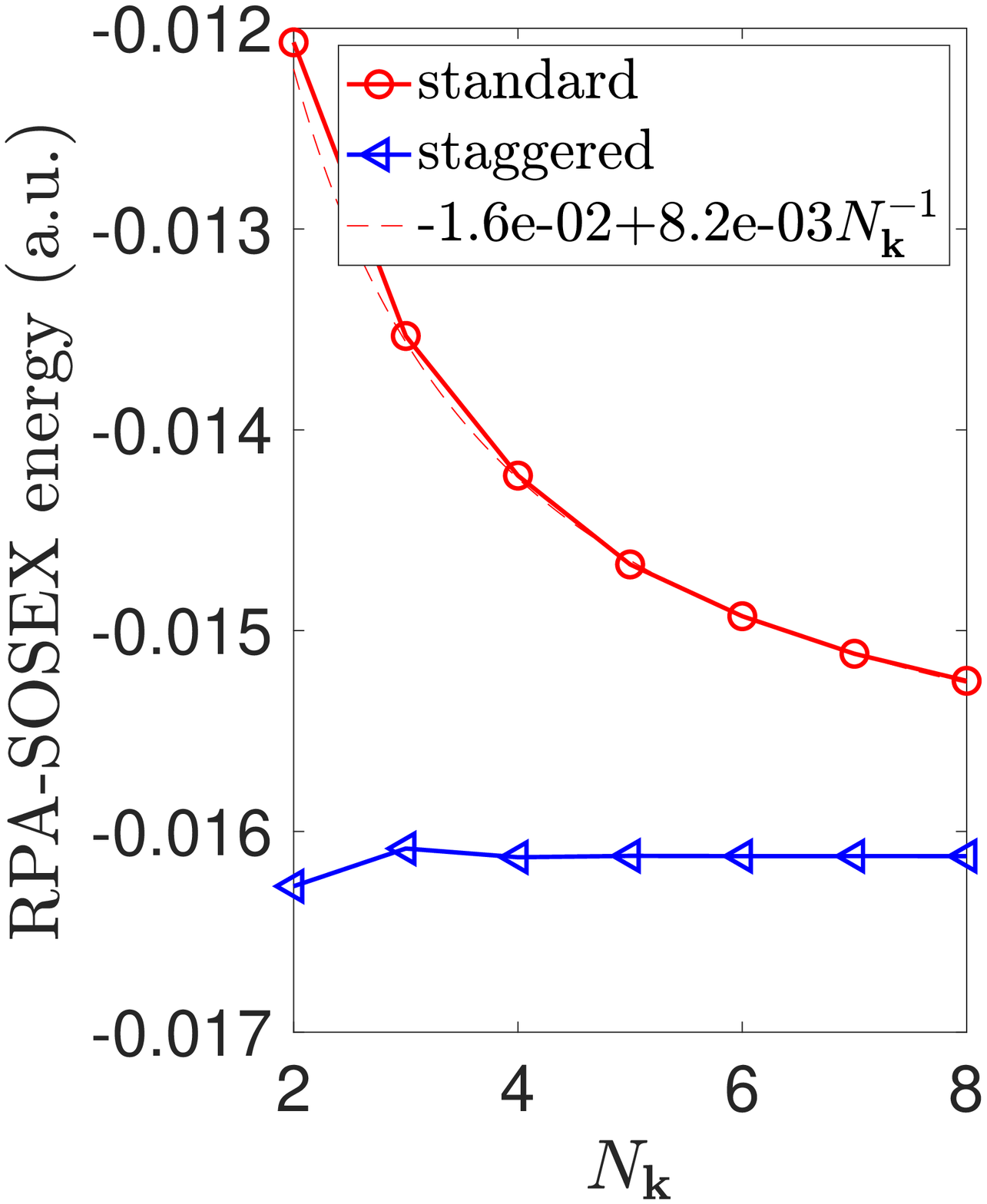}
        }
        \subfloat[LiH quasi-1D]{
                \includegraphics[width=0.24\textwidth]{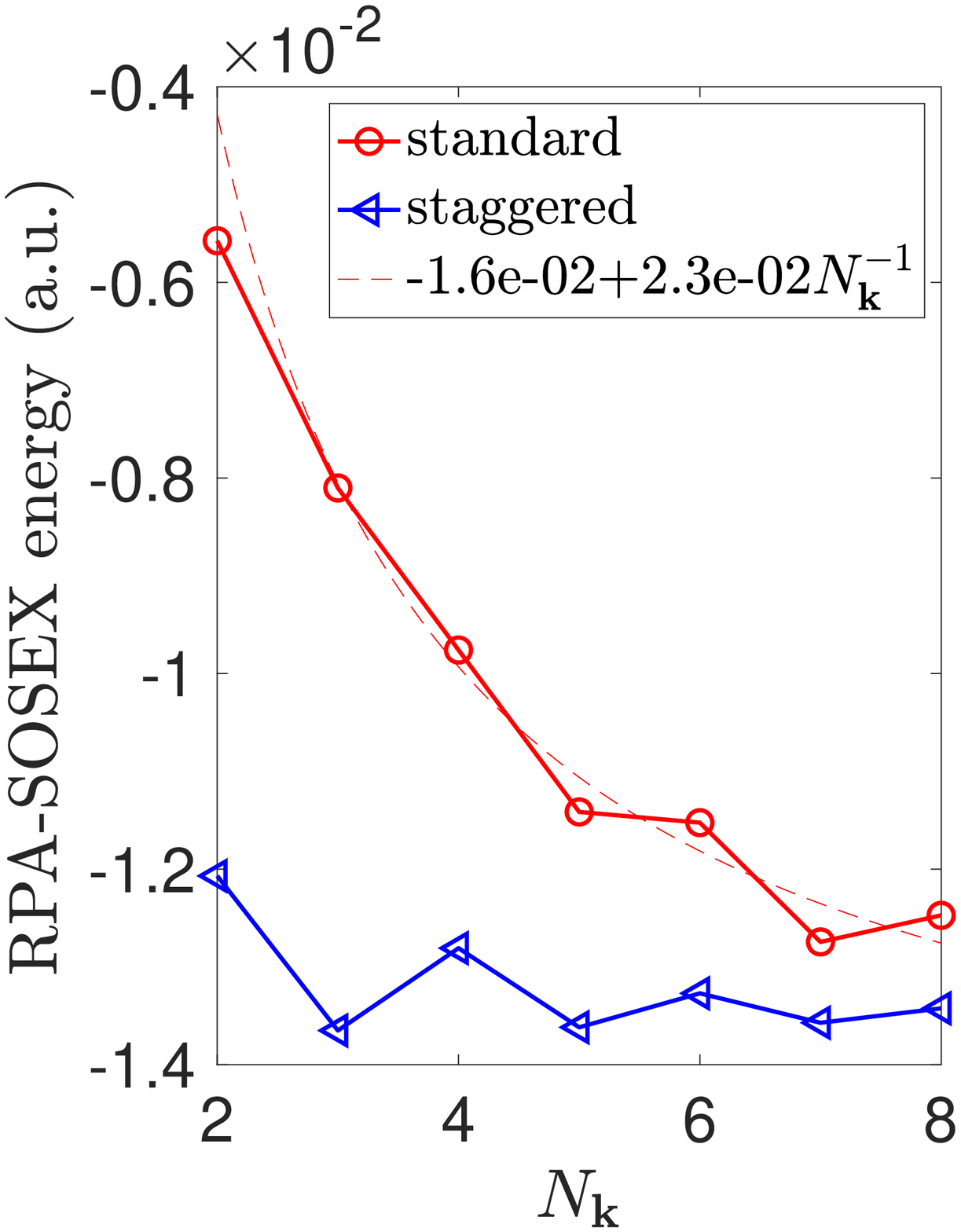}
        }
        \subfloat[Si quasi-1D]{
                \includegraphics[width=0.24\textwidth]{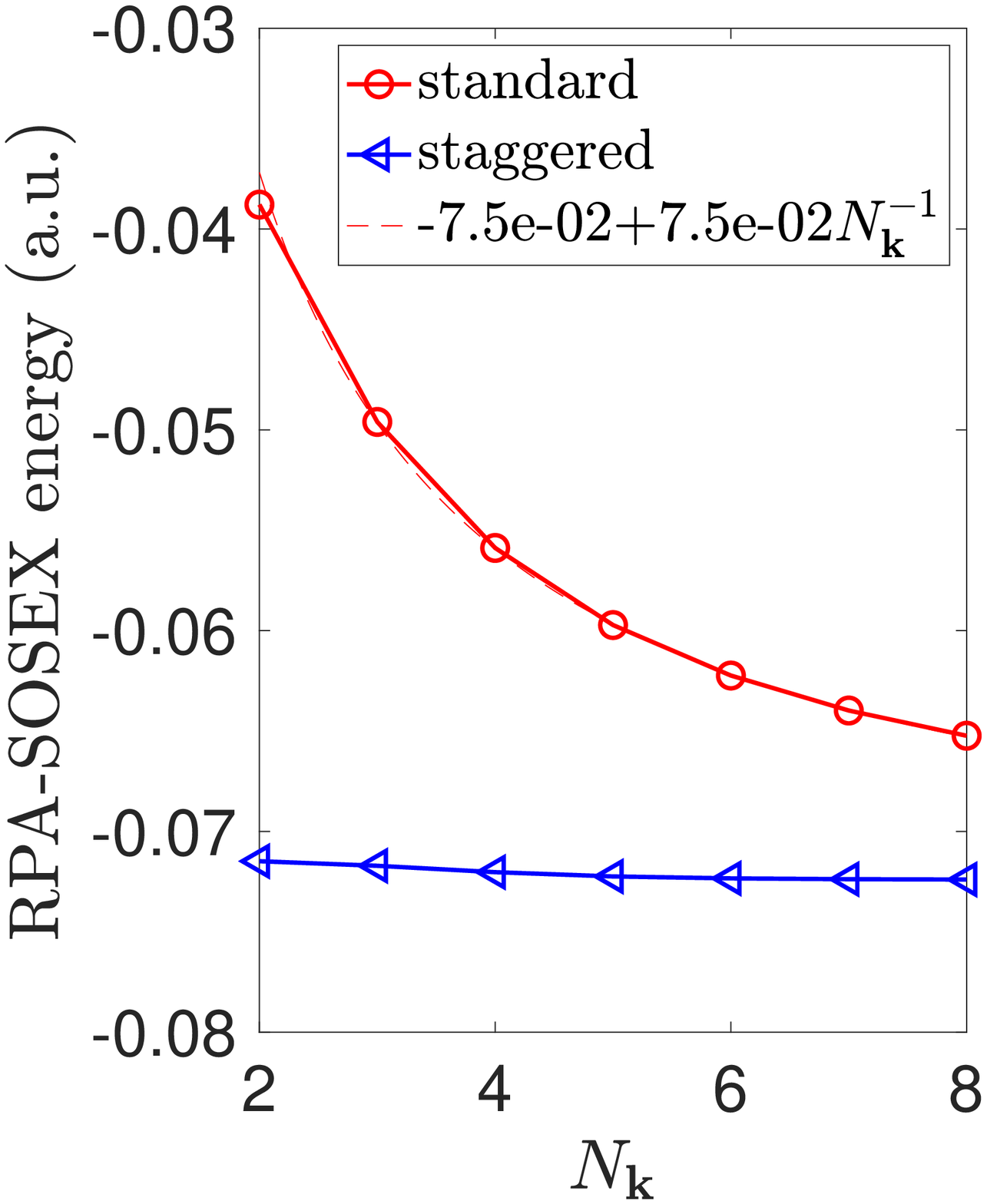}
        }
        \subfloat[Diamond quasi-1D]{
                \includegraphics[width=0.24\textwidth]{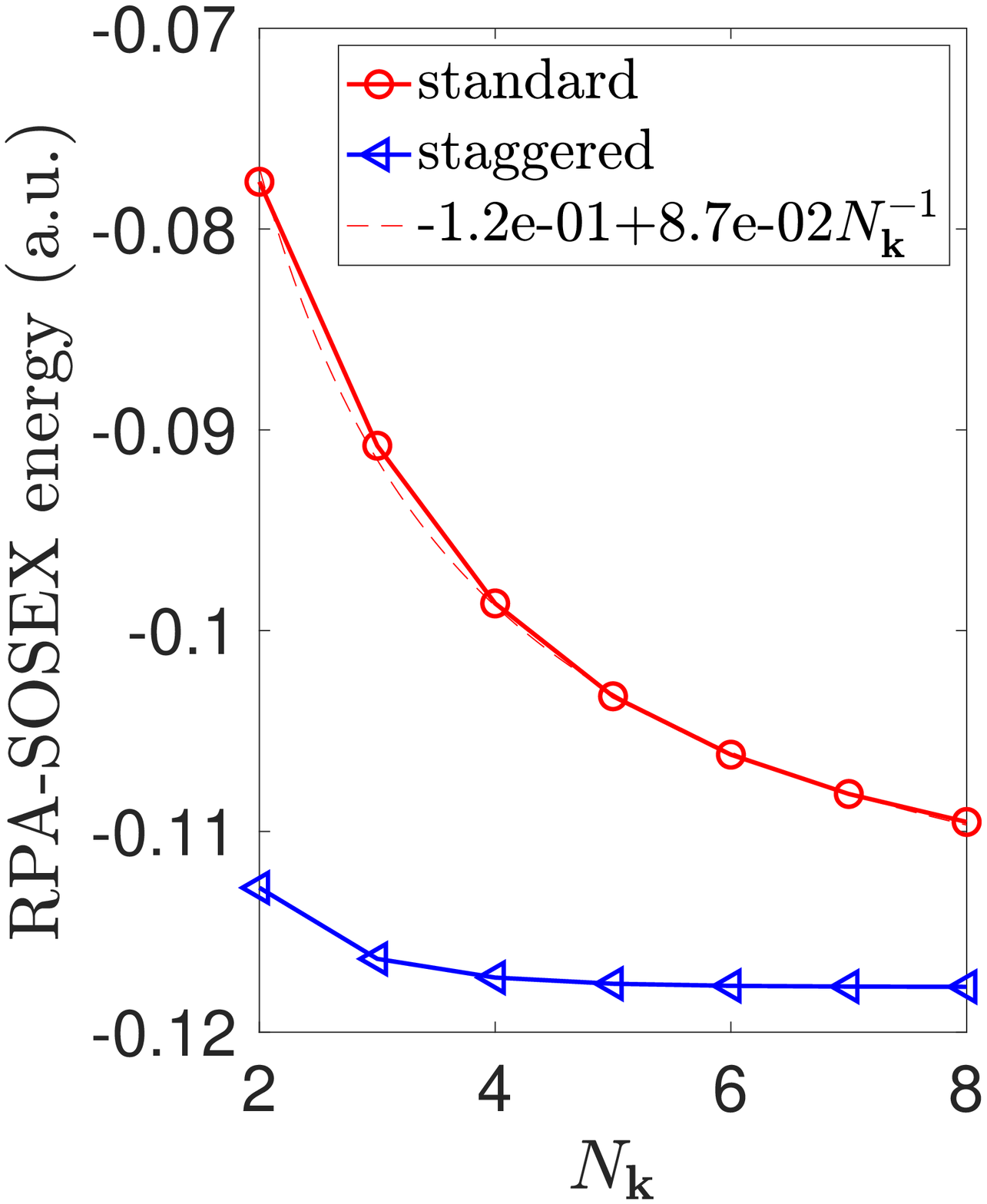}
        }
        
        \subfloat[H2 quasi-2D]{
                \includegraphics[width=0.24\textwidth]{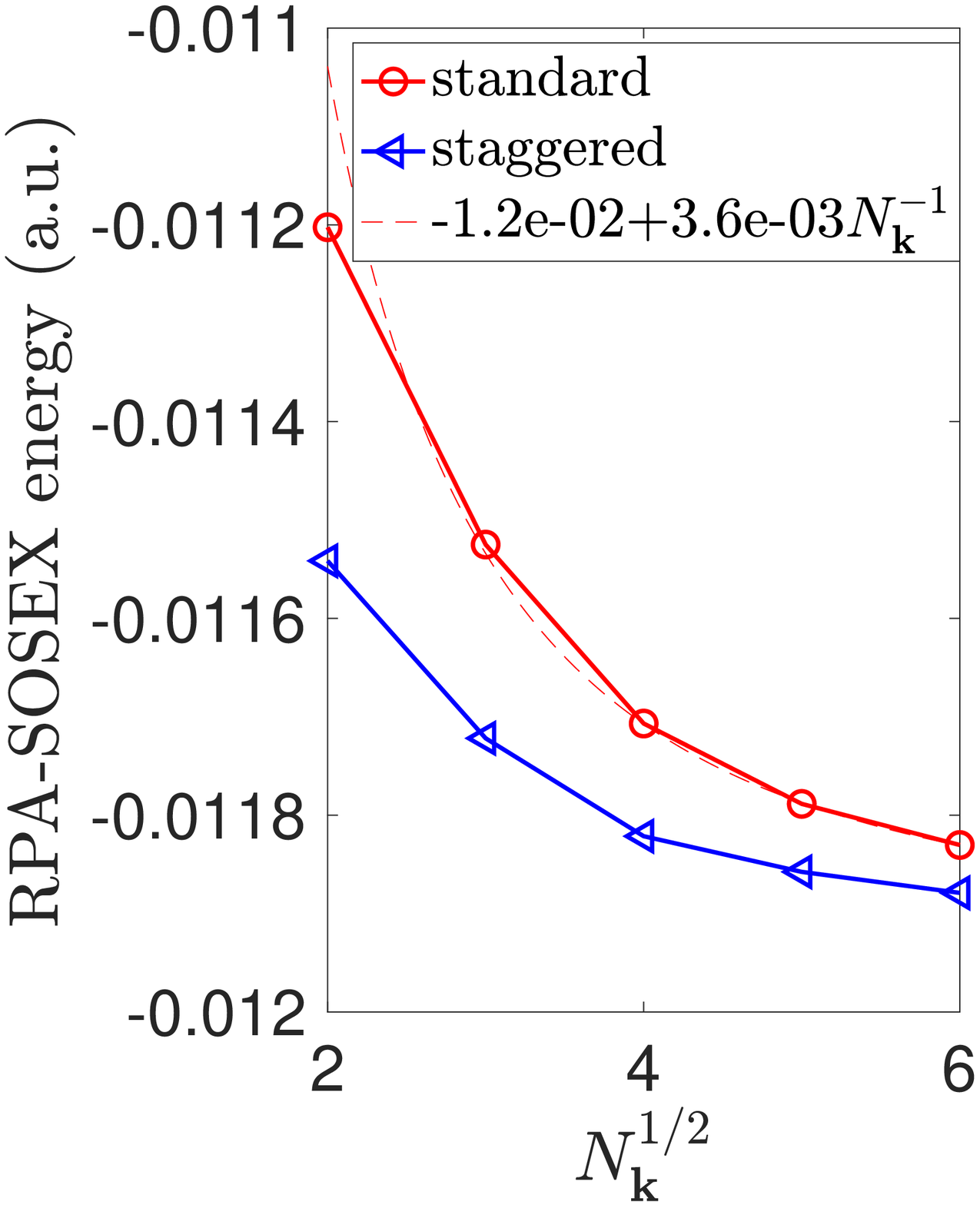}
        }
        \subfloat[LiH quasi-2D]{
                \includegraphics[width=0.24\textwidth]{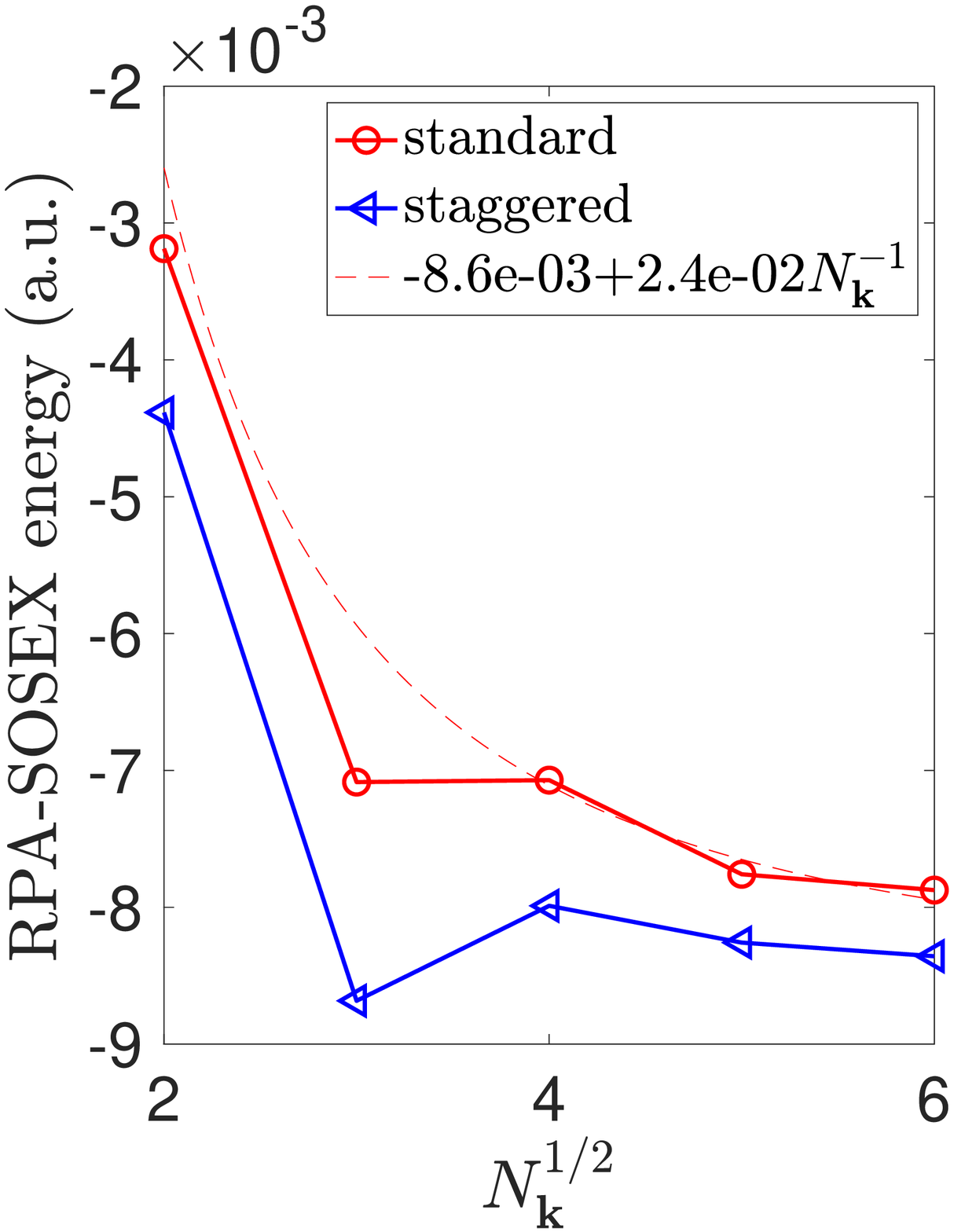}
        }
        \subfloat[Si quasi-2D]{
                \includegraphics[width=0.24\textwidth]{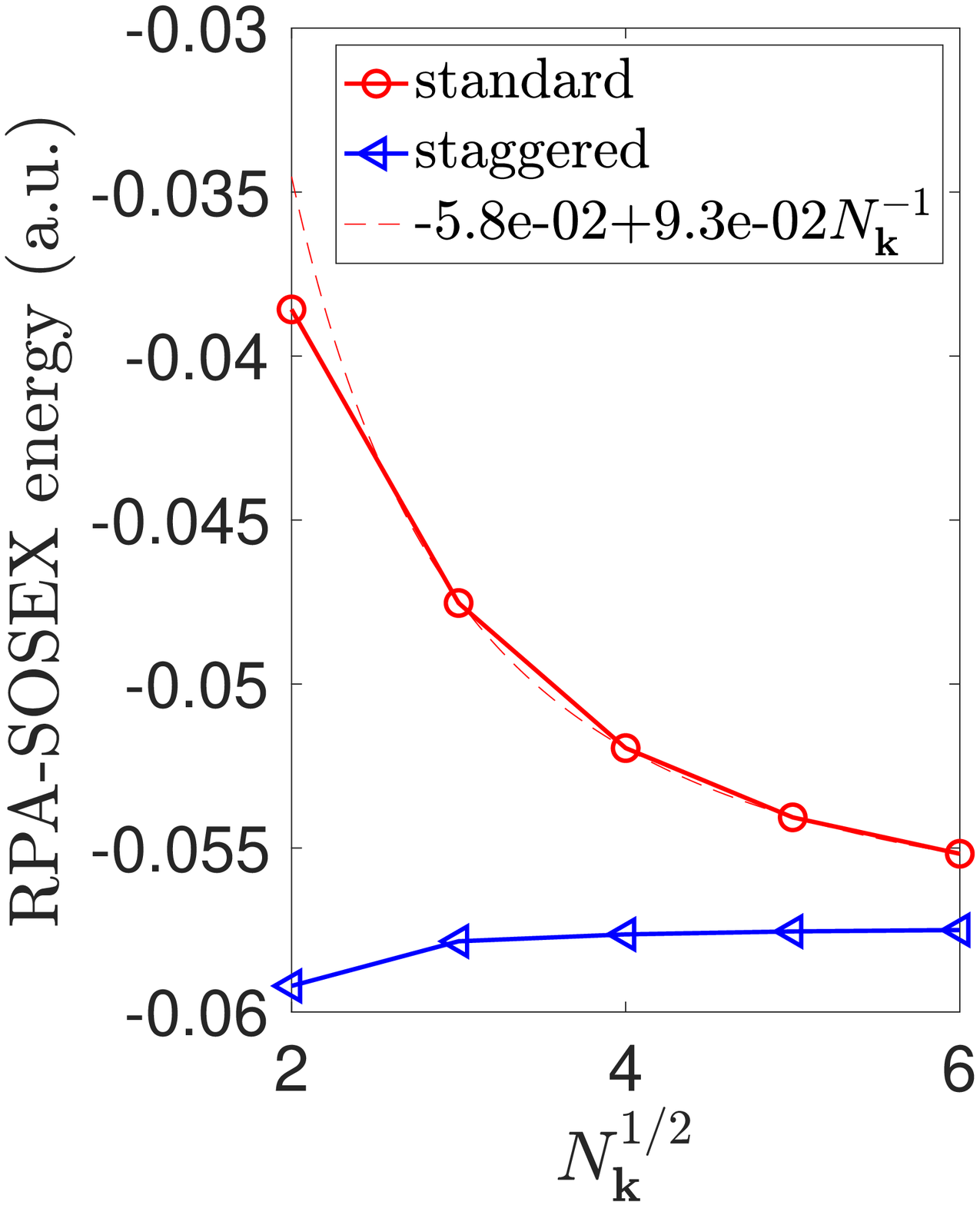}
        }
        \subfloat[Diamond quasi-2D]{
                \includegraphics[width=0.24\textwidth]{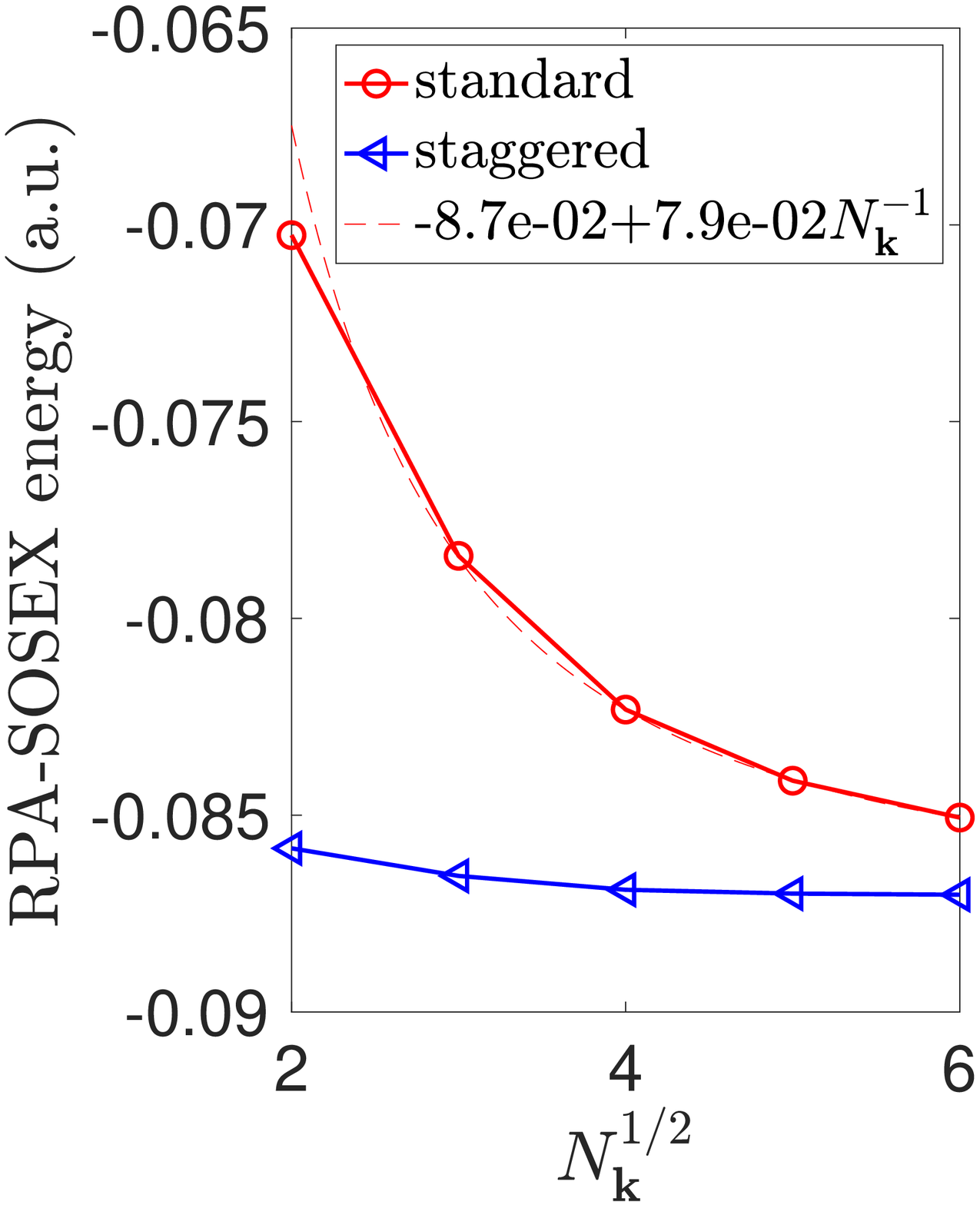}
        }

        \caption{RPA-SOSEX correlation energy per unit cell  computed by the standard and the staggered mesh methods using the drCCD formalism. }
        \label{fig:rpa_sosex}
\end{figure}

\Cref{fig:rpa} and \Cref{fig:rpa_sosex} plot  the RPA and RPA-SOSEX correlation energies,  computed by the standard and the staggered mesh methods for quasi-1D, quasi-2D, and 3D systems, respectively. 
Using the standard mesh, the finite-size errors of RPA and RPA-SOSEX correlation energies calculations scale as $\Or(N_\vk^{-1})$. 
In almost all cases, the finite-size error is significantly reduced using the staggered mesh method. 
Specifically, the staggered mesh method outperforms the standard one in all the quasi-1D systems. 
The finite-size errors decay rapidly with respect to $N_\vk$ and the curves are nearly flat.
For quasi-2D and 3D cases, the reduction of the error in the staggered mesh method is more pronounced for isotropic systems (silicon and diamond) than for anisotropic systems (hydrogen dimer and lithium hydride).  
Such a behavior agrees with the behavior of the staggered mesh method for MP2 calculations in the previous study \cite{XingLiLin2021}, and can be explained based on our  analysis in the next section.

%
%

\newpage

\section{Analysis}\label{sec:analysis}

In this section, we analyze the finite-size error in staggered mesh method for RPA and RPA-SOSEX energy calculations proposed in \cref{sec:method}, and provide explanation of its performance in different systems as numerically observed in \cref{sec:numerical}. 
Due to the equivalence between the drCCD and AC formalisms for RPA, we choose to focus on the analysis using the drCCD formalism.

\Cref{subsec:expansion} first expands the RPA and RPA-SOSEX correlation energies into two infinite partial summations based on fixed point iterations for solving the drCCD amplitude equation. 
In particular, the leading terms of the expansions contain the direct and the exchange terms of the MP2 correlation energy. 
\Cref{subsec:mp2} briefly reviews the basic idea of the original staggered mesh method for MP2. 
\Cref{subsec:rpa} further shows that the method for analyzing the performance of the staggered mesh method for MP2 calculations can be generalized to analyze the performance of  the staggered mesh method for RPA and RPA-SOSEX correlation energies, in an order-by-order fashion. 
\Cref{subsec:ac} provides a similar finite-size error analysis for RPA in the AC formalism. 
Interestingly, our analysis as well as numerical results indicate that for RPA correlation energy calculations, the ``head/wing'' correction to the dielectric operator is not needed in the staggered mesh method.


\subsection{Perturbative  expansions of RPA and RPA-SOSEX energy}\label{subsec:expansion}
\begin{figure}[htbp]
\centering
\includegraphics[width=0.8\textwidth]{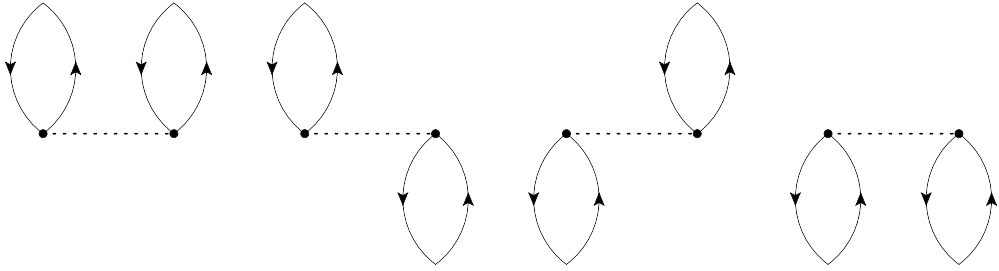}
\caption{Four forms of interaction vertices that constitute a ring diagram (in the Goldstone diagram formalism\cite{ShavittBarlett2009}).}
\label{fig:ring_vertex}
\end{figure}

For brevity of notation, we use the capital letter $P$ to denote an index pair $(p, \vk_p)$.
The standard method for RPA and RPA-SOSEX correlation energies with one MP mesh $\mathcal{K}$ requires solving the nonlinear drCCD amplitude equation in  \cref{eqn:rdccd_amplitude}.
This equation can be solved, for instance, by the fixed point iteration method as
\begin{equation}\label{eqn:fixedpoint}
        \begin{split}
        (t_{IJ}^{AB})^{(n+1)}
        & = \dfrac{1}{\epsilon_{IJ}^{AB}}
        \Big(
        \braket{AB|IJ } +  2\sum_{KC}\braket{KB|CJ} (t_{I K}^{A C})^{(n)}\\
        & \qquad + 2\sum_{KC}\braket{AK|IC} (t_{KJ}^{CB})^{(n)}+ 4\sum_{KLCD}\braket{KL|CD} (t_{I K}^{A C})^{(n)} (t_{L J}^{D B})^{(n)}
        \Big),
        \end{split}
\end{equation}
starting with the initial guess  $(t_{IJ}^{AB})^{(0)} = \braket{AB|IJ }/ \epsilon_{IJ}^{AB}$. Here, $\epsilon_{IJ}^{AB} =  \epsilon_I + \epsilon_J - \epsilon_A - \epsilon_B$.
After $n$ iterations,  the two correlation energies can be approximated by
\begin{align}
        E_\text{rpa}^{(n)}(N_\vk)                 & =  \dfrac{1}{N_\vk}\sum_{IJAB} 2\braket{ I J| A B}(t_{I J}^{A B})^{(n)},
        \label{eqn:rpa_fixedpoint}\\
        E_\text{rpa-sosex}^{(n)}(N_\vk) & =  \dfrac{1}{N_\vk}\sum_{IJAB} (2\braket{ I J| A B} - \braket{IJ|BA})(t_{I J}^{A B})^{(n)}.
        \label{eqn:rpa_sosex_fixedpoint}
\end{align}
Expanding the representation of $(t_{IJ}^{AB})^{(n)}$ shows that $E_\text{rpa}^{(n)}(N_\vk)$  consists of a subset of finite order many body perturbation energies computed with MP mesh $\mathcal{K}$
that are associated with ring diagrams, i.e., diagrams constructed by combining interaction vertices in the four forms in \cref{fig:ring_vertex}. 
Similarly, $E_\text{rpa-sosex}^{(n)}(N_\vk)$ contains additional perturbation  terms that are associated with specific exchange terms of ring diagrams, i.e., with $\braket{IJ|AB}$ replaced by $\braket{IJ|BA}$. 

\begin{figure}[htbp]
\centering
\includegraphics[width=\textwidth]{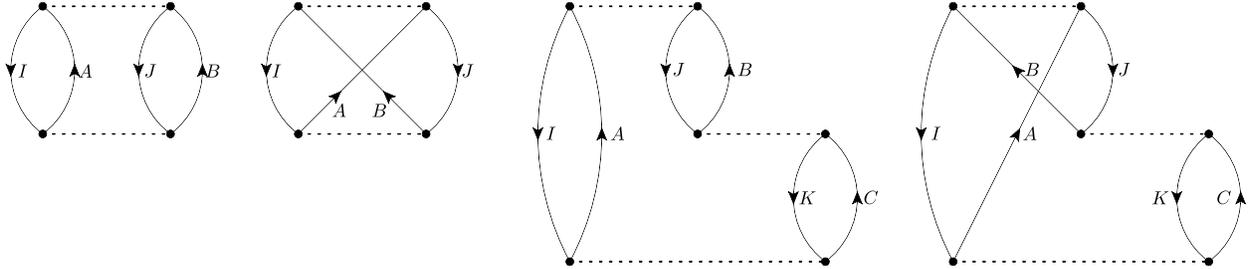}
\caption{Second and third order perturbation energies $E_\text{2d}$, $E_\text{2x}$, $E_\text{3d,1}$, and $E_\text{3x,1}$ included in the RPA and RPA-SOSEX correlation energies.}
\label{fig:diagram}
\end{figure}

For example, $E_\text{rpa}^{(n)}(N_\vk)$ and $E_\text{rpa-sosex}^{(n)}(N_\vk)$ with $n \geqslant 1$ contains two second order terms
\begin{equation}
        \label{eqn:rpa_mp2}
        \begin{split}
        E_\text{2d}(N_\vk) & = 2\dfrac{1}{N_\vk}\sum_{IJAB}\dfrac{1}{\epsilon_{IJ}^{AB}} \braket{IJ|AB}\braket{AB|IJ},
        \\
        E_\text{2x}(N_\vk) & = -\dfrac{1}{N_\vk} \sum_{IJAB}\dfrac{1}{\epsilon_{IJ}^{AB}} \braket{IJ|BA}\braket{AB|IJ},
        \end{split}
\end{equation}
and four third order terms as
\begin{equation}
        \label{eqn:rpa_mp3}
        \begin{split}
                E_\text{3d,1}(N_\vk) & = 4\dfrac{1}{N_\vk}\sum_{IJABKC} \braket{IJ|AB} \dfrac{1}{\epsilon_{IJ}^{AB}}\dfrac{1}{\epsilon_{IK}^{AC}}\braket{AC|IK}        \braket{KB|CJ},
                \\
                E_\text{3d,2}(N_\vk) & = 4\dfrac{1}{N_\vk}\sum_{IJABKC} \braket{IJ|AB} \dfrac{1}{\epsilon_{IJ}^{AB}}\dfrac{1}{\epsilon_{JK}^{BC}}\braket{AK|IC}        \braket{CB|KJ},
                \\
                E_\text{3x,1}(N_\vk) & = -2\dfrac{1}{N_\vk}\sum_{IJABKC} \braket{IJ|BA} \dfrac{1}{\epsilon_{IJ}^{AB}}\dfrac{1}{\epsilon_{IK}^{AC}}\braket{AC|IK} \braket{KB|CJ},
                \\
                E_\text{3x,2}(N_\vk) & = -2\dfrac{1}{N_\vk}\sum_{IJABKC} \braket{IJ|BA} \dfrac{1}{\epsilon_{IJ}^{AB}}\dfrac{1}{\epsilon_{JK}^{BC}}\braket{AK|IC} \braket{CB|KJ}.
        \end{split}
\end{equation}
Note that $E_\text{3d,2}$ and $E_\text{3x,2}$ are the complex conjugates of $E_\text{3d,1}$ and $E_\text{3x,1}$, respectively (up to the permutation of dummy variables in the summation).
The subscript ``d'' refers to direct terms that are contained in $E_\text{rpa}^{(n)}(N_\vk)$ and ``x'' refers to exchange terms that are contained in the SOSEX correction $E_\text{rpa-sosex}^{(n)}(N_\vk) - E_\text{rpa}^{(n)}(N_\vk)$. 
\Cref{fig:diagram} plots the diagrams associated with these second and third order energy terms.

\subsection{Staggered mesh method for MP2 energy}\label{subsec:mp2}
Before analyzing the performance of the staggered mesh method for RPA calculations, we first briefly review the method for analyzing the performance of MP2 calculations using the standard and the staggered mesh methods in Refs.\citenum{XingLiLin2021, XingLiLin2021_2}.
This is also because the standard MP2 energy calculation with one MP mesh $\mathcal{K}$ is defined as 
\begin{equation}\label{eqn:mp2}
        E_\text{mp2}(N_\vk) =\dfrac{1}{N_\vk}\sum_{\substack{ijab\\\vk_i,\vk_j, \vk_a \in \mathcal{K}}}(2\braket{i\vk_i,j\vk_j|a\vk_a,b\vk_b} - \braket{i\vk_i,j\vk_j|b\vk_b, a\vk_a})\dfrac{ \braket{a\vk_a,b\vk_b|i\vk_i,j\vk_j} }{\epsilon_{i\vk_i,j\vk_j}^{a\vk_a, b\vk_b}}, 
\end{equation}
which is exactly  the two second order terms $E_\text{2d}$ and $E_\text{2x}$ in \cref{eqn:rpa_mp2} included in the RPA and RPA-SOSEX correlation energies. 

Assuming that all orbitals and orbital energies are exact, both analytical and numerical results in Refs.\citenum{XingLiLin2021, XingLiLin2021_2} show that the finite-size error in the staggered mesh method decays super-algebraically (i.e., faster than any polynomial rate) for general quasi-1D systems, and scales as $\Or(N_\vk^{-2})/\Or(N_\vk^{-\frac53})$ for special  quasi-2D/3D systems of high symmetries. 
For general quasi-2D/3D systems, the method still has $\Or(N_\vk^{-1})$ error but avoids a significant portion of the $\Or(N_\vk^{-1})$ error  in the standard method.

Below, we use the direct term $E_\text{2d}$ to demonstrate the basic idea of the analysis of the staggered mesh method. 
In the standard method, this direct energy term is computed as 
\begin{align*}
                E_\text{2d}(N_\vk)  
                & = \dfrac{1}{N_\vk^3}\sum_{\vk_i,\vk_j, \vk_a \in \mathcal{K}} \left(2\sum_{ijab}\braket{i\vk_i,j\vk_j|a\vk_a,b\vk_b}_\# \dfrac{ \braket{a\vk_a,b\vk_b|i\vk_i,j\vk_j}_\# }{\epsilon_{i\vk_i,j\vk_j}^{a\vk_a, b\vk_b}}\right)
                \\
                & = \dfrac{1}{N_\vk^3}\sum_{\vk_i,\vk_j, \vk_a \in \mathcal{K}}F_\text{2d}(\vk_i, \vk_j, \vk_a),
\end{align*}
where $\braket{n_1\vk_1,n_2\vk_2|n_3\vk_3,n_4\vk_4}_\#
:= 
N_\vk   \braket{n_1\vk_1,n_2\vk_2|n_3\vk_3,n_4\vk_4}$ denotes a normalized ERI term. 
Recall that $\vk_b\in \mathcal{K}$ is uniquely determined by $\vk_i,\vk_j,\vk_a$. 
In the TDL, the summation $\frac{1}{N_\vk}\sum_{\vk\in \mathcal{K}}$ converges to the integral $\frac{1}{|\Omega^*|}\int_{\Omega^*}\ud\vk$, and $E_\text{2d}(N_\vk)$ converges to an integral
\[
E_\text{2d}^\text{TDL}= \dfrac{1}{|\Omega_*|^3}
\int_{\Omega^*} \ud\vk_i\int_{\Omega^*}\ud\vk_j\int_{\Omega^*}\ud\vk_a F_\text{2d}(\vk_i, \vk_j, \vk_a). 
\]
Thus, $E_\text{2d}(N_\vk)$ can be interpreted as a numerical quadrature for $E_\text{2d}^\text{TDL}$ using a trapezoidal rule with uniform mesh $\mathcal{K}\times\mathcal{K}\times \mathcal{K}$ for variables $(\vk_i,\vk_j,\vk_a)$.
If the integrand $F_\text{2d}(\vk_i, \vk_j, \vk_a)$ can be evaluated exactly, then the finite-size error  $E_\text{2d}^\text{TDL} - E_\text{2d}(N_\vk)$ is simply the quadrature error .

Note that $\braket{n_1\vk_1,n_2\vk_2|n_3\vk_3,n_4\vk_4}_\#$ as a function of $\vk_1,\vk_2,\vk_3$ is discontinuous at $\vk_3 - \vk_1 \in \mathbb{L}^*$.
This is because the fraction in \cref{eqn:eri} is ill-defined when $\vk_3 - \vk_1 + \vG = \bm{0}$. 
This difference $\vk_3 - \vk_1$ is referred to as the momentum transfer. 
Thus, $F_\text{2d}(\vk_i, \vk_j, \vk_a)$ is only discontinuous at $\vk_a - \vk_i \in \mathbb{L}^*$ due to the two  ERIs in its definition. 
Ref.~\citenum{XingLiLin2021_2} shows that the quadrature error of $F_\text{2d}$, thus the finite-size error in $E_\text{2d}(N_\vk)$,  has two main sources: the lack of overall smoothness and the placement of quadrature nodes at points where $F_\text{2d}$ is discontinuous. 
In general, both sources  lead to $\Or(N_\vk^{-1})$ quadrature error. 
The standard method samples $\vk_i$ and $\vk_a$ on $\mathcal{K}$, and thus always has many quadrature nodes at $\vk_a - \vk_i = \bm{0}$, i.e., the zero momentum transfer, resulting $\Or(N_\vk^{-1})$ quadrature error from the second source.

The staggered mesh method avoids the second error source. 
Specifically, the quadrature nodes for $F_\text{2d}$ over $(\vk_i,\vk_j,\vk_a)$ are $\mathcal{K}_\text{occ}\times\mathcal{K}_\text{occ}\times\mathcal{K}_\text{vir}$ which satisfies $\vk_a - \vk_i \not\in \mathbb{L}^*$ for any $\vk_i\in\mathcal{K}_\text{occ}, \vk_a \in \mathcal{K}_\text{vir}$. 
As a result, the  quadrature error of the method only comes from the lack of overall smoothness in $F_\text{2d}$. 
A detailed analysis  shows that the nonsmooth terms in $F_\text{2d}$ are in the following two  forms
\begin{equation}\label{eqn:fraction_form}
\dfrac{f(\vq)}{|\vq|^2}\ \text{with}\ f(\vq) = \Or(|\vq|^2) \quad \text{and}\quad \dfrac{f(\vq)}{|\vq|^4}\ \text{with}\ f(\vq) = \Or(|\vq|^4) 
\end{equation}
where $f$ denotes a generic smooth function compactly supported in $\Omega^*$  and $\vq$ is the minimum image of $\vk_a - \vk_i$ in $\Omega^*$.  

Unfortunately, standard Euler-Maclaurin type of analysis of the quadrature error applied to integrand of the form in \cref{eqn:fraction_form} gives overly pessimistic result, and in particular provides no meaningful result of the convergence rate. 
A major improvement has been obtained by Corollary 16 in Ref.~\citenum{XingLiLin2021_2}, which proves that these nonsmooth terms lead to dominant $\Or(N_\vk^{-1})$ quadrature error in the staggered mesh method, and the convergence rate is sharp for general systems. 
However, for quasi-1D systems and certain quasi-2D/3D systems, the discontinuity of these terms is removable  and the error from the first source can be $o(N_\vk^{-1})$.
In these cases, the staggered mesh method has asymptotically smaller finite-size error than the standard method for computing $E_\text{2d}^\text{TDL}$. 
Below is a simple example to illustrate the connections between removable discontinuity and improved quadrature error. 

Consider $h(\vq) = \frac{\vq^TM\vq}{|\vq|^2}H(\vq)$ in a hypercube $V$ where $M$ is a matrix and $H(\vq)$ is smooth and compactly supported in $V$. 
Here $h(\bm{0})$ is indeterminate and usually  set to $0$ in numerical calculation. 
For a trapezoidal rule with $N_\vk$ quadrature nodes that include $\vq = \bm{0}$, setting $h(\bm{0}) = 0$ introduces $\Or(N_\vk^{-1})$ error.
If $M$ is a scaled identity matrix $\alpha I$, then $\lim_{\vq \rightarrow \bm{0}}h(\vq) = \alpha H(\bm{0})$. 
We can then redefine $h(\bm{0}) = \lim_{\vq \rightarrow \bm{0}}h(\vq)$,  and the resulting $h(\vq)$ is a smooth, compactly supported function. 
In this case, $h(\vq)$ is said to have \emph{removable discontinuity} and a trapezoidal rule without $\vq = \bm{0}$ has super-algebraically decaying  error according to the Euler-Maclaurin formula. 
Otherwise, $\lim_{\vq \rightarrow \bm{0}}h(\vq)$ does not exist and the quadrature error can be shown to scale as $\Or(N_\vk^{-1})$\cite{XingLiLin2021_2} even if $\vq = \bm{0}$ is not a quadrature node.
Similar discussion can give quantitative conditions \cite{XingLiLin2021_2} for the removable discontinuity of the nonsmooth terms \cref{eqn:fraction_form} in MP2 calculations. 
Numerical observations suggest that systems with higher symmetries, such as these silicon and diamond systems tested in \cref{sec:numerical}, are more likely to satisfy these conditions and thus have asymptotically smaller  finite-size errors using staggered mesh method.

Similar discussion also applies to the exchange term in MP2 energy but with additional nonsmooth terms in the form
\[
\dfrac{f(\vq,\vq')}{|\vq|^2 |\vq'|^2}\ \text{with}\ f(\vq,\vq') = \Or(|\vq|^2|\vq'|^2),
\]
where $\vq'$ denotes the minimum image of $\vk_a - \vk_j$ in $\Omega^*$ and $f(\vq, \vq')$ is compactly supported in $\Omega^*\times \Omega^*$. We refer readers to Ref.~\citenum{XingLiLin2021_2} for details.


\subsection{Staggered mesh method for RPA/RPA-SOSEX in the drCCD formalism}\label{subsec:rpa}


Now consider an $m$-th order perturbation energy term included in RPA/RPA-SOSEX energy expansions discussed in \cref{subsec:expansion}.
Denote its standard calculation using one MP mesh $\mathcal{K}$ as $E_m(N_\vk)$. 
The TDL energy $E_m^\text{TDL}$ can be formulated as an integral over $(m+1)$ momentum vectors in $\Omega^*$, and 
$E_m(N_\vk)$ corresponds to a trapezoidal quadrature rule for estimating $E_m^\text{TDL}$. 
If the integrand can be accurately evaluated, 
the finite-size error $E_m^\text{TDL} - E_m(N_\vk)$ is given by the quadrature error. 
Similar to the case of MP2 calculations, the quadrature error for $E_m^\text{TDL}$ has two main sources: the lack of overall smoothness in the integrand and the placement of quadrature nodes at points of discontinuity.

For gapped systems, the integrand for $E_m^\text{TDL}$ has its points of discontinuity described by a simple rule: it is discontinuous at $\vk_3 - \vk_1 \in \mathbb{L}^*$ for each involved ERI $\braket{n_1\vk_1,n_2\vk_2|n_3\vk_3,n_4\vk_4}$.
For example, the integrand for $E_\text{3x,1}^\text{TDL}$ in \cref{eqn:rpa_mp3} is a function of  $\vk_i,\vk_j,\vk_a, \vk_k$, and is discontinuous at $\vk_a - \vk_i \in \mathbb{L}^*$,
$\vk_i - \vk_a \in \mathbb{L}^*$,  or $\vk_c - \vk_k \in \mathbb{L}^*$ due to the three involved ERIs ($\vk_c = \vk_a + \vk_k - \vk_i$ by crystal momentum conservation). 
For the standard method with all $\vk$ on the same mesh $\mathcal{K}$, the calculation of $E_m(N_\vk)$ triggers zero momentum transfer $\vk_3 - \vk_1 = \bm{0}$ in many involved ERIs, and the error due to the placement of quadrature nodes at points of discontinuity can be shown to be $\Or(N_\vk^{-1})$. 

Following the analysis in Ref.~\citenum{XingLiLin2021_2}, it can be shown similarly that the nonsmooth terms in the integrand associated with ring diagrams for $E_m^\text{TDL}$ are in the following form 
\begin{equation}\label{eqn:rpa_direct_fraction}
\dfrac{f(\vq)}{|\vq|^{2s}}\ \text{with}\ f(\vq) = \Or(|\vq|^{2s}), \quad s = 1,2,\ldots, m.
\end{equation}
Here $f(\vq)$ denotes a generic smooth function compactly supported in $\Omega^*$ and $\vq$ is the minimum image of $\vk_a - \vk_i$ in $\Omega^*$. 
If $E_m^\text{TDL}$ is associated with an exchange term, there are additional nonsmooth terms of form: 
\begin{equation}\label{eqn:rpa_exchange_fraction}
\dfrac{f(\vq, \vq')}{|\vq|^{2s}|\vq'|^2}\ \text{with}\ f(\vq,\vq') = \Or(|\vq|^{2s}|\vq'|^2), \quad s = 1,2,\ldots, m-1,
\end{equation}
where $\vq'$ denotes the minimum image of $\vk_a - \vk_j$ in $\Omega^*$. 
It is important to realize that Corollary 16  in Ref.~\citenum{XingLiLin2021_2} is also applicable to the nonsmooth forms in \cref{eqn:rpa_direct_fraction,eqn:rpa_exchange_fraction}. Therefore their resulting quadrature errors still scale as $\Or(N_\vk^{-1})$. 

Overall, the quadrature error in the standard calculation of each finite order energy term in RPA and RPA-SOSEX correlation energies, i.e., $E_m^\text{TDL} - E_m(N_\vk)$, scales as $\Or(N_\vk^{-1})$. 
Using the drCCD formalism, it can be shown that the numerical calculation of each finite order energy term  induced by the standard calculation of RPA and RPA-SOSEX uses exactly the same mesh $\mathcal{K}$.
This error estimate  is consistent with the numerical results in \cref{sec:numerical}.

\begin{figure}[htbp]
\centering
\includegraphics[width=0.8\textwidth]{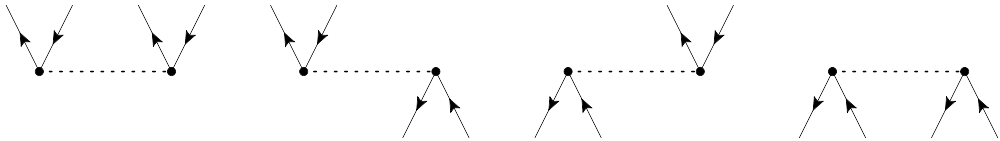}
\caption{Four types of interaction vertices included in any finite order energies included in RPA and RPA-SOSEX correlation energies (in the Goldstone diagram formalism).}
\label{fig:eri_vertex}
\end{figure}

In order to reduce the finite-size error, we first note that from the diagram representation of $E_m(N_\vk)$, all the interaction vertices (including both ring diagrams and their exchanges) are in the four forms in \cref{fig:eri_vertex}, corresponding to ERIs of form $\braket{vv|oo}$, $\braket{vo|ov}$, $\braket{ov|vo}$, and $\braket{oo|vv}$ where ``$v$'' and ``$o$'' stand for virtual and occupied orbitals, respectively.
Thus, all the ERIs in the integrand for $E_m^\text{TDL}$ belong to these four forms, 
and the points of discontinuity of the integrand, or equivalently the zero momentum transfers of all the involved ERIs,  are always associated with the difference between a pair of occupied and virtual momentum vectors being in $\mathbb{L}^*$. 

In the standard calculation, all occupied and virtual momentum vectors belong to $\mathcal{K}$, and there are always quadrature nodes at those points of discontinuity (with zero momentum transfer), resulting $\Or(N_\vk^{-1})$ finite-size error. 
In the staggered mesh method, the difference between any two sampled occupied and virtual momentum vectors is guaranteed not to be in $\mathbb{L}^*$. 
The staggered mesh method thus completely avoids the second error source above, and its quadrature error only comes from the lack of overall smoothness in the integrand. 
Similar to the discussion in Ref.~\citenum{XingLiLin2021_2}, the discontinuity of the nonsmooth forms in \cref{eqn:rpa_direct_fraction,eqn:rpa_exchange_fraction} can 
be removable for quasi-1D and certain quasi-2D/3D systems with high symmetries. 
In these cases, the quadrature error of the staggered mesh method can be $o(N_\vk^{-1})$. 


\subsection{Staggered mesh method for RPA in the AC formalism}\label{subsec:ac}
The analysis of the finite-size errors in \cref{subsec:rpa} is mainly based on the special structure of ring diagrams. 
Now we demonstrate a preliminary analysis of the finite-size error in the AC formalism.
For convenience of the discussion of the ``head/wing'' correction later, we use a planewave basis set. The AC formula in \cref{eqn:acfdt_energy} for RPA can be rewritten as
\begin{equation}\label{eqn:acfdt_energy_q}
        E_\text{rpa}(N_\vk)=\dfrac{1}{N_\vk}\sum_{\vq \in \mathcal{K}_\vq}\frac{1}{4 \pi} \int_{-\infty}^{\infty} \operatorname{Tr}\left(  \log(\varepsilon(\mathrm{i}\omega, \vq))\right)  + \sum_{\vG\in\mathbb{L}^*}(1- \varepsilon_{\vG,\vG}(\mathrm{i}\omega, \vq))\mathrm{d} \omega, 
\end{equation}
where the symmetrized dielectric operator $\varepsilon(\I\omega, \vq)$ indexed by $\vG, \vG'\in \mathbb{L}^*$ is defined as
\begin{equation}\label{eqn:dielectric}
\varepsilon_{\mathbf{G}, \mathbf{G}^{\prime}}(\mathrm{i} \omega, \mathbf{q})=\delta_{\mathbf{G}, \mathbf{G}^{\prime}}-\frac{\sqrt{4 \pi}}{|\mathbf{q}+\mathbf{G}|} \Pi_{\mathbf{G}, \mathbf{G}^{\prime}}(\mathrm{i} \omega, \mathbf{q}) \frac{\sqrt{4 \pi}}{\left|\mathbf{q}+\mathbf{G}^{\prime}\right|},
\end{equation}
and
\begin{equation}\label{eqn:acfdt_Pi}
\Pi_{\mathbf{G}, \mathbf{G}^{\prime}}(\mathrm{i} \omega, \vq)=\frac{4}{N_\vk|\Omega|} \sum_{\vk_i\in\mathcal{K}} \sum_{i a} \hat{\varrho}_{i \vk_i, a (\vk_i + \vq)}(\mathbf{G}) \hat{\varrho}^*_{i \mathbf{k}_{i}, a(\vk_i+\vq)}\left(\mathbf{G}^{\prime}\right) \frac{\epsilon_{i \mathbf{k}_{i}}-\epsilon_{a (\vk_i+\vq)}}{\omega^{2}+\left(\epsilon_{a (\vk_i+\vq)}-\epsilon_{i \mathbf{k}_{i}}\right)^{2}}. 
\end{equation}
Here, $\vq$ is defined as the minimum image of $\vk_a - \vk_i$ in the first Brillouin zone, and $\mathcal{K}_\vq$ is an MP mesh for $\vq$ induced by sampling $\vk_i,\vk_a \in \mathcal{K}$ in the standard method. 
In this case, $\mathcal{K}_\vq$ contains the $\Gamma$ point $\vq = \bm{0}$ and is of the same size as $\mathcal{K}$. 

In the TDL, the summations $\frac{1}{N_\vk}\sum_{\vq \in \mathcal{K}_\vq}$ in \cref{eqn:acfdt_energy_q} and $\frac{1}{N_\vk}\sum_{\vk_i \in \mathcal{K}}$ in \cref{eqn:acfdt_Pi} converge to the integrals $\frac{1}{|\Omega^*|}\int_{\Omega^*}\ud\vq$ and $\frac{1}{|\Omega^*|}\int_{\Omega^*}\ud\vk_i$, respectively. 
Assuming all the orbitals and orbital energies are obtained exactly, the finite-size error $E_\text{rpa}(N_\vk) - E_\text{rpa}^\text{TDL}$ comes from the quadrature errors in the integrals over $\vq$ and $\vk_i$. 
First note that, given $\omega, \vG, \vG',$ and $\vq$, the integrand for  $\Pi_{\mathbf{G}, \mathbf{G}^{\prime}}(\mathrm{i}\omega, \vq)$ in \cref{eqn:acfdt_Pi} is smooth and periodic with respect to $\vk_i$. 
Thus the quadrature error for computing each  $\Pi_{\mathbf{G}, \mathbf{G}^{\prime}}(\mathrm{i}\omega, \vq)$ decays super-algebraically according to the Euler-Maclaurin formula for trapezoidal quadrature rules. 
In other words, the dominant finite-size error in RPA energy calculation comes from the integral over $\vq$ in \cref{eqn:acfdt_energy_q}. 

For the integration over $\vq$ in \cref{eqn:acfdt_energy_q}, we note that $\varepsilon_{\vG, \vG'}(\text{i}\omega, \vq)$ is discontinuous and indeterminate at $\vq = \bm{0}$ when $\vG$ or $\vG'$ equals  $\bm{0}$ by the definition in \cref{eqn:dielectric}. 
More specifically, $\varepsilon_{\bm{0},\bm{0}}(\mathrm{i}\omega, \vq) , \varepsilon_{\bm{0},\vG}(\mathrm{i}\omega, \vq)$, and $\varepsilon_{\vG,\bm{0}}(\mathrm{i}\omega, \vq)$ with $\vG\neq \bm{0}$ are $\Or(1)$ for $\vq$ near $\bm{0}$ but do not converge when $\vq \rightarrow \bm{0}$. 
The standard method (which has $\bm{0} \in \mathcal{K}_\vq$) neglects these $\Or(1)$ discontinuous terms in \cref{eqn:dielectric} at $\vq = \bm{0}$ and sets $\varepsilon_{\bm{0},\bm{0}}(\mathrm{i}\omega, \bm{0}) = 1$, $\varepsilon_{\bm{0},\vG}(\mathrm{i}\omega, \bm{0}) = \varepsilon_{\vG,\bm{0}}(\mathrm{i}\omega, \bm{0}) = 0$.
As a result, there is $\Or(1)$ quadrature error in the $\frac{|\Omega^*|}{N_\vk}$-sized volume element centered at $\vq = \bm{0}$, leading to overall $\Or(N_\vk^{-1})$ quadrature error for the integration over $\vq$. 
In the staggered mesh method, $\mathcal{K}_\vq$ induced by $\vk_i \in \mathcal{K}_\text{occ}, \vk_a \in \mathcal{K}_\text{vir}$ does not contain $\vq = \bm{0}$ and thus the above $\Or(N_\vk^{-1})$ quadrature error is completely avoided. 
The remaining quadrature error of the staggered mesh method (which also presents in the standard method) is due to the lack of overall smoothness of the integrand with respect to $\vq$  in \cref{eqn:acfdt_energy_q}, i.e.,  
\[
\frac{1}{4 \pi} \int_{-\infty}^{\infty} \operatorname{Tr}\left(  \log(\varepsilon(\mathrm{i}\omega, \vq))\right)  + \sum_{\vG\in\mathbb{L}^*}(1- \varepsilon_{\vG,\vG}(\mathrm{i}\omega, \vq))\mathrm{d} \omega.
\]
The nonsmoothness of this integrand with respect to $\vq$ can be difficult to analyze directly due to the $\text{Tr}(\text{log}(\cdot))$ operation and the integral over $\omega$. 
Existing quadrature error analysis results cannot be applied in this case, and further studies are needed.
One possible approach is to apply the Taylor expansion of $\log(\varepsilon(\mathrm{i}\omega, \vq))=\log(I+(\varepsilon(\mathrm{i}\omega, \vq)-I))$ with respect to $(\varepsilon(\mathrm{i}\omega, \vq)-I)$, and then expand the trace operation. Each resulting term takes the form of a ring diagram, which is similar to that of the perturbative expansion in the drCCD formalism in \cref{subsec:expansion}.
After taking into account the imaginary energy $\I\omega$, we could then apply a similar error analysis over each finite order term.

It is worth noting that in periodic GW calculation\cite{ZhuChan2021,WilhelmHutter2017,FreysoldtEggertEta2007}, the ``head/wing'' correction to the dielectric operator $\varepsilon_{\vG, \vG'}(i\omega, \vq)$ is often used to reduce the finite-size error of the self energy. 
In this correction scheme, $\varepsilon_{\bm{0},\bm{0}}(\mathrm{i}\omega, \bm{0})$ and $\varepsilon_{\bm{0},\vG}(\mathrm{i}\omega, \bm{0}), \varepsilon_{\vG,\bm{0}}(\mathrm{i}\omega, \bm{0})$ (which are called the head and wings, respectively) are replaced by their values at some $\vq$ points that are very close to $\bm{0}$ in numerical calculations. 
This correction can also be used in the standard RPA correlation energy calculation. 
\REV{In this case, the head/wing correction is similar to the staggered mesh method in that it also
 compensates the omitted integrand at $\vq = \bm{0}$ (but by computing the integrand value at some neighboring $\vq$ points).}
It is worth mentioning that our analysis above shows that the staggered mesh method completely avoids the need of performing such ``head/wing'' corrections. Numerical results suggest that such corrections do not provide significant improvements in standard RPA energy calculations. We refer readers to \cref{fig:rpa_correction} in Appendix. 

\section{Conclusion}

We develop a staggered mesh method to reduce finite-size errors for RPA and RPA-SOSEX correlation energy calculations for insulating systems. 
This is demonstrated using the drCCD and AC formalisms. 
The two formalisms have significant differences in terms of their implementation. Nonetheless, they are equivalent formulations of the RPA correlation energy, and the numerical values of the energies can very well match with each other with proper choice of parameters (e.g., number of quadrature points along the imaginary energy axis).
Numerical evidence and analytic discussions both show that the staggered mesh method can significantly reduce the finite-size error compared to the standard method for quasi-1D and certain quasi-2D and 3D systems with high symmetries.


Our analysis of the finite-size error is conducted for each term of the perturbative expansion as ring diagrams. 
Such a treatment can be justified when the expansion converges. 
However, it would be more desirable if the analysis can be directly performed for the infinite series (such as \cref{eqn:acfdt_energy_q} in the AC formalism).
This can be particularly important for the treatment of metallic systems, of which each individual perturbative term may diverge as in the case of the electron gas. 
The treatment of metallic systems also introduces other difficulties, such as the singularity in the orbital energy fraction terms $1/\epsilon^{AB}_{IJ}$, which are beyond the scope of the paper.


In the RPA and RPA-SOSEX energy calculations, the non-smoothness of the integrand is entirely due to the transition between a pair of occupied and virtual orbitals, which is also the foundation of the effectiveness of the staggered mesh method. 
An immediate question is whether the method can be applied to arbitrary perturbation energies included in other correlation energy calculations such as MP3 and CCSD. 
Unfortunately, even in the MP3 correlation energy, there are terms with momentum transfer between two occupied orbitals or two virtual orbitals, e.g., the four-hole-two-particle energy term. 
\REV{For instance, the leading nonsmooth term of $\braket{ab\|cd}$ scales as $1/|\vq|^2$ when $\vq\to\bm{0}$. 
This is similar to the Fock exchange energy calculation, and the quadrature error of this term is $\mathcal{O}(N_\vk^{-\frac13})$. 
It may be necessary to introduce an additional correction term to remove the singular $1/|\vq|^2$ term before applying the staggered mesh method. Another technical difficulty is that Corollary 16 in Ref.~\citenum{XingLiLin2021_2} cannot be readily used to analyze the finite-size error of such four-hole-two-particle energy terms.} 
Nonetheless, based on the error analysis of the drCCD formalism, the staggered mesh method may still be effective for certain CCD calculations that include more than the particle-hole ring contractions in drCCD, as long as the involved ERIs are in the four forms in \cref{fig:eri_vertex}. 
On the other hand, from the perspective of the AC formalism, the staggered mesh method naturally avoids the need of performing ``head/wing'' corrections, and therefore can be useful in reducing the finite-size errors of other physical quantities, such as the quasi-particle energies in GW calculations.

\section*{Acknowledgement}
The work of L.L. was in part supported by the Air Force Office of Scientific Research under award number FA9550-18-1-0095, by the U.S. Department of Energy, Office of Science, Office of Advanced Scientific Computing Research under Grant No. DE-SC0017867 and by the National Science Foundation under Grant No. DMS-1652330. This material is based upon work supported by the U.S. Department of Energy, Office of Science, Office of Advanced Scientific Computing Research and Office of Basic Energy Sciences, Scientific Discovery through Advanced Computing (SciDAC) program under Award Number DE‐SC0022198 (X.X.). We thank Garnet Chan and Tianyu Zhu for their helpful discussions and for providing the PyRPA software package.


\appendix
\section*{Appendix}
\begin{figure}[htbp]
        \centering
        \captionsetup[subfigure]{labelformat=empty}
        \subfloat[H2 quasi-1D]{
                \includegraphics[width=0.24\textwidth]{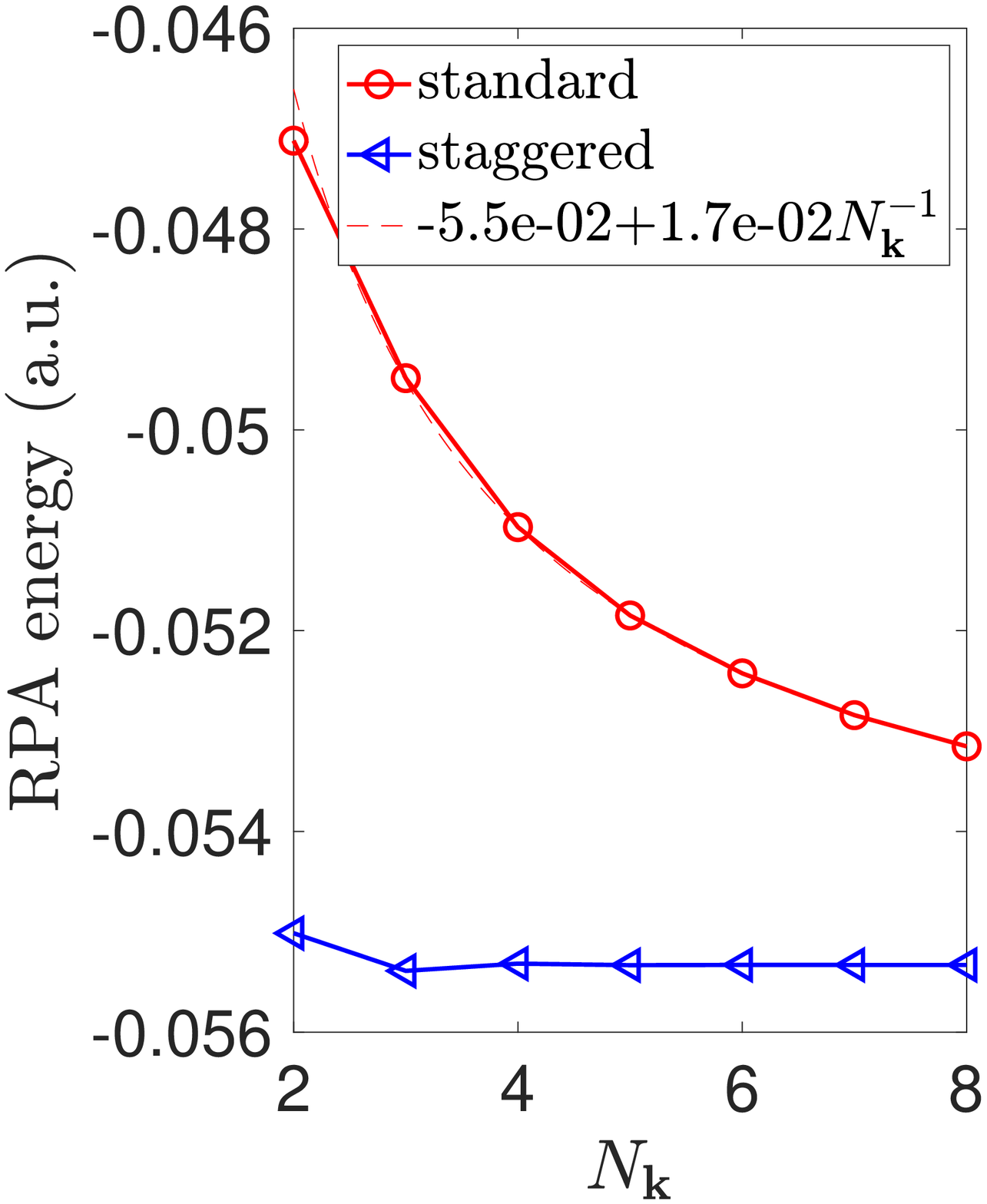}
        }
        \subfloat[LiH quasi-1D]{
                \includegraphics[width=0.24\textwidth]{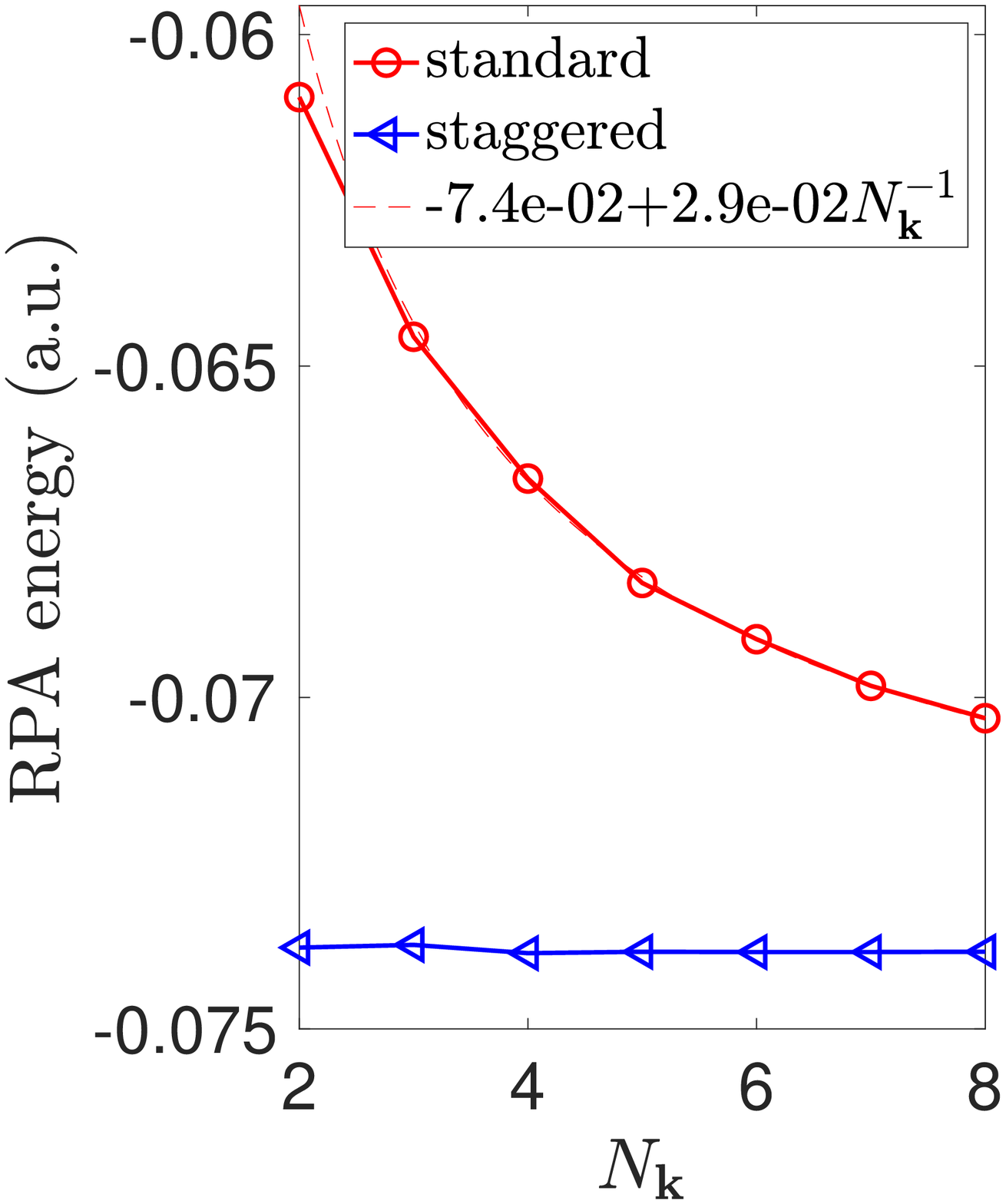}
        }
        \subfloat[Si quasi-1D]{
                \includegraphics[width=0.24\textwidth]{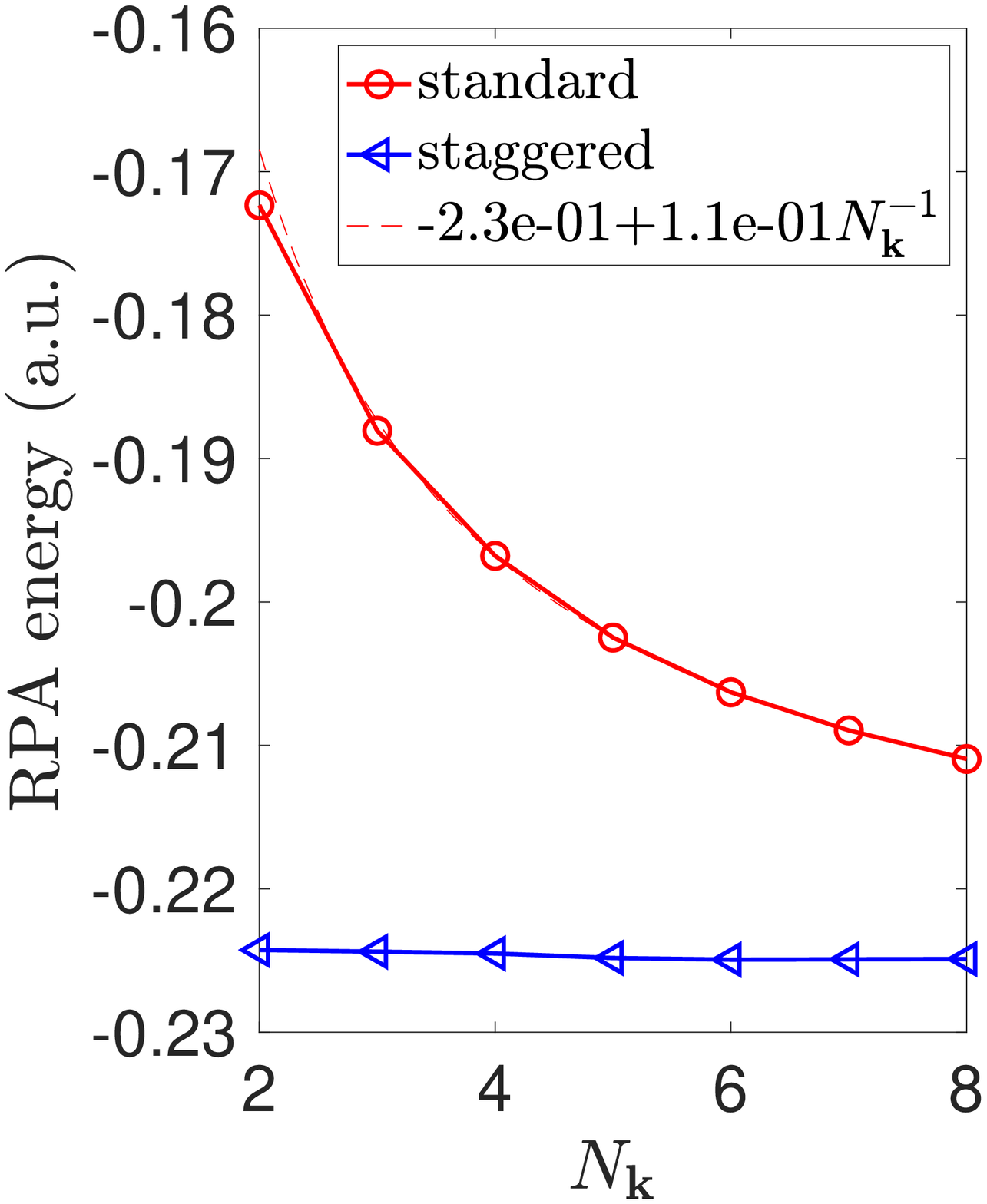}
        }
        \subfloat[Diamond quasi-1D]{
                \includegraphics[width=0.24\textwidth]{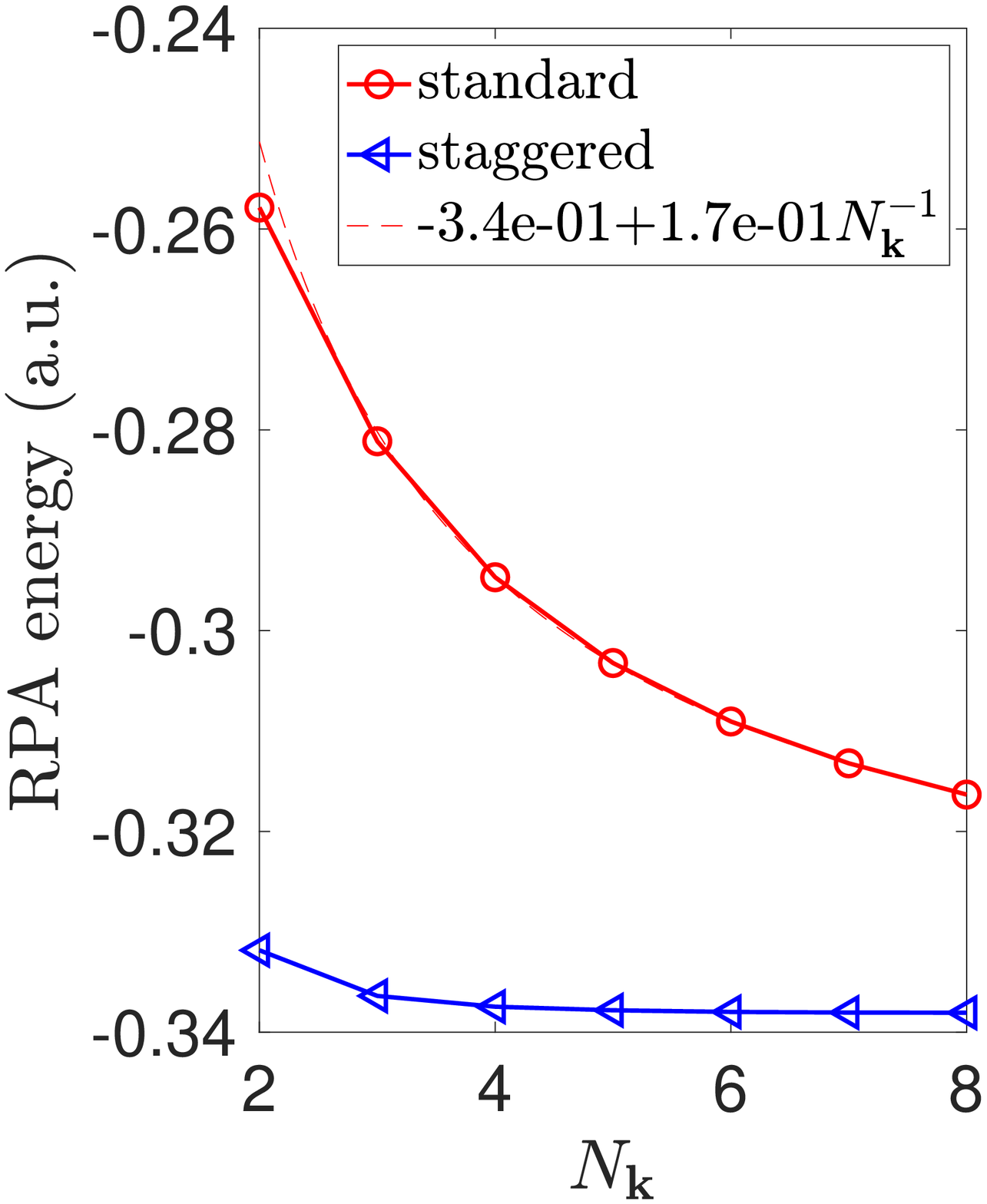}
        }
        
        \subfloat[H2 quasi-2D]{
                \includegraphics[width=0.24\textwidth]{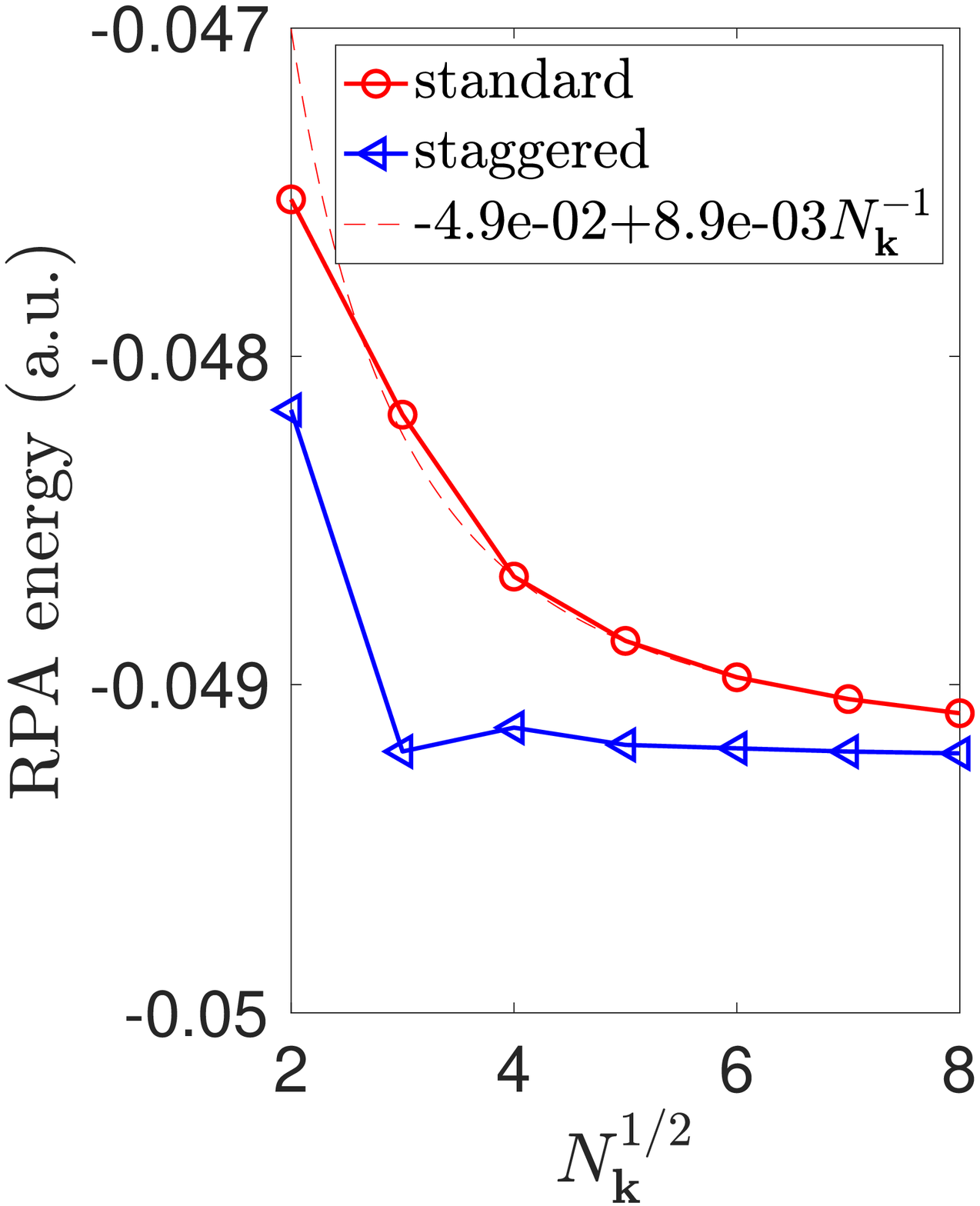}
        }
        \subfloat[LiH quasi-2D]{
                \includegraphics[width=0.24\textwidth]{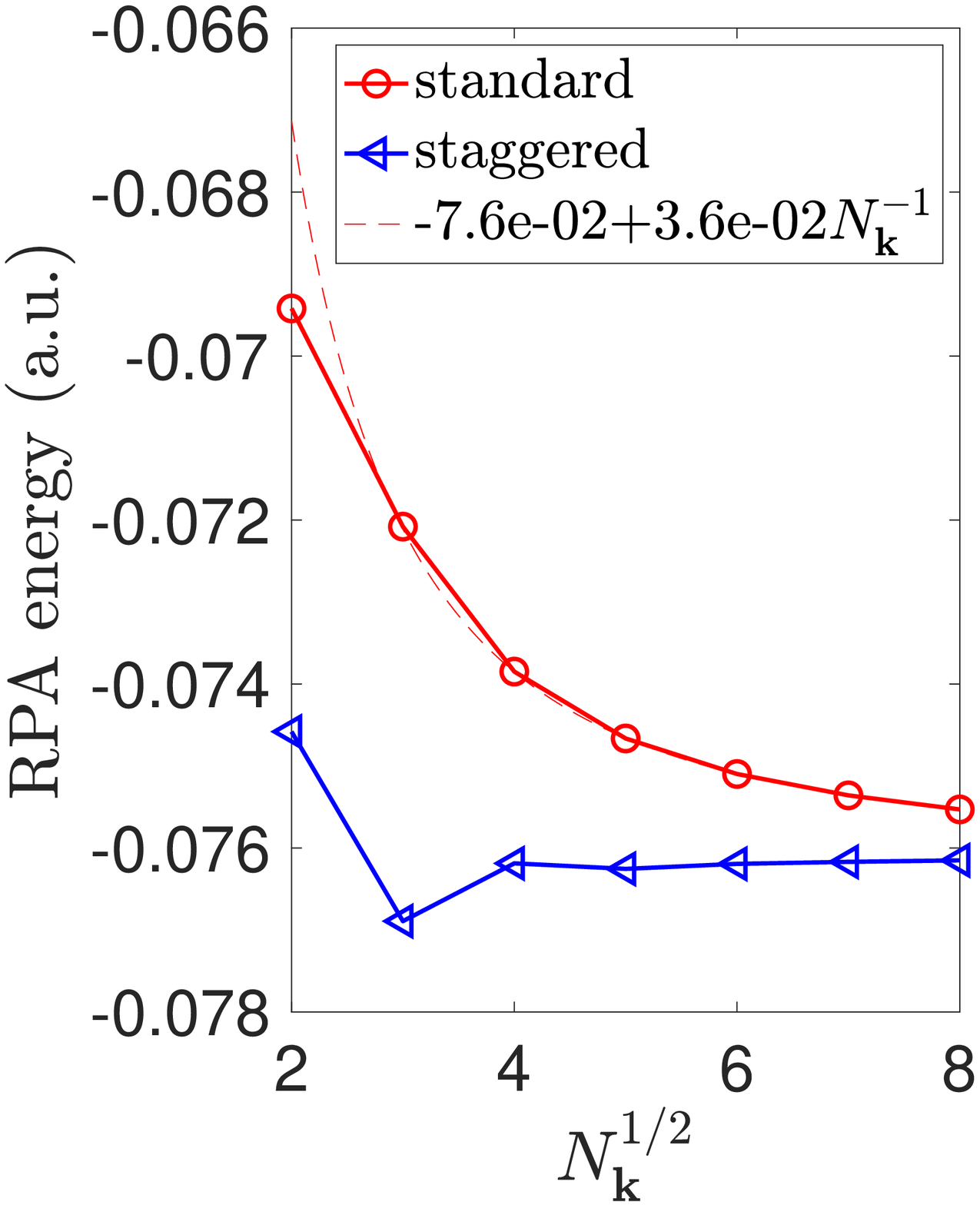}
        }
        \subfloat[Si quasi-2D]{
                \includegraphics[width=0.24\textwidth]{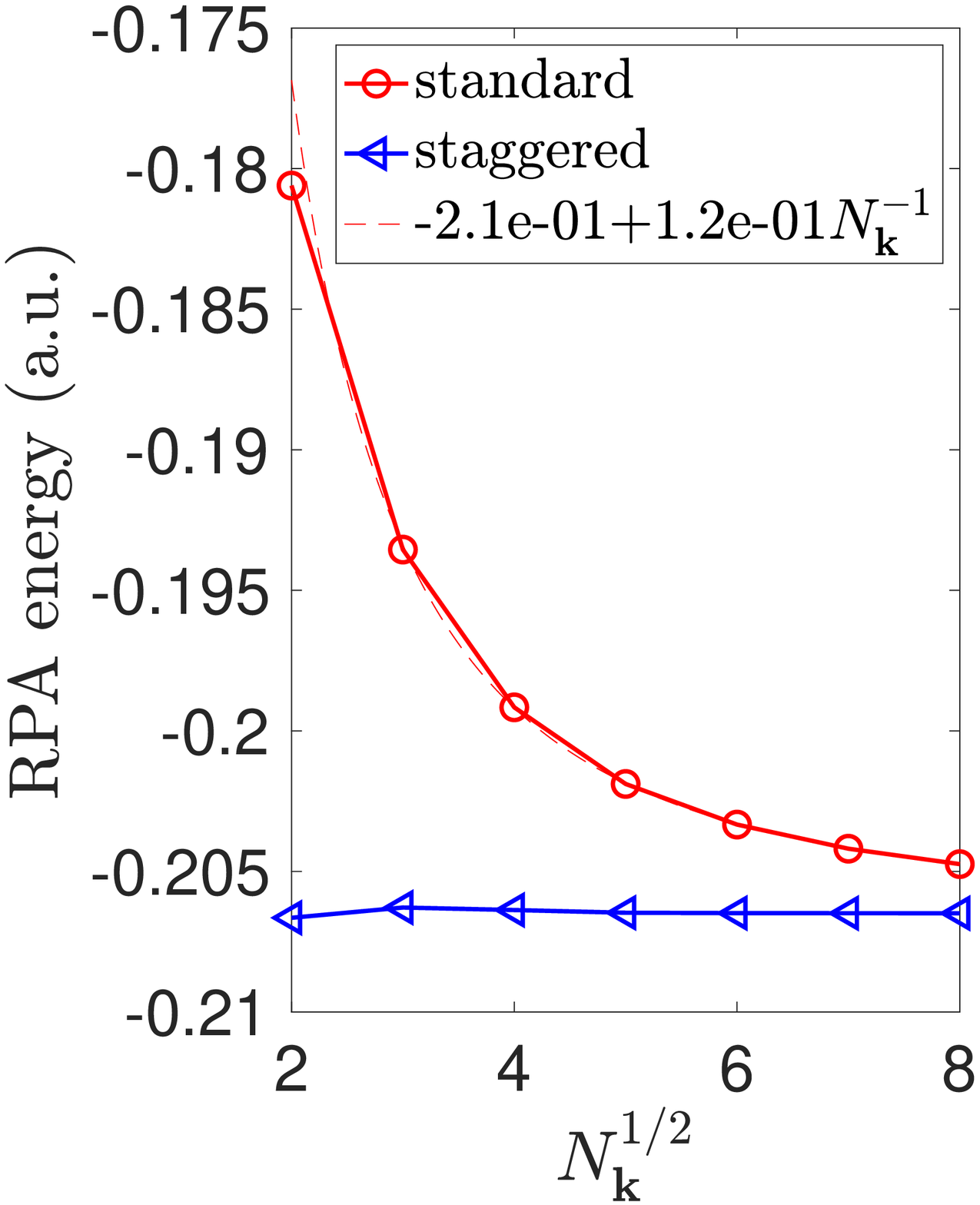}
        }
        \subfloat[Diamond quasi-2D]{
                \includegraphics[width=0.24\textwidth]{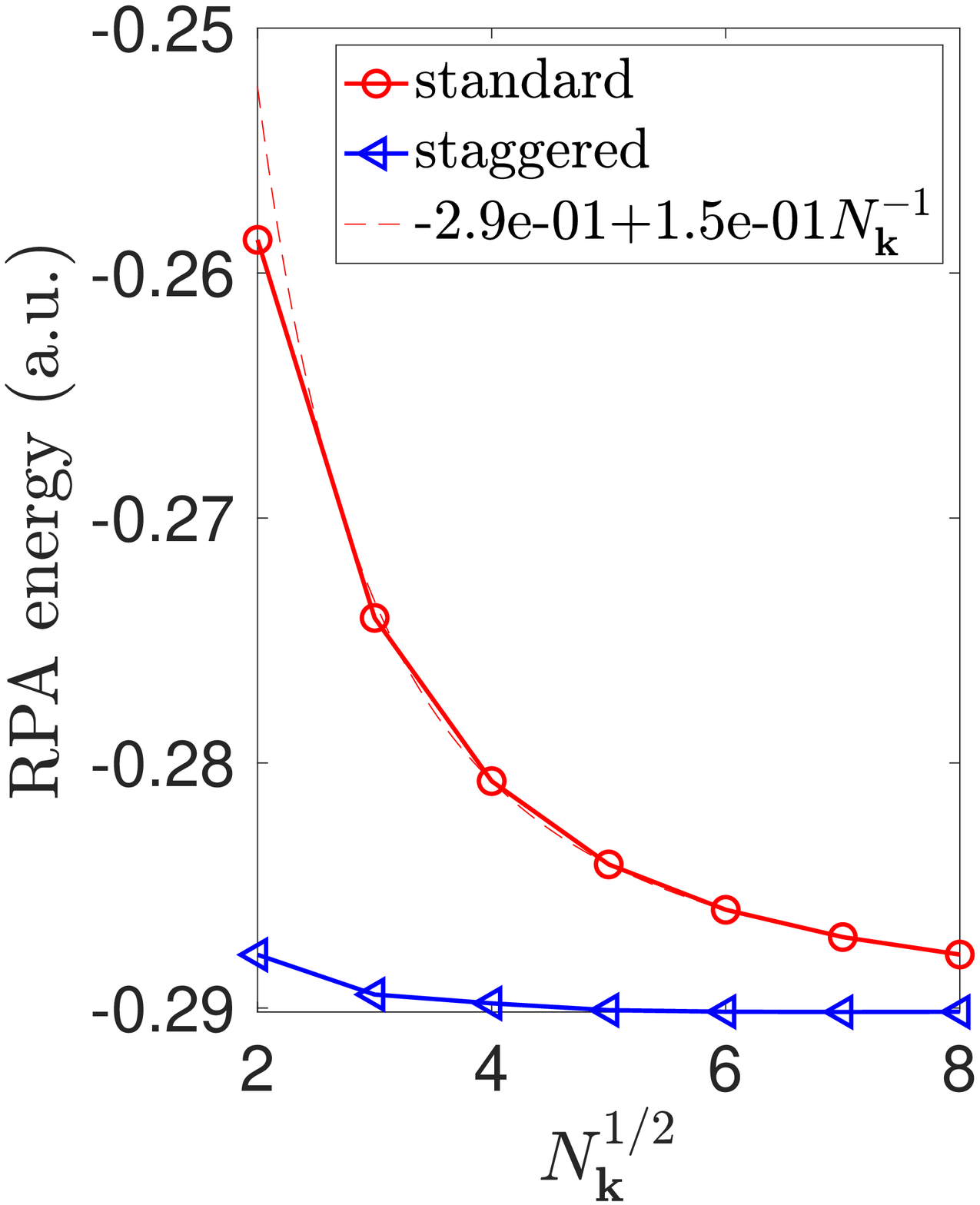}
        }
        
        \subfloat[H2 3D]{
                \includegraphics[width=0.24\textwidth]{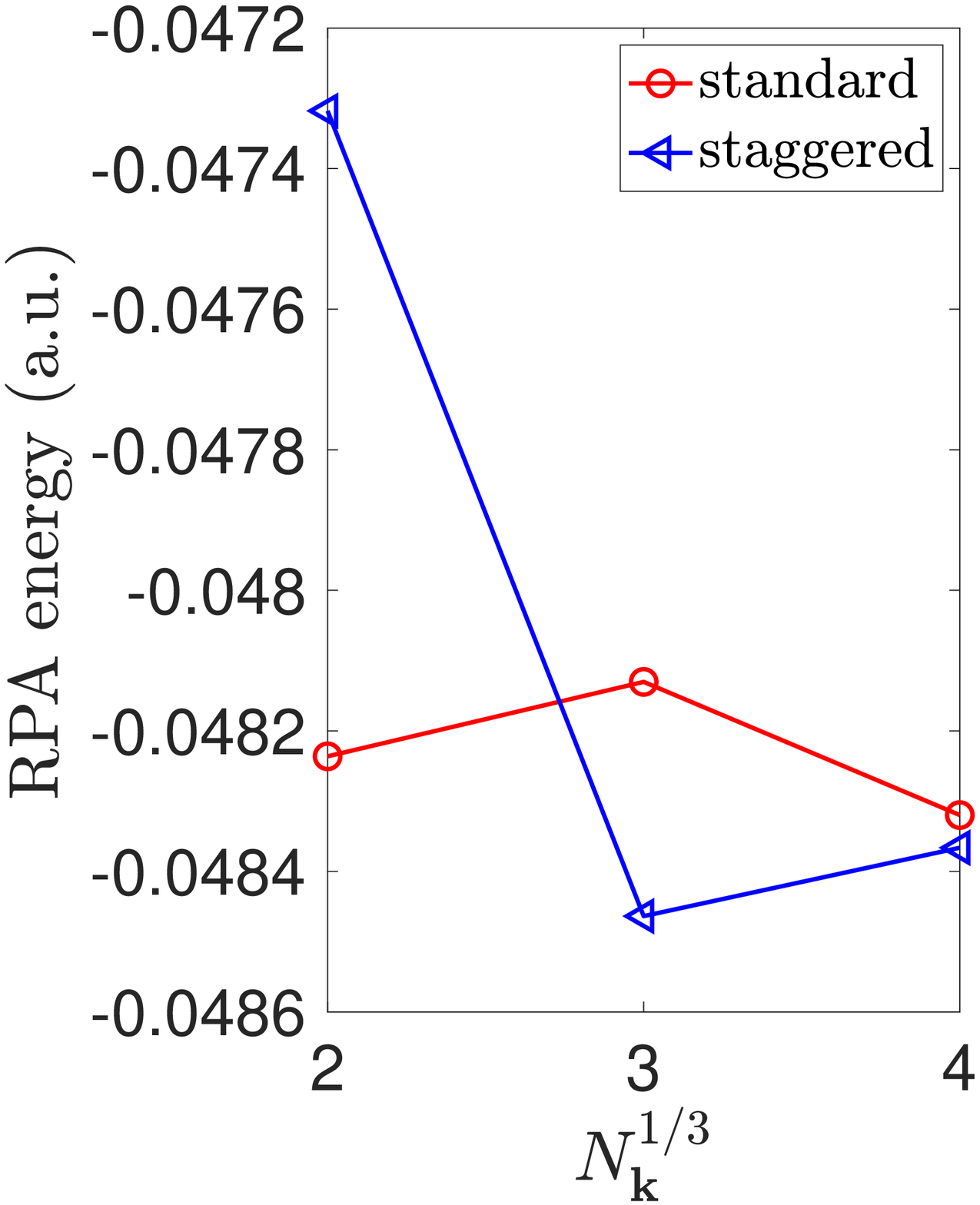}
        }
        \subfloat[LiH 3D]{
                \includegraphics[width=0.24\textwidth]{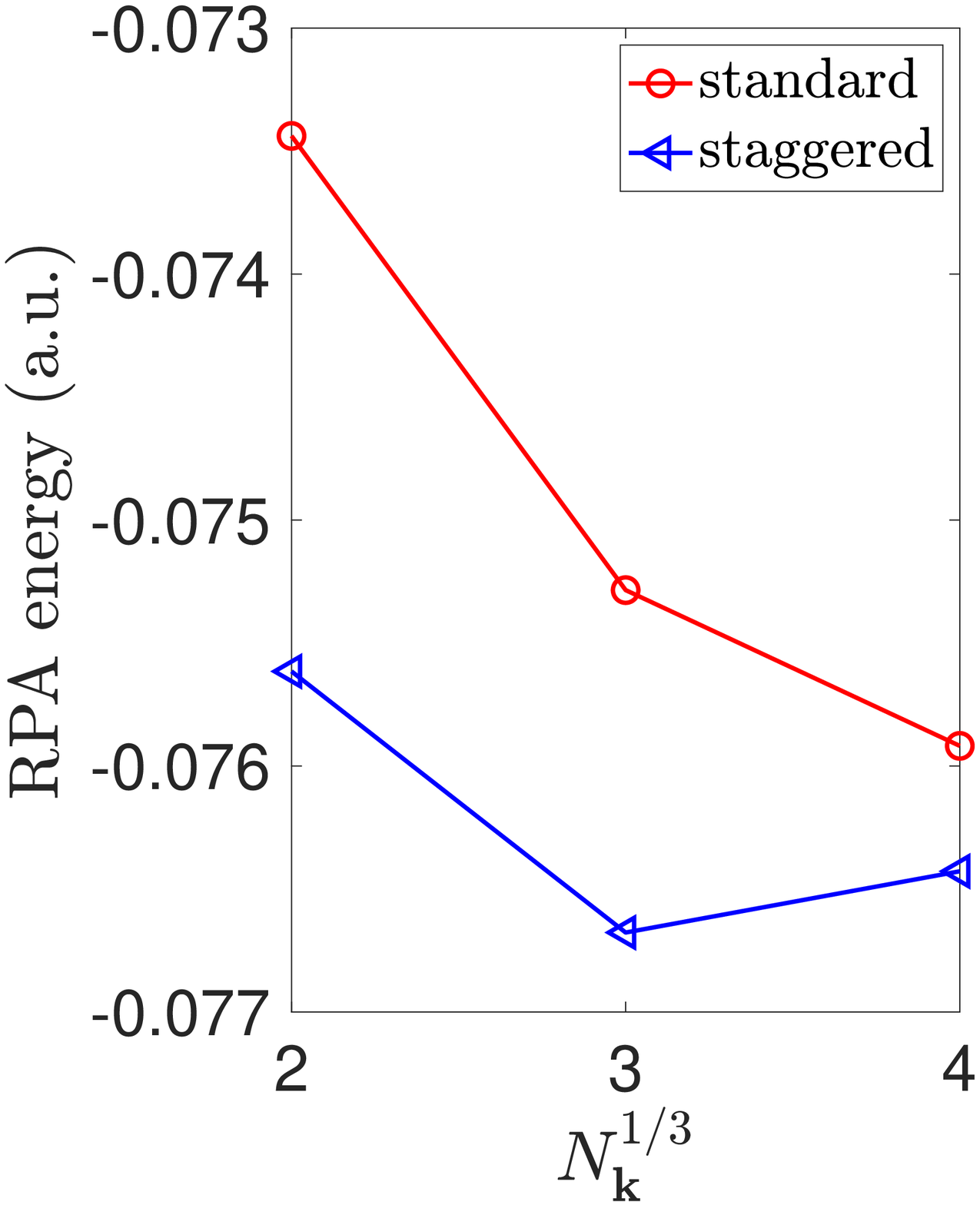}
        }
        \subfloat[Si 3D]{
                \includegraphics[width=0.24\textwidth]{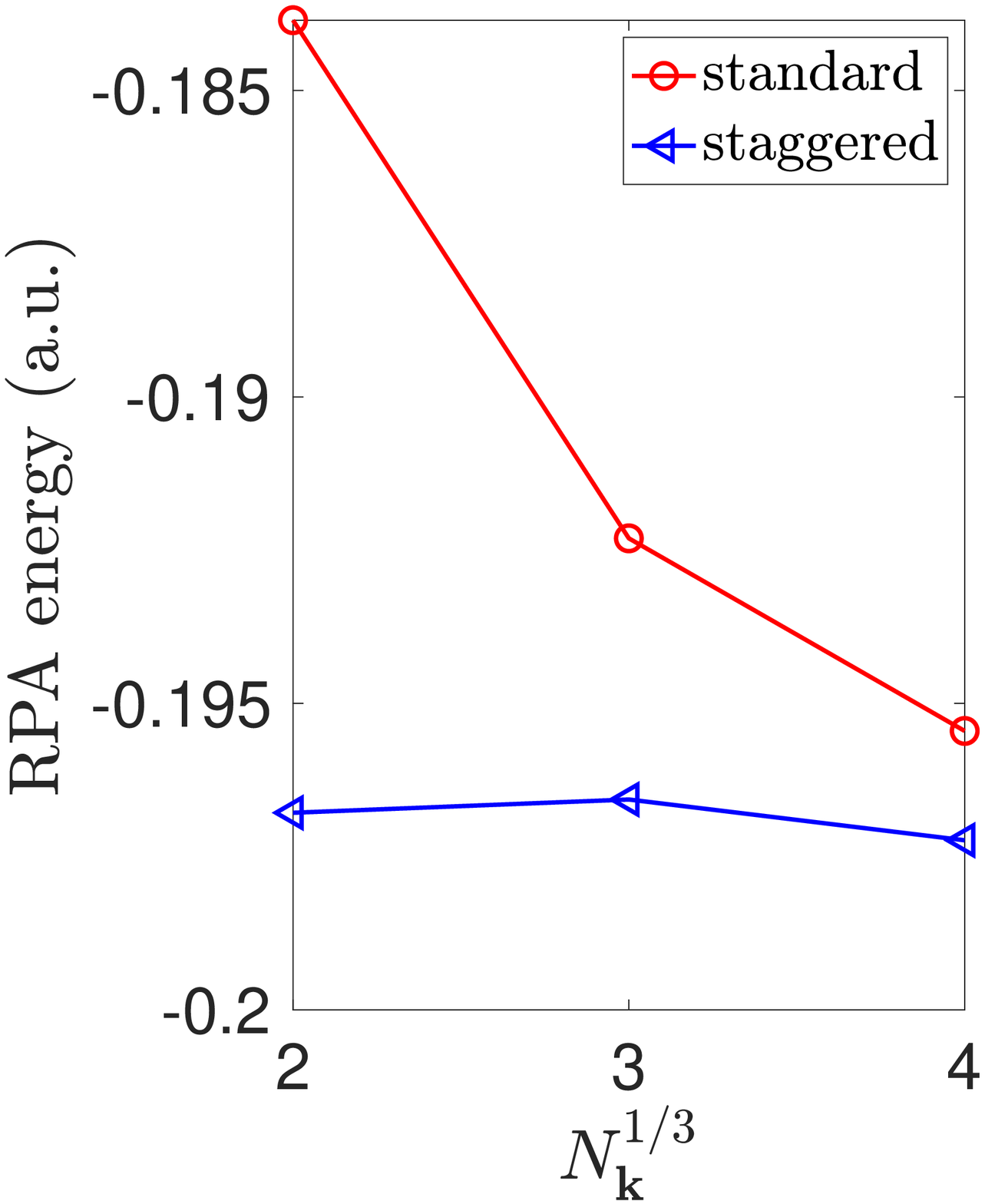}
        }
        \subfloat[Diamond 3D]{
                \includegraphics[width=0.24\textwidth]{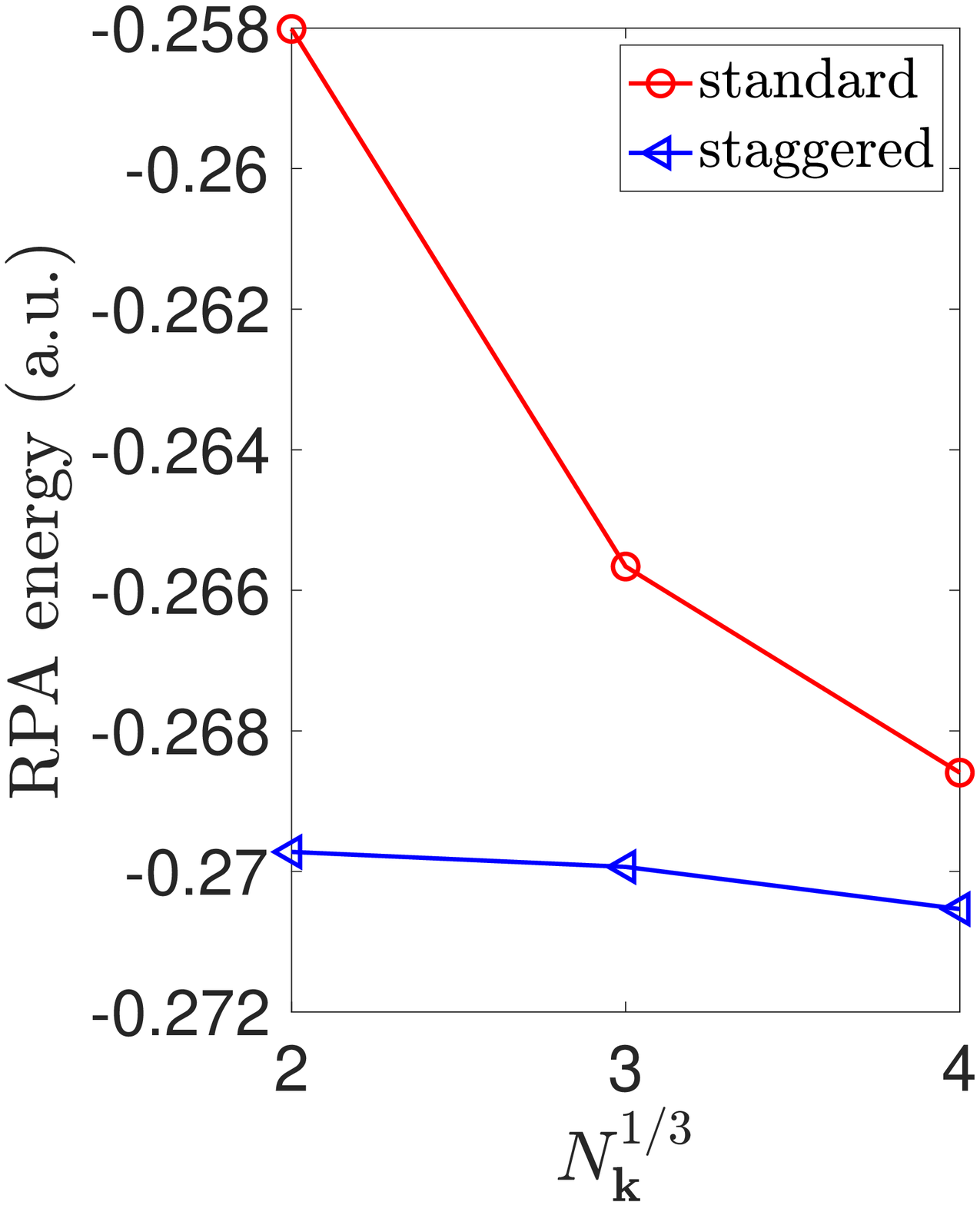}
        }
        
        \caption{RPA correlation energy per unit cell computed by the standard and the staggered mesh methods using the AC formalism and the larger gth-dzvp basis set.
        }
        \label{fig:rpa_dzvp}
\end{figure}

\begin{figure}[htbp]
        \centering
        \captionsetup[subfigure]{labelformat=empty}
        \subfloat[H2 gth-szv]{
                \includegraphics[width=0.24\textwidth]{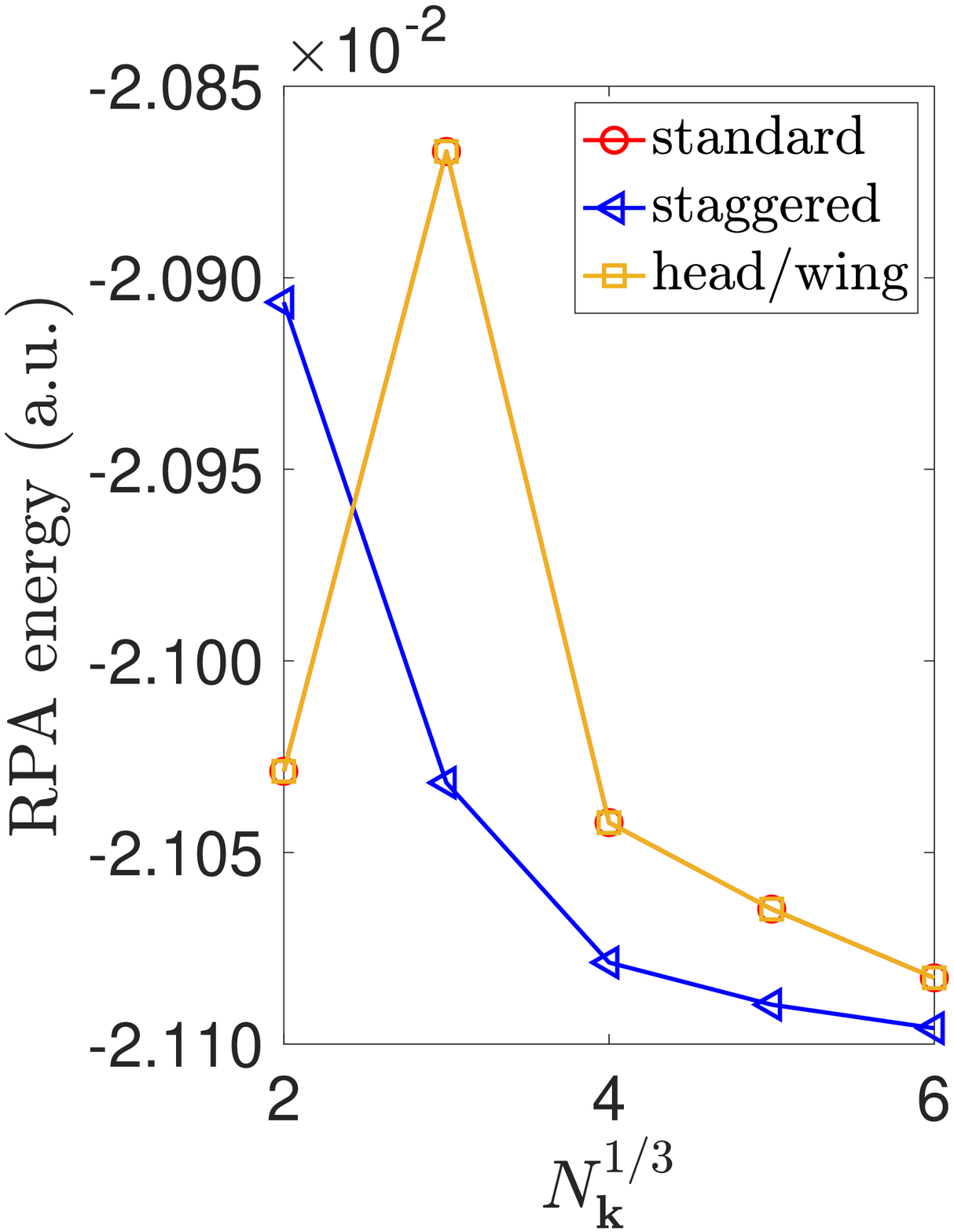}
        }
        \subfloat[LiH gth-szv]{
                \includegraphics[width=0.24\textwidth]{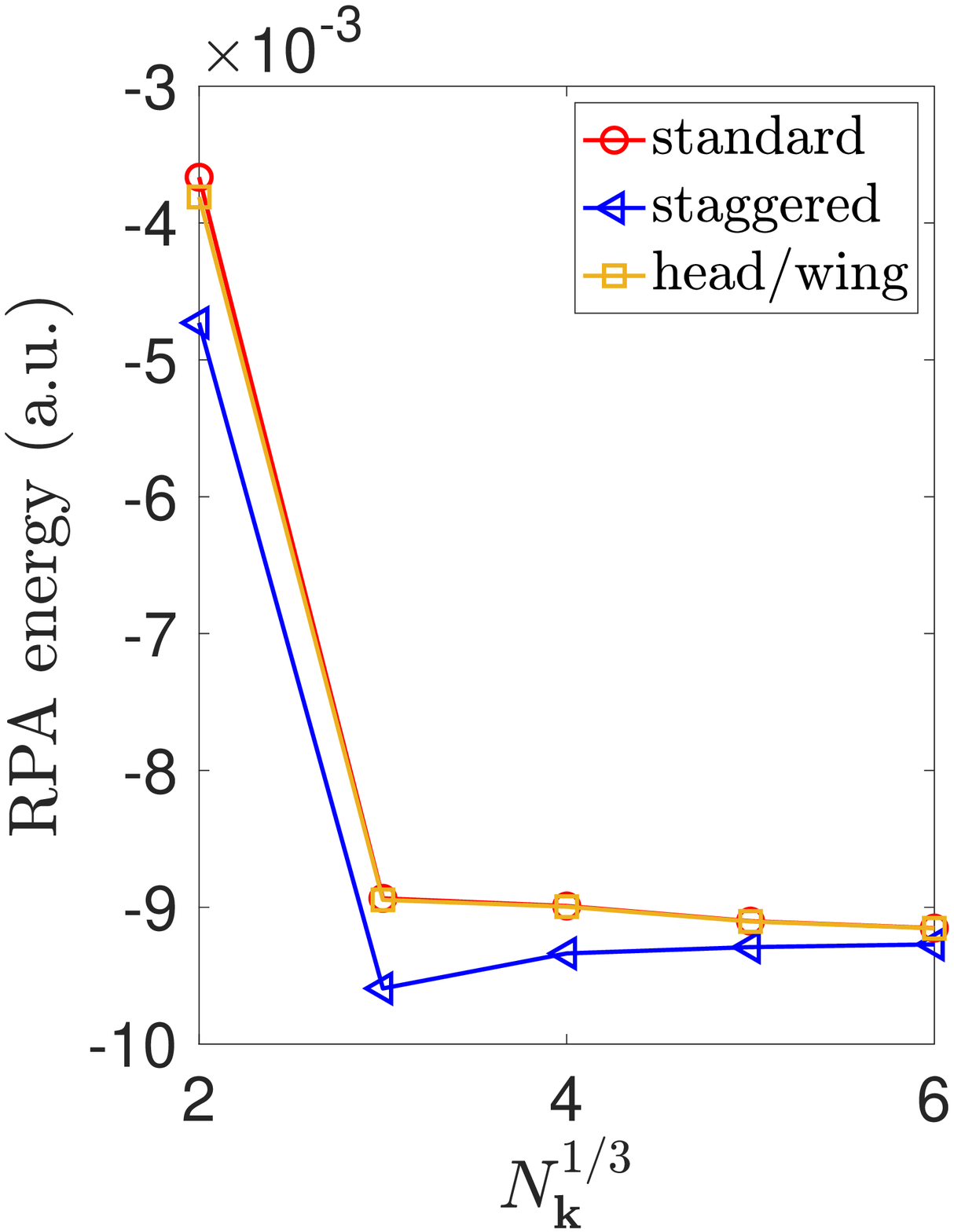}
        }
        \subfloat[Si gth-szv]{
                \includegraphics[width=0.24\textwidth]{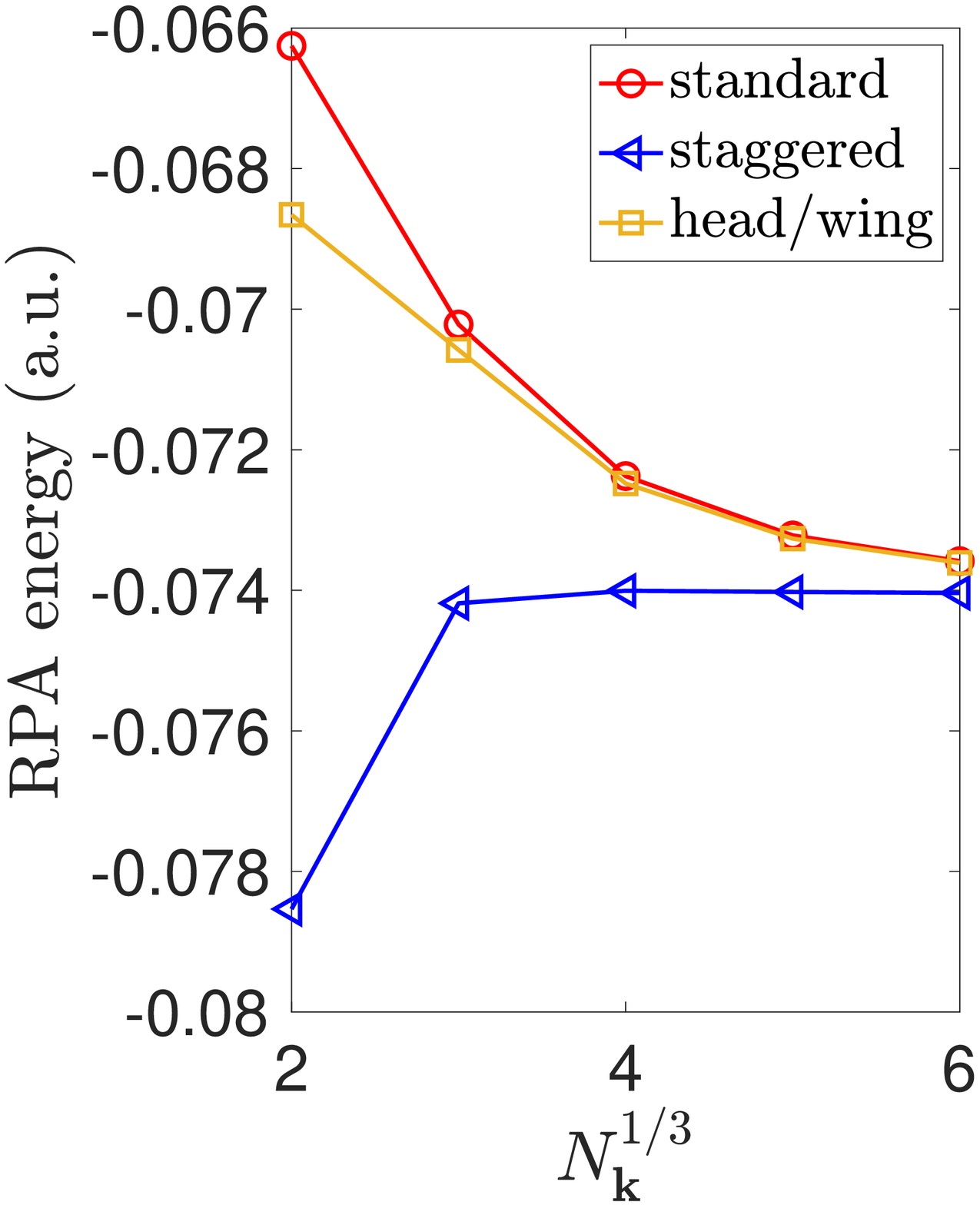}
        }
        \subfloat[Diamond gth-szv]{
                \includegraphics[width=0.24\textwidth]{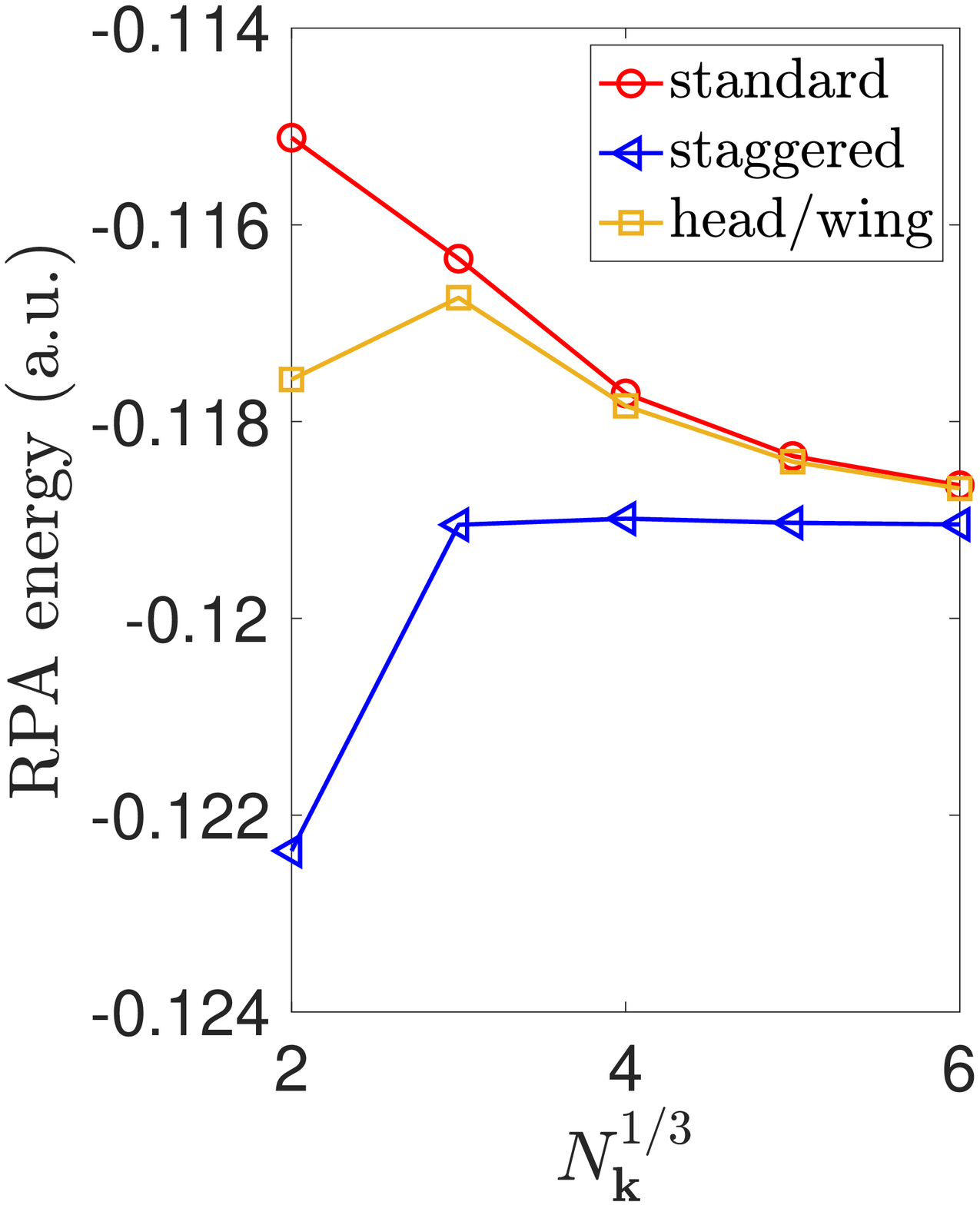}
        }
        
        \subfloat[H2 gth-dzvp]{
                \includegraphics[width=0.24\textwidth]{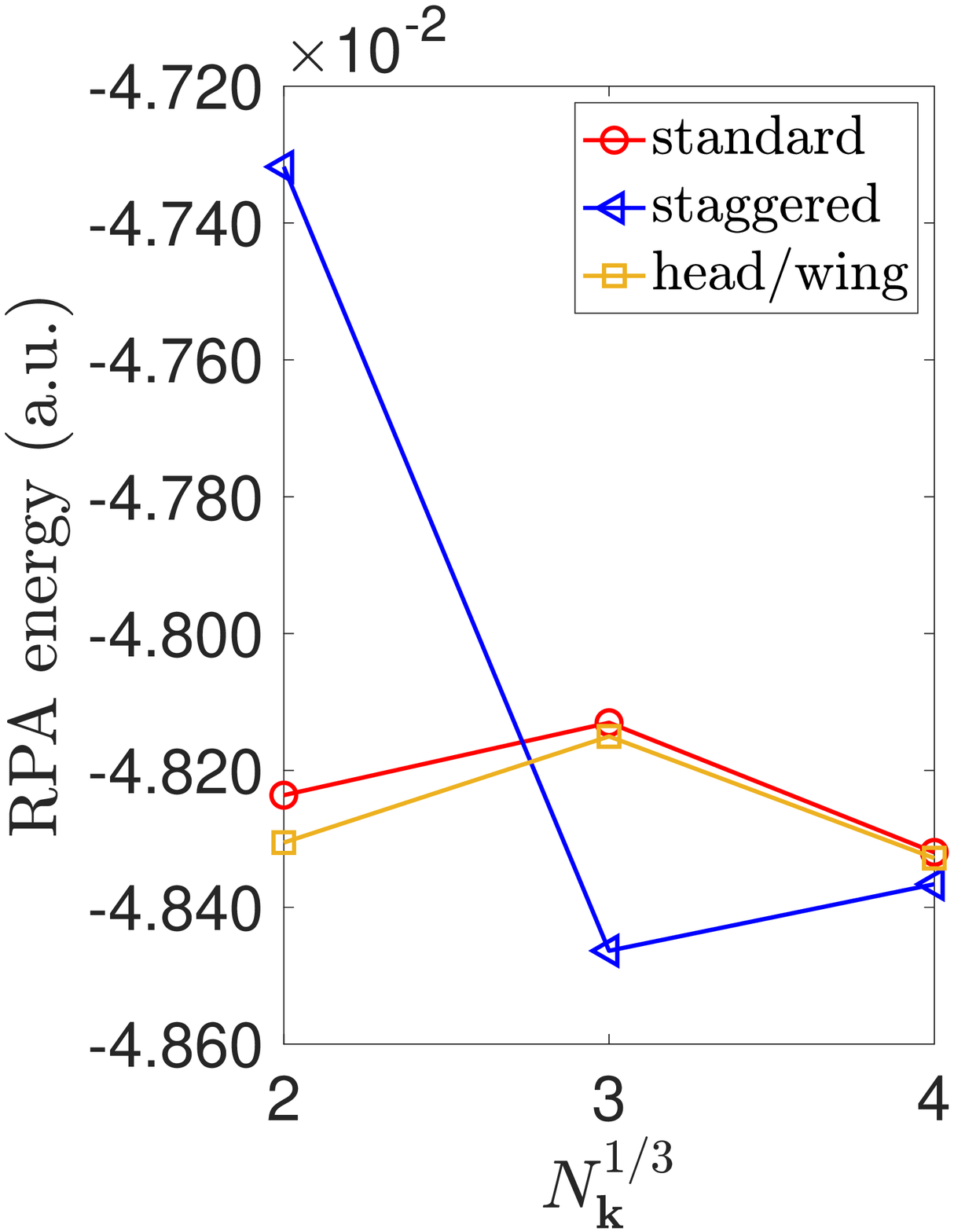}
        }
        \subfloat[LiH gth-dzvp]{
                \includegraphics[width=0.24\textwidth]{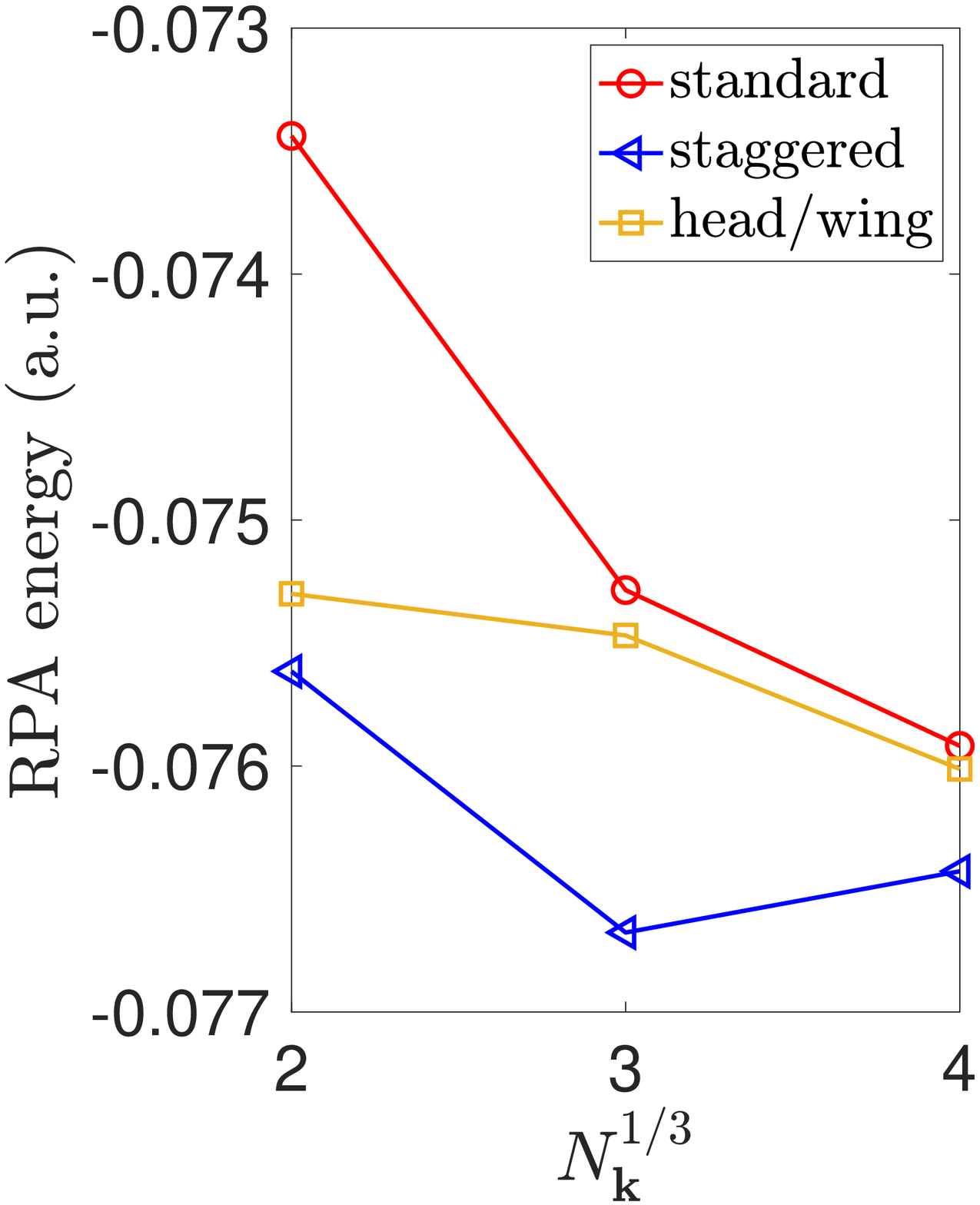}
        }
        \subfloat[Si gth-dzvp]{
                \includegraphics[width=0.24\textwidth]{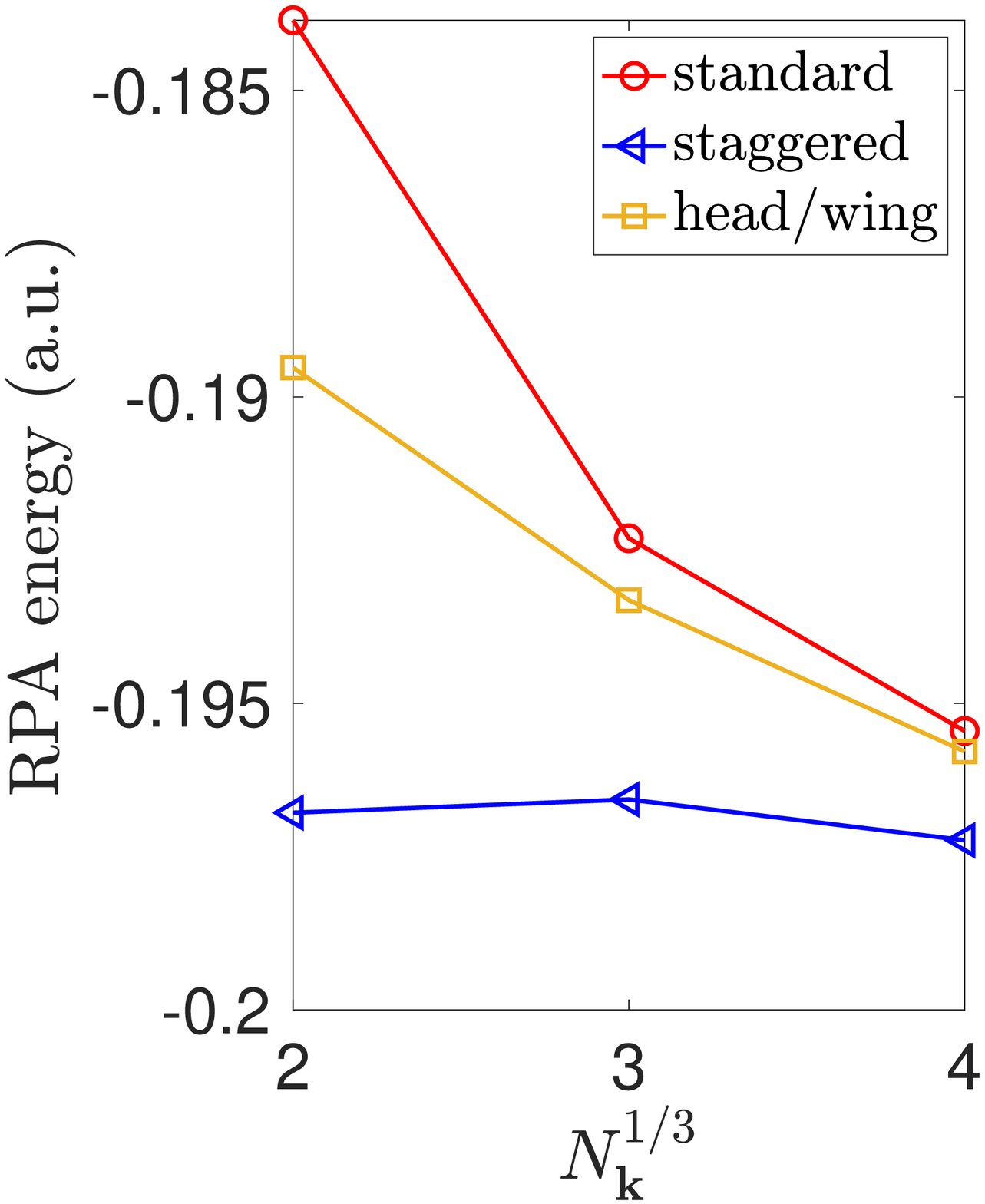}
        }
        \subfloat[Diamond gth-dzvp]{
                \includegraphics[width=0.24\textwidth]{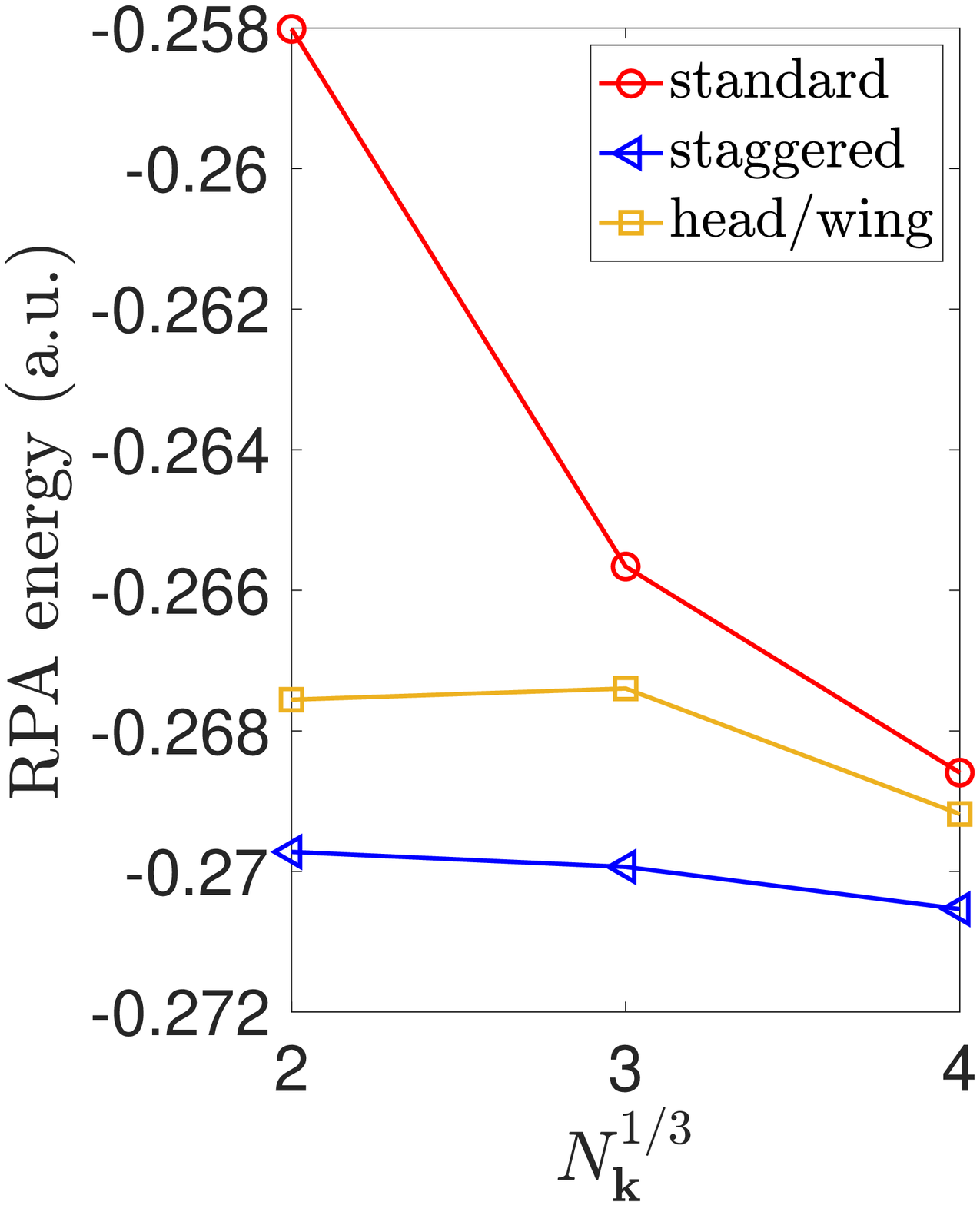}
        }
        
        \caption{
                RPA correlation energy with the head/wing correction for 3D systems using the AC formalism with two different basis sets.
        For the head/wing correction, $\varepsilon_{\bm{0},\bm{0}}(\mathrm{i}\omega, \vq), \varepsilon_{\bm{0},\vG}(\mathrm{i}\omega, \vq)$, and $\varepsilon_{\vG,\bm{0}}(\mathrm{i}\omega, \vq)$ at $\vq = \bm{0}$ are replaced by their values at a single $\vq$ point with relative coordinate $(0.001,0,0)$ in $\Omega^*$ as used in Ref.~\citenum{ZhuChan2021}. 
}
        \label{fig:rpa_correction}
\end{figure}

\Cref{fig:rpa_dzvp} plots the RPA correlation energies computed using the AC formalism for quasi-1D, quasi-2D, and 3D systems with the larger gth-dzvp basis set.
\Cref{fig:rpa_correction} plots the RPA correlation energy with the head/wing correction for 3D systems.

\newpage
\begin{figure}
 \centering
 \includegraphics[width = 0.6\textwidth]{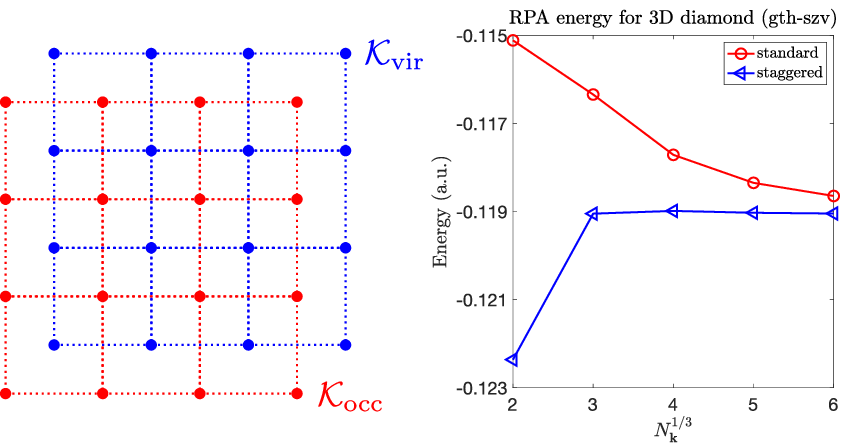}
 \caption{TOC}
 \end{figure}

\bibliography{rpa}

\end{document}